\documentclass[12pt]{iopart}
\usepackage{iopams}
\usepackage{rotating}
\usepackage{epsfig}
\bibstyle{plain}
\newcommand{\be}{\begin{equation}}
\newcommand{\ee}{\end{equation}}
\newcommand{\bea}{\begin{eqnarray}}
\newcommand{\eea}{\end{eqnarray}}
\eqnobysec

\begin{document}
\review[The Casimir Effect]{The Casimir effect: Recent controversies and 
progress}
\author{Kimball A. Milton}
\address{Oklahoma Center for High Energy Physics and
Department of Physics and Astronomy, The University of
Oklahoma, Norman, OK 73019 USA}
\ead{milton@nhn.ou.edu}
\date\today

\begin{abstract}
The phenomena implied by the existence of quantum vacuum fluctuations, grouped
under the title of the Casimir effect, are reviewed, with emphasis on new
results discovered in the past four years.  The Casimir force between
parallel plates is rederived as the strong-coupling limit of $\delta$-function
potential planes.  The role of surface divergences is clarified.  A
summary of effects relevant to measurements of the Casimir force between
real materials is given, starting from a geometrical optics derivation
of the Lifshitz formula, and including a rederivation of the Casimir-Polder
forces.  A great deal of attention is given to the recent controversy
concerning temperature corrections to the Casimir force between real metal
surfaces.  A summary of new improvements to the proximity force approximation
is given, followed by a synopsis of the current experimental situation.
New results on Casimir self-stress are reported, again based on $\delta$-%
function potentials.  Progress in understanding divergences in the
self-stress of dielectric bodies is described, in particular the status
of a continuing calculation of the self-stress of a dielectric cylinder.
Casimir effects for solitons, and the status of the so-called dynamical
Casimir effect, are summarized.  The possibilities of understanding
dark energy, strongly constrained by both cosmological and terrestrial
experiments, in terms of quantum fluctuations are discussed.
Throughout, the centrality of quantum vacuum energy in fundamental physics
in emphasized. 
\end{abstract}

\maketitle

\section{Introduction}
\label{sec:intro}
The essence of quantum physics is fluctuations.  That is, knowing the
position of a particle precisely means losing all knowledge about its
momentum, and vice versa, and generally the product
of uncertainties of a generalized coordinate $q$ and its corresponding
momentum $p$ is bounded below:
\be
\Delta q\Delta p\ge\frac\hbar 2,
\ee
which reflects the fundamental commutation relation
\be
[q,p]=i\hbar.
\ee
The Hamiltonian commutes with neither $q$ nor $p$ in general;
this means that in an energy eigenstate the fluctuations in $q$ and $p$
are both nonzero:
\be
\Delta q> 0, \quad \Delta p>0.
\ee
Moreover, a harmonic oscillator has correspondingly a ground-state energy
which is nonzero:
\be
E_{{\rm ho},n}=\hbar\omega\left(n+\frac12\right).
\ee
The apparent implication of this is that a crystal, which may be thought
of, roughly, as a collection of atoms held in harmonic potentials, should
have a large zero-point energy at zero temperature:
\be
E_{\rm ZP}=\sum_{\rm atoms}\frac12\hbar \omega,
\ee
$\omega$ being the characteristic frequency of each potential.

The vacuum of quantum field theory may similarly be regarded as
an enormously large collection of harmonic oscillators, representing the
fluctuations of, for quantum electrodynamics, the electric and magnetic
fields at each point in space.  (Canonically, the momentum-coordinate
pair correspond to the electric field and the vector potential.)  Put 
otherwise, the QED vacuum is a sea of virtual photons.  Thus the zero-point
energy density of the vacuum is
\be
U=\sum \frac12\hbar\omega=2\int\frac{(d\bi{k})}{(2\pi)^3}
\frac12\hbar c|\bi{k}|,
\ee
where $\bi{k}$ is the wavevector of the photon, and the factor of 2
reflects the two polarization states of the photon.   

This is an enormously large quantity.  If we say that the largest wavevector
appearing in the integral is $K$, say $\hbar cK\sim 10^{19}$ GeV,
the Planck scale, then $U\sim 10^{115}$ GeV/cm$^3$.  So it is no surprise
that Dirac suggested that this zero-point energy be simply discarded, as
some irrelevant constant \cite{dirac34} (yet he became increasingly concerned about the
inconsistency of doing so throughout his life
\cite{diracbook}).  Pauli recognized that this
energy surely coupled to gravity, and it would then give rise to a large
cosmological constant, so large that the size of the universe could not
even reach the distance to the moon \cite{straumann02,straumann02a}.  
This cosmological constant problem
is with us to the present \cite{weincc,weinccc}.  
But this was not the most perplexing issue
confronting quantum electrodynamics in the 1930s.

Renormalization theory, that is, a consistent theory of quantum 
electrodynamics, was invented first by Schwinger
\cite{Schwinger:1948iu} and then Feynman \cite{Feynman:1948fi} in 1948;
yet remarkably, across the Atlantic, Casimir in the same year predicted the
direct macroscopically observable consequence of vacuum fluctuations
that now bears his name \cite{casimir}.  
This is the attraction between parallel 
uncharged conducting plates that has been so convincingly demonstrated
by many experiments in the last few years \cite{newreview}.  
Lifshitz and his group generalized
the theory to include dielectric materials in the 1950s
\cite{lifshitz,dzyaloshinskii0,dzyaloshinskii,landauandlifshitz}. 
There were many
experiments to detect the effect in the 1950s and 1960s, but most were 
inconclusive, because the forces were so small, and it was very difficult
to keep various interfering phenomena from washing out the effect
\cite{sparnaayrev}.  However,
there could be very little doubt of the reality of the phenomenon, since it
was intimately tied to the theory of van der Waals forces between
molecules, the retarded
version of which had been worked out by Casimir 
\cite{casimirandpolder} just before he discovered
(with a nudge from Bohr \cite{casimir50}) the force between plates.  
Finally, in 1973,
the Lifshitz theory was vindicated by an experiment by Sabisky and Anderson
\cite{sabisky}.

But by and large field theorists were unaware of the effect until Glashow's
student Boyer carried out a remarkable calculation of the Casimir self-energy
of a perfectly conducting spherical shell in 1968 \cite{boyersphere}.  
Glashow was aware of
Casimir's proposal \cite{casimir2} that a classical electron could be stablized by zero-point
attraction, and thought the calculation made a suitable thesis project.
Boyer's result was a surprise: The zero-point force was repulsive for the
case of a sphere.  Davies improved on the calculation \cite{davis}; 
then a decade later
there were two independent reconfirmations of Boyer's result, one based
on multiple scattering techniques \cite{balian}
(now undergoing a renaissance, for example, see \cite{Jaffe:2003mb}) and one
on Green's functions techniques \cite{mildersch} (dubbed source theory
\cite{schwinger}).  Applications to
hadronic physics followed in the next few years
\cite{johnsondpf,miltonbag,miltoncond}, and in the last two decades,
there has been something of an explosion of interest in the field, with
many different calculations being carried out \cite{miltonbook,newreview}.

However, fundamental understanding has been very slow in coming.  Why is
the cosmological constant neither large nor zero?  Why is the Casimir force
on a sphere repulsive, when it is attractive between two plates?  And is
it possible to make sense of Casimir force calculations between two
bodies, or of the Casimir self-energy of a single body, in terms of supposedly
better understood techniques of perturbative quantum field theory
\cite{Graham:2003ib}?
As we will see, none of these questions yet has a definitive answer, yet
progress has been coming.  Even the temperature corrections to the Casimir
effect, which were considered by Sauer \cite{sauer}, Mehra \cite{mehra}, 
and Lifshitz \cite{lifshitz} in the 1950s and 1960s,
have become controversial \cite{Svet,Svet2,bostrom,bostrom2,sernelius01,
sernelius01a,klim2,Bordag:2001as,genet,lamoreaux01,klim01,Brevik:2002bi,
bezerra66,genetijmpa,reynaudqfext}.  
Thus recent conferences on the Casimir effect
have been quite exciting events \cite{itamp,QFEXT03}.  
It is the aim of the present review
to bring the various issues into focus, and suggest paths toward the solutions
of the difficulties.  It is a mark of the vitality and even centrality of
this field that such a review is desirable on the heels of two significant
meetings on the subject, and less than three years after the appearance of
two major monographs \cite{newreview,miltonbook} on Casimir phenomena.
There are in addition a number of earlier, excellent reviews
\cite{mostbook,krech,milonni}, as well as more specialized
treatments \cite{Od,El,kirstenbook,Vassilevich:2003xt}.
Throughout this review Gaussian units are employed.

This review is organized in the following manner.
In \sref{Sec1} we compute Casimir energies and pressures between
parallel $\delta$ function planes, which in the limit of large
coupling reproduce the results for a scalar field satisfying Dirichlet
boundary conditions on those surfaces. 
 Although these results have
been described before, clarification of the nature of surface energy and
divergences is provided. TM modes are also discussed here for
the first time. Then, in \sref{secreal}
we rederive the Lifshitz formula for the Casimir force between parallel
dielectric slabs using a multiple reflection technique.  The Casimir-Polder
forces between two atoms, and between an atom and a plate, are rederived.
After reviewing roughness and conductivity corrections, a detailed
discussion of the temperature controversy is given, with the conclusion that the
TE zero-mode absence must be taken seriously, which will imply that
large temperature corrections should be seen experimentally.
New approaches to moving beyond the proximity approximation in computing
forces between nonparallel plane surface are reviewed.  A discussion
of the remarkable progress experimentally since 1997 is provided.
In \sref{sec:selfstress} after a review of the general situation with respect
to surface divergences, TE and TM forces on $\delta$-function spheres
are described in detail, which in the limit of strong coupling reduce
to the corresponding finite electromagnetic contributions.  For weak coupling,
Casimir energies are finite in second order in the coupling strength,
but divergent in third order, a fact which has been known for several
years.  This mirrors the corresponding result for a dilute dielectric
sphere, which diverges in third order in the deviation of the
permittivity from its vacuum value.  Self-stresses on cylinders are
also treated, with a detailed discussion of the status of a new calculation
for a dielectric cylinder, which should give a vanishing self-stress in
second order in the relative permittivity.  \Sref{sec:solitions} briefly
summarizes recent work on quantum fluctuation phenomena in solitonic
physics, which has provided the underlying basis for much of the interest 
in Casimir phenomena over the years.  Dynamical Casimir effects,
ranging from sonoluminescence through the Unruh effect, are the subject
of \sref{sec:dce}.  The presumed basis for understanding the cosmological
dark energy in terms of the Casimir fluctuations is treated in
\sref{sec:cc}, where there may be a tight constraint emerging between
terrestrial measurements of deviations from Newtonian gravity and
the size of extra dimensions.  The review ends with a summary of
perspectives for the future of the field.

\section{Casimir Effect Between Parallel Plates: A $\delta$-Potential
Derivation}\label{Sec1}

In this section, we will rederive the classic Casimir result for the
force between parallel conducting plates \cite{casimir}.  Since the
usual Green's function derivation may be found in monographs \cite{miltonbook},
and was recently reviewed in connection with current controversies over
finiteness of Casimir energies \cite{Milton:2002vm}, we will here present
a different approach, based on $\delta$-function potentials, which in the
limit of strong coupling reduce to the appropriate Dirichlet or Robin
boundary conditions of a perfectly conducting surface, as appropriate to
TE and TM modes, respectively.  Such potentials were first considered
by the Leipzig group  \cite{hennig,bkv}, but recently have been the focus
of the program of the MIT group \cite{graham2,Graham:2002fw,Graham:2003ib}.
The discussion here is based on a recent paper by the author 
\cite{Milton:2004vy}.  We first consider two $\delta$-function potentials in
$1+1$ dimensions.

\subsection{$1+1$ dimensions}
\label{Sec1.1}
We consider a massive scalar field (mass $\mu$)
 interacting with two $\delta$-function
potentials, one at $x=0$ and one at $x=a$, which has an interaction
Lagrange density
\be
\mathcal{L}_{\rm int}=-\frac12\frac{\lambda}a\delta(x)\phi^2(x)
-\frac12\frac{\lambda'}a\delta(x-a)\phi^2(x),
\label{oneplusonelag}
\ee
where we have chosen the coupling constants $\lambda$ and $\lambda'$
to be dimensionless.  (But see the following.)  In the limit as both
couplings become infinite, these potentials enforce Dirichlet boundary
conditions at the two points:
\be
\lambda,\lambda'\to\infty:\qquad \phi(0),\phi(a)\to0.
\ee
  
The Casimir energy for this
situation may be computed in terms of the Green's function $G$,
\be
G(x,x')=\rmi\langle T\phi(x)\phi(x')\rangle,
\label{fgf}
\ee
which has a time Fourier transform,
\be
G(x,x')=\int\frac{\rmd\omega}{2\pi}\rme^{-\rmi\omega(t-t')}g(x,x';\omega).
\label{ft}
\ee
Actually, this is a somewhat symbolic expression, for the Feynman Green's
function (\ref{fgf}) implies that the frequency contour of integration
here must pass below the singularities in $\omega$ on the negative real
axis, and above those on the positive real axis \cite{kantowski,Brevik:2000hk}.
The reduced Green's function in (\ref{ft}) in turn satisfies
\be
\left[-\frac{\partial^2}{\partial x^2}+\kappa^2+\frac{\lambda}a\delta(x)
+\frac{\lambda'}a\delta(x-a)\right]g(x,x')=\delta(x-x').
\ee
Here $\kappa^2=\mu^2-\omega^2$.
This equation is easily solved, with the result
\numparts
\label{gee0}
\bea
g(x,x')=\frac1{2\kappa}e^{-\kappa|x-x'|}+
\frac1{2\kappa\Delta}\Bigg[\
\frac{\lambda\lambda'}{(2\kappa a)^2}2\cosh\kappa|x-x'|\nonumber\\
\mbox{}-\frac{\lambda}{2\kappa a}\left(1+\frac{\lambda'}
{2\kappa a}\right)\rme^{2\kappa a}
\rme^{-\kappa(x+x')}-\frac{\lambda'}{2\kappa a}\left(1+\frac{\lambda}
{2\kappa a}\right)\rme^{\kappa(x+x')}\Bigg]
\label{gin}
\eea
for both fields inside, $0<x,x'<a$, while if both field points are outside,
$a<x,x'$, 
\bea
\fl g(x,x')=\frac1{2\kappa}\rme^{-\kappa|x-x'|}+\frac1{2\kappa\Delta}\rme^{-\kappa
(x+x'-2a)}\left[-\frac{\lambda}{2\kappa a}\left(1-\frac{\lambda'}
{2\kappa a}\right)
-\frac{\lambda'}{2\kappa a}\left(1+\frac{\lambda}
{2\kappa a}\right)\rme^{2\kappa a}\right].
\label{gout}\nonumber\\
\eea
For $x,x'<0$,
\bea
\fl g(x,x')=\frac1{2\kappa}\rme^{-\kappa|x-x'|}+\frac1{2\kappa\Delta}
\rme^{\kappa(x+x')}\left[-\frac{\lambda'}{2\kappa a}\left(1-\frac{\lambda}
{2\kappa a}\right)
-\frac{\lambda}{2\kappa a}\left(1+\frac{\lambda'}
{2\kappa a}\right)\rme^{2\kappa a}\right].\nonumber\\
\label{gleft}
\eea
\endnumparts
Here, the denominator is
\be
\Delta=\left(1+\frac{\lambda}{2\kappa a}\right)\left(1+\frac{\lambda'}
{2\kappa a}\right)\rme^{2\kappa a}-\frac{\lambda\lambda'}{(2\kappa a)^2}.
\ee
Note that in the strong coupling limit we recover the familiar results, 
for example, inside
\be
\lambda,\lambda'\to\infty:
\quad g(x,x')\to-\frac{\sinh\kappa x_<\sinh\kappa(x_>-a)}
{\kappa\sinh\kappa a}.
\ee
Evidently, this Green's function vanishes at $x=0$ and at $x=a$.

\begin{figure}
\begin{center}
\begin{turn}{270}
\epsfig{figure=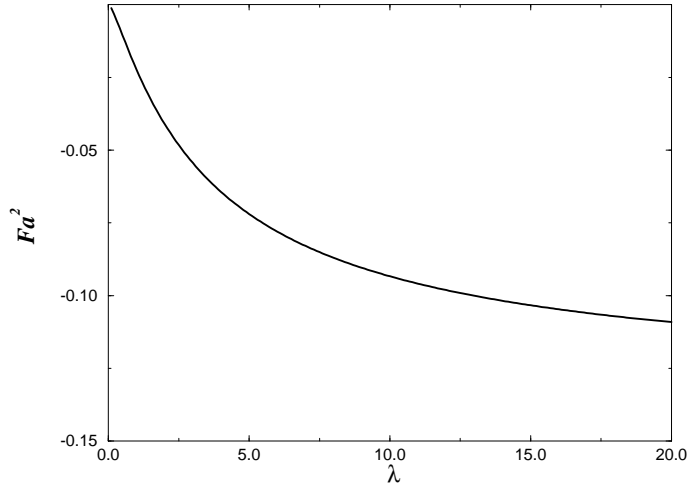,height=10cm}
\end{turn}
\end{center}
\caption{Casimir force (\ref{cepoints})
 between two $\delta$-function points having strength
$\lambda$ and separated by a distance $a$.}
\label{fig1}
\end{figure}

We can now calculate the force on one of the $\delta$-function points by
calculating the discontinuity of the stress tensor,  obtained from the
Green's function (\ref{fgf}) by
\be
\langle T^{\mu\nu}\rangle=\left(\partial^\mu\partial^{\nu\prime}-\frac12
g^{\mu\nu}\partial^\lambda\partial'_\lambda\right)
\frac1{\rmi}G(x,x')\bigg|_{x=x'}.
\ee
Writing a reduced stress tensor by 
\be
\langle T^{\mu\nu}\rangle=\int\frac{\rmd\omega}{2\pi} t^{\mu\nu},
\ee
we find inside
\bea
t_{xx}=\frac1{2\rmi}(\omega^2+\partial_x\partial_{x'})g(x,x')
\bigg|_{x=x'}\nonumber\\
=\frac1{4\rmi\kappa\Delta}
\Bigg\{(2\omega^2-\mu^2)\left[\left(1+\frac{\lambda}
{2\kappa a}\right)\left(1+\frac{\lambda'}{2\kappa a}\right)
\rme^{2\kappa a}+\frac{\lambda\lambda'}{(2\kappa
a)^2}\right]\nonumber\\
\mbox{}-\mu^2\left[\frac{\lambda}{2\kappa a}\left(1+\frac{\lambda'}
{2\kappa a}\right)
\rme^{-2\kappa(x- a)}+\frac{\lambda'}{2\kappa a}\left(1+\frac{\lambda}
{2\kappa a}\right)\rme^{2\kappa x}\right]\Bigg\}.
\label{parstin}
\eea
Let us henceforth simplify the considerations by taking the massless limit,
$\mu=0$.  Then the stress tensor just to the left of the point $x=a$ is
\numparts
\be
t_{xx}\bigg|_{x=a-}=-\frac\kappa{2\rmi}\left\{
1+2\left[\left(\frac{2\kappa a}\lambda+1\right)\left(\frac{2\kappa
a}{\lambda'}+1\right)\rme^{2\kappa a}-1\right]^{-1}\right\}.
\label{stressin}
\ee
From this we must subtract the stress just to the right of the point at
$x=a$, obtained from (\ref{gout}), which turns out to be in the massless
limit
\be
t_{xx}\bigg|_{x=a+}=-\frac\kappa{2\rmi},\label{stressout}
\ee
\endnumparts
which just cancels the 1 in braces in (\ref{stressin}).
Thus the force on the point $x=a$ due to the quantum fluctuations in the scalar
field is given by the simple, finite expression
\be
\fl F=\langle T_{xx}\rangle\bigg|_{x=a-}-\langle T_{xx}\rangle\bigg|_{x=a+}
=-\frac1{4\pi a^2}\int_0^\infty
\rmd y\,y\,\frac1{(y/\lambda+1)(y/\lambda'+1)\rme^y-1}.
\label{cepoints}
\ee
This reduces to the well-known, L\"uscher result \cite{luscher,luscher2}
in the limit $\lambda,\lambda'\to
\infty$,
\be
\lim_{\lambda=\lambda'\to\infty}F=-\frac\pi{24a^2},
\label{eq:luscher}
\ee
and for $\lambda=\lambda'$ is plotted in Fig.~\ref{fig1}.

Recently, Sundberg and Jaffe \cite{sundberg} have used their background
field method to calculate the Casimir force due to fermion fields 
between two $\delta$-function spikes in $1+1$ dimension. Apart from quibbles
about infinite energies, in the limit $\lambda\to\infty$ they recover the
same result as for scalar, (\ref{eq:luscher}), which is as expected
\cite{johnson}, since in the ideal limit the relative factor between
scalar and spinor energies is $2(1-2^{-D})$ in $D$ spatial dimensions,
i.e., 7/4 for three dimensions and 1 for one.

We can also compute the energy density.  In this simple 
massless case, the calculation
appears identical, because $t_{xx}=t_{00}$ (reflecting the
conformal invariance of the free theory).  
The energy density is constant [(\ref{parstin}) with $\mu=0$]
and subtracting from it the $a$-independent part that would be present if
no potential were present, we immediate see that the total energy is
$E=Fa$, so $F=-\partial E/\partial a$.  This result  differs
significantly from that given in 
Refs.~\cite{Graham:2002fw,graham2,Jaffe:2003ji}, which is a 
divergent expression in the massless limit, not  transformable into the 
expression found by this naive procedure.  However, that result may be easily
derived from the following expression for the total energy,
\bea
E&=\int(\rmd\bi{r})\,\langle T^{00}\rangle=\frac1{2\rmi}\int(\rmd\bi{r})
(\partial^0\partial^{\prime0}-\nabla^2)G(x,x')\bigg|_{x=x'}\nonumber\\
&=\frac1{2\rmi}\int(\rmd\bi{r})\int\frac{\rmd\omega}{2\pi}
2\omega^2\mathcal{G}(\bi{r,r}),
\label{casenergy}
\eea  
if we integrate by parts and omit the surface term.
Integrating over the Green's functions in the three regions,
given by (\ref{gin}), (\ref{gout}), and (\ref{gleft}), we obtain for
$\lambda=\lambda'$,
\be
E=\frac1{2\pi a}\int_0^\infty \rmd y\frac1{1+y/\lambda}-\frac1{4\pi a}
\int_0^\infty \rmd y\,y\frac{1+2/(y+\lambda)}{(y/\lambda+1)^2\rme^y-1},
\label{11energy}
\ee
where the first term is regarded as an irrelevant constant ($\lambda/a$ is
constant), and the second is the same as that given by equation~(70) of
Ref.~\cite{graham2} upon integration by parts.

The origin of this discrepancy with the naive energy
is the existence of a surface contribution
to the energy.  Because $\partial_\mu T^{\mu\nu}=0$, we have, for a region
$V$ bounded by a surface $S$,
\be
0=\frac{\rmd}{\rmd t}\int_V (\rmd\bi{r}) T^{00}+\oint_S \rmd S_i T^{0i}.
\ee
Here $T^{0i}=\partial^0\phi\partial^i\phi$, so we conclude that there is
an additional contribution to the energy,
\numparts
\bea
E_s&=-\frac1{2i}\int \rmd\bi{S}\cdot\bnabla G(x,x')\bigg|_{x'=x}
\label{es1}\\
&=-\frac1{2i}\int_{-\infty}^\infty\frac{\rmd\omega}{2\pi}\sum\frac{\rmd}{\rmd x}
g(x,x')\bigg|_{x'=x},
\label{es2}
\eea
\endnumparts
where the derivative is taken at the boundaries (here $x=0$, $a$) in the
sense of the outward normal from the region in question.  When this surface
term is taken into account the extra terms in (\ref{11energy}) 
are supplied.  The integrated formula (\ref{casenergy}) 
automatically builds in this 
surface contribution, as the implicit surface term in the integration
by parts.  (These terms are slightly unfamiliar because they do not arise
in cases of Neumann or Dirichlet boundary conditions.)  See Fulling 
\cite{Fulling:2003zx} for further discussion.  That the surface
energy of an interface arises from the volume energy of a smoothed
interface is demonstrated in Ref.~\cite{Milton:2004vy}, and elaborated
in \sref{sec:2.4}.

It is interesting to
consider the behavior of the force or energy for small coupling $\lambda$.
It is clear that, in fact, (\ref{cepoints}) is not analytic at $\lambda=0$.
(This reflects an infrared divergence in the Feynman diagram calculation.)
If we expand out the leading $\lambda^2$ term we are left with a divergent
integral.  A correct asymptotic evaluation leads to the behavior
\be
F\sim \frac{\lambda^2}{4\pi a^2}\left(\ln 2\lambda+\gamma\right),\qquad 
E\sim-\frac{\lambda^2}{4\pi a}(\ln2\lambda+\gamma-1),\quad
\lambda\to 0.\label{paragsmall}
\ee
This behavior indeed was anticipated in earlier perturbative analyses.
In Ref.~\cite{Milton:2002vm} the general result was given for the Casimir
energy for a $D$ dimensional spherical $\delta$-function potential
(a factor of $1/4\pi$ was inadvertently omitted)
\be
E=-\frac{\lambda^2}{\pi a}
\frac{\Gamma\left(\frac{D-1}2\right)\Gamma(D-3/2)
\Gamma(1-D/2)}{2^{1+2D}[\Gamma(D/2)]^2}.\label{Ed}
\ee
This possesses an infrared divergence as $D\to1$:
\be
E^{(D=1)}=\frac{\lambda^2}{4\pi a}\Gamma(0),\label{d=1div}
\ee
which is consistent with the nonanalytic behavior seen in (\ref{paragsmall}).

\subsection{Parallel Planes in $3+1$ Dimensions}
\label{Sec:2.5}
It is trivial to extract the expression for the Casimir pressure between two
$\delta$ function planes in three spatial dimensions, where the background
lies at $x=0$ and $x=a$.  We merely have to insert into the above a
transverse momentum transform,
\be
G(x,x')=\int\frac{\rmd\omega}{2\pi}\rme^{-\rmi\omega(t-t')}
\int\frac{(\rmd\bi{k})}
{(2\pi)^2}\rme^{\rmi\bi{k\cdot(r-r')_\perp}}g(x,x';\kappa),
\label{redgreen}
\ee
where now $\kappa^2=\mu^2+k^2-\omega^2$.  Then $g$ has exactly the same form
as in (\ref{gin})--(\ref{gleft}).
The reduced stress tensor is given by, for the massless case,
\be
t_{xx}=\frac12(\partial_x\partial_{x'}-\kappa^2)\frac1\rmi 
g(x,x')\bigg|_{x=x'},\label{txxfromg}
\ee
so we immediately see that the attractive pressure on the planes is given by
($\lambda=\lambda'$)
\be
P=-\frac1{32\pi^2 a^4}\int_0^\infty \rmd y\,y^3\,\frac1{(y/\lambda+1)^2
\rme^y-1},
\label{31pressure}
\ee
which coincides with the result given in 
Refs.~\cite{Graham:2003ib,Weigel:2003tp}.  The leading behavior
for small $\lambda$ is
\numparts
\be
P^{\rm TE}\sim-\frac{\lambda^2}{32\pi^2 a^4},\qquad \lambda\ll 1,
\label{pertte}
\ee
while for large $\lambda$ it approaches half of Casimir's result \cite{casimir}
for perfectly conducting parallel plates,
\be
P^{\rm TE}\sim -\frac{\pi^2}{480 a^4},\qquad\lambda\gg1.
\label{strongte}
\ee
\endnumparts

The Casimir energy per unit area again might be expected to be
\be
\mathcal{E}=-\frac1{96\pi^2a^3}\int_0^\infty \rmd y\frac{y^3}{(y/\lambda+1)^2
\rme^y-1}=\frac13\frac{P}a,
\label{naivee}
\ee
because then $P=-\frac\partial{\partial a}\mathcal{E}$.  In fact, however,
it is straightforward to compute the energy density $\langle T^{00}\rangle$ 
is the three
regions, $x<0$, $0<x<a$, and $a<x$, and then integrate it over $x$ to
obtain the energy/area, which differs from (\ref{naivee}) because, now,
there exists transverse momentum. We also must include the surface term
(\ref{es1}), which is of opposite sign, and of double magnitude, to the
$k^2$ term. The net extra term is
\be
\mathcal{E}'=\frac1{48\pi^2a^3}\int_0^\infty \rmd y\,y^2\frac1{1+y/\lambda}
\left[1-\frac{y/\lambda}{(y/\lambda+1)^2\rme^y-1}\right].
\ee
If we regard $\lambda/a$ as constant (so that the strength of the coupling
is independent of the separation between the planes) we may drop the first,
divergent term here as irrelevant, being independent of $a$,
because $y=2\kappa a$,  and then the
total energy is
\be
\mathcal{E}=-\frac1{96\pi^2a^3}\int_0^\infty \rmd y\,y^3\frac{1+2/(\lambda+y)}
{(y/\lambda+1)^2\rme^y-1},
\label{31energy}
\ee
which coincides with the massless limit of the energy first found by Bordag
\etal~\cite{hennig}, and given in Refs.~\cite{Graham:2003ib,Weigel:2003tp}.  
As noted in \sref{Sec1.1}, this result may also readily be derived through
use of (\ref{casenergy}).  When differentiated with respect to $a$,
(\ref{31energy}), with $\lambda/a$ fixed, yields the pressure 
(\ref{31pressure}).

In the limit of strong coupling, we obtain
\be
\lim_{\lambda\to\infty}\mathcal{E}=-\frac{\pi^2}{1440 a^3},
\label{1/2casimir}
\ee
which is exactly one-half the energy found by Casimir for 
perfectly conducting plates \cite{casimir}.
Evidently, in this case, the TE modes (calculated here) and
the TM modes (calculated in the following subsection) give equal contributions.

\subsection{TM Modes}
\label{sec:tm11}
To verify this claim, we solve a similar problem with boundary conditions
that the derivative of $g$ is continuous at $x=0$ and $a$,
\numparts
\be
\frac\partial{\partial x}g(x,x')\bigg|_{x=0,a} \mbox{ is continuous},
\label{tmbc1}
\ee
but the function itself is discontinuous,
\be
g(x,x')\bigg|_{x=a-}^{x=a+}=\lambda a \frac\partial{\partial x}g(x, x')\bigg|
_{x=a},
\label{tmbc2}
\ee
and similarly at $x=0$.
These boundary conditions reduce, in the limit of strong coupling, to
Neumann boundary conditions on the planes, appropriate to electromagnetic
TM modes:
\be
\lambda\to\infty:\qquad\frac\partial{\partial x}g(x,x')\bigg|_{x=0,a}=0.
\label{nbc}
\ee
\endnumparts

It is completely straightforward to work out the reduced Green's function
in this case.  When both points are between the planes, $0<x,x'<a$,
\numparts
\bea
g(x,x')=\frac1{2\kappa}\rme^{-\kappa|x-x'|}+\frac1{2\kappa\tilde \Delta}
\Bigg\{\left(\frac{\lambda \kappa a}2\right)^2 2\cosh\kappa(x-x')\nonumber\\
\mbox{}+
\frac{\lambda \kappa a}2\left(1+\frac{\lambda\kappa a}2\right)
\left[
\rme^{\kappa(x+x')}+\rme^{-\kappa(x+x'-2a)}\right]\Bigg\},
\eea
while if both points are outside the planes, $a<x,x'$,
\bea
g(x,x')=\frac1{2\kappa}\rme^{-\kappa|x-x'|}\nonumber\\
\mbox{}+\frac1{2\kappa\tilde\Delta}\frac{\lambda\kappa a}2
\rme^{-\kappa(x+x'-2a)}\left[\left(1-\frac{\lambda\kappa a}2\right)+
\left(1+\frac{\lambda\kappa a}2\right)\rme^{2\kappa a}\right],
\eea
\endnumparts
where the denominator is
\be
\tilde\Delta=\left(1+\frac{\lambda\kappa a}2\right)^2\rme^{2\kappa a}-
\left(\frac{\lambda\kappa a}2\right)^2.
\ee

It is easy to check that in the strong-coupling limit, the appropriate
Neumann boundary condition (\ref{nbc}) is recovered.  For example, in the
interior region, $0<x,x'<a$,
\be
\lim_{\lambda\to\infty}g(x,x')=\frac{\cosh\kappa x_<\cosh\kappa(x_>-a)}{\kappa
\sinh\kappa a}.
\ee

\begin{figure}
\begin{center}
\begin{turn}{270}
\epsfig{figure=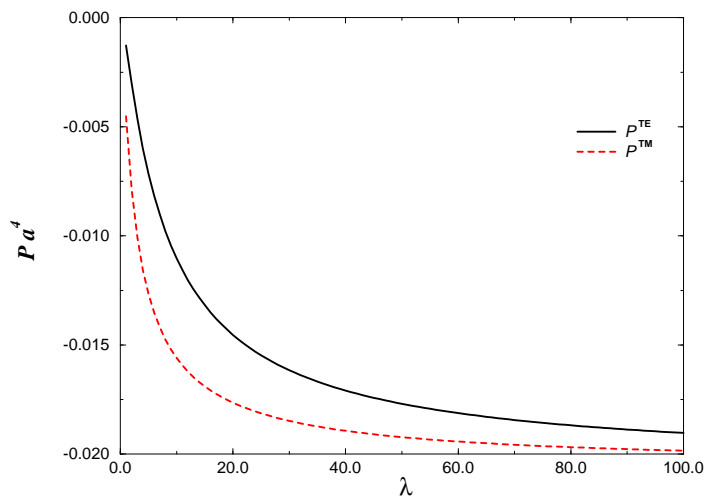,height=10cm}
\end{turn}
\end{center}
\caption{TE and TM Casimir pressures between
 $\delta$-function planes having strength
$\lambda$ and separated by a distance $a$.}
\label{fig2}
\end{figure}

Now we can compute the pressure on the plane by computing the $xx$ component
of the stress tensor,  which is given by (\ref{txxfromg}),
\be
t_{xx}=\frac1{2\rmi}(-\kappa^2+\partial_x\partial'_x) g(x,x')\bigg|_{x=x'}.
\ee
The action of derivatives on exponentials is very simple, so we find
\numparts
\bea
t_{xx}\bigg|_{x=a-}=\frac1{2\rmi}\left[-\kappa-\frac{2\kappa}{\tilde\Delta}
\left(\frac{\lambda\kappa a}2\right)^2\right],\\
t_{xx}\bigg|_{x=a+}=-\frac1{2\rmi}\kappa,
\eea
\endnumparts
so the flux of momentum deposited in the plane $x=a$ is
\be
t_{xx}\bigg|_{x=a-}-t_{xx}\bigg|_{x=a+}=\frac{\rmi \kappa}{\left(\frac2{\lambda
\kappa a}+1\right)^2\rme^{2\kappa a}-1},
\ee
and then by integrating over frequency and transverse momentum we obtain
the pressure:
\be
P^{\rm TM}=-\frac1{32\pi^2 a^4}\int_0^\infty \rmd y\,y^3\frac1{
\left(\frac4{\lambda y}+1\right)^2\rme^y-1}.
\label{ptm}
\ee
In the limit of weak coupling, this behaves as follows:
\be
P^{\rm TM}\sim -\frac{15}{64\pi^2 a^4}\lambda^2,
\ee
which is to be compared with (\ref{pertte}).
In strong coupling, on the other hand, it has precisely the same limit as
the TE contribution, (\ref{strongte}), which confirms the expectation
given at the end of the previous subsection.  Graphs of the two functions
are given in Fig.~\ref{fig2}.

For calibration purposes we give the Casimir pressure in practical units
between ideal perfectly conducting parallel plates at zero temperature:
\be
P=-\frac{\pi^2}{240 a^4}\hbar c=-\frac{1.30\mbox{ mPa}}{(a/1\mu\mbox{m})^4}.
\label{casresult}
\ee

\subsection{Surface energy as bulk energy of boundary layer}
\label{sec:2.4}

Here we show that the surface energy can be interpreted as the bulk
energy of the boundary layer.  We do this by considering a scalar field
in $1+1$ dimensions interacting with the background
\be
\mathcal{L}_{\rm int}=-\frac\lambda 2\phi^2\sigma,
\label{bkgdpot}
\ee
where
\be
\sigma(x)=\left\{\begin{array}{cc}
h,&-\frac\delta2<x<\frac\delta 2,\\
0,&\mbox{otherwise},
\end{array}\right.
\ee
with the property that $h\delta=1$.
The reduced Green's function satisfies
\be
\left[-\frac{\partial^2}{\partial x^2}
+\kappa^2+\lambda\sigma(x)\right]g(x,x')=\delta(x-x').
\ee
This may be easily solved in the region of the slab, $-\frac\delta2<x<\frac
\delta2$,
\bea
g(x,x')=\frac1{2\kappa'}\bigg\{\rme^{-\kappa'|x-x'|}
+\frac1{\hat\Delta}\bigg[
(\kappa^{\prime2}-\kappa^2)\cosh\kappa'(x+x')\nonumber\\
\qquad\mbox{}+(\kappa'-\kappa)^2\rme^{-\kappa'\delta}\cosh\kappa'(x-x')\bigg]
\bigg\}.
\label{slabgf}
\eea
Here $\kappa'=\sqrt{\kappa^2+\lambda h}$, and
\be
\hat\Delta=2\kappa\kappa'\cosh\kappa'\delta+
(\kappa^2+\kappa^{\prime2})\sinh\kappa'\delta.
\ee
This result may also easily be derived from the multiple reflection
formulas given in \sref{sec:lifshitz}, and agrees with that given
by Graham and Olum \cite{Graham:2002yr}.  The energy of the slab now is
obtained by integrating the energy density
\be
t^{00}=\frac1{2i}(\omega^2+\partial_x\partial_{x'}+\lambda h)g\bigg|_{x=x'}
\ee
over frequency and the width of the slab.  This gives the vacuum energy of the
slab
\bea
E_{s}&=&\frac12\int_{-\infty}^\infty \frac{\rmd\kappa}{2\pi}\frac1{2\kappa'\hat
\Delta}\bigg[(\kappa'-\kappa)^2(-\kappa^2-\kappa^{\prime2}+\lambda h)
\rme^{-\kappa'\delta}\delta\nonumber\\
&&\quad\mbox{}+(\kappa^{\prime2}-\kappa^2)(-\kappa^2+\kappa^{\prime2}+\lambda h\
)
\frac{\sinh\kappa'\delta}\delta\bigg].
\eea
If we now take the limit $\delta\to0$ and $h\to \infty$ so that $h\delta=1$,
we immediately obtain
\be
E_s=\frac1{2\pi}\int_0^\infty \rmd\kappa\frac{\lambda}{\lambda+2\kappa},
\label{eslab}
\ee
which precisely coincides with one-half the constant term in (\ref{11energy}),
with $\lambda$ there replaced by $\lambda a$ here.

There is no surface term in the total Casimir energy as long as the
slab is of finite width, because we may easily check that $\frac{d}{dx}g
\big|_{x=x'}$ is continuous at the boundaries $\pm\frac\delta2$.  However,
if we only consider the energy internal to the slab we encounter not
only the energy (\ref{casenergy}) but a surface term from the integration
by parts.  It is only this boundary term that gives rise to $E_s$, 
(\ref{eslab}),
in this way of proceeding.

Further insight is provided by examining the local energy density.
In this we follow the work of Graham and Olum \cite{Graham:2002yr,Olum:2002ra}.
  However, let us
proceed here with more generality, and consider the stress tensor with
an arbitrary conformal term,
\be
T^{\mu\nu}=\partial^\mu\phi\partial^\nu\phi-\frac12 g^{\mu\nu}(\partial_\lambda
\phi\partial^\lambda\phi+\lambda h\phi^2)-\alpha(\partial^\mu\partial^\nu
-g^{\mu\nu}\partial^2)\phi^2,
\ee
in $d+2$ dimensions, $d$ being the number of transverse dimensions.
Applying the corresponding differential operator to the Green's function
(\ref{slabgf}),  introducing polar coordinates in the $(\zeta,k)$ plane,
with $\zeta=\kappa\cos\theta$, $k=\kappa\sin\theta$, and
\be
\langle\sin^2\theta\rangle=\frac{d}{d+1},
\ee
we get the following form for the energy density within the slab,
\bea
\fl T^{00}=\frac{2^{-d-2}\pi^{-(d+1)/2}}
{\Gamma((d+3)/2)}\int_0^\infty \frac{\rmd 
\kappa\,\kappa^d}{\kappa'\hat\Delta}\bigg\{(\kappa^{\prime2}-\kappa^2)
\left[(1-4\alpha)(1+d)\kappa^{\prime2}-\kappa^2\right]\cosh2\kappa'x\nonumber\\
\mbox{}-(\kappa'-\kappa)^2\rme^{-\kappa'\delta}\kappa^2\bigg\}.
\eea  
From this we can calculate the behavior of the energy density as the
boundary is approached from the inside:
\be
T^{00}\sim \frac{\Gamma(d+1)\lambda h}{2^{d+4}\pi^{(d+1)/2}\Gamma((d+3)/2)}
\frac{1-4\alpha(d+1)/d}{(\delta-2|x|)^d},\quad |x|\to\delta/2.
\label{slabsurfdiv}
\ee  For $d=2$ for example, this agrees with the result found in 
Ref.~\cite{Graham:2002yr} for $\alpha=0$:
\be
T^{00}\sim\frac{\lambda h}{96\pi^2}\frac{(1-6\alpha)}{(\delta/2-|x|)^d},
\qquad|x|\to\frac\delta2.
\ee
Note that, as we expect, this surface divergence vanishes for the conformal
stress tensor \cite{ccj}, where $\alpha=d/4(d+1)$.  (There will be subleading
divergences if $d>2$.)  

We can also calculate the energy density on the other side of the boundary,
from the Green's function for $x,x'<-\delta/2$,
\be
g(x,x')=\frac1{2\kappa}\left[\rme^{-\kappa|x-x'|}-\rme^{\kappa(x+x'+\delta)}
(\kappa^{\prime2}-\kappa^2)\frac{\sinh\kappa'\delta}{\hat\Delta}\right],
\ee 
and the corresponding energy density is given by
\be
\fl T^{00} =-\frac{d(1-4\alpha (d+1)/d)}{2^{d+2}\pi^{(d+1)/2}\Gamma((d+3)/2)}
\int_0^\infty \rmd \kappa\,\kappa^{d+1}\frac1{\hat\Delta}(\kappa^{\prime2}
-\kappa^2)\rme^{2\kappa (x+\delta/2)}\sinh\kappa'\delta,
\ee
which vanishes if the conformal value of $\alpha$ is used.
The divergent term, as $x\to-\delta/2$, is just the negative of that found in
(\ref{slabsurfdiv}).  This is why, when the total energy is computed by
integrating the energy density, it is  finite for $d<2$, and independent
of $\alpha$. The divergence encountered for $d=2$ may be handled by 
renormalization of the interaction potential \cite{Graham:2002yr}.
 In the limit as $h\to\infty$, $h\delta=1$, we recover
the divergent expression (\ref{eslab}) for $d=0$, or in general
\be
\lim_{h\to\infty}E_s=\frac1{2^{d+2}\pi^{(d+1)/2}\Gamma((d+3)/2)}
\int_0^\infty \rmd \kappa \,\kappa^d\frac\lambda{\lambda+2\kappa}.
\ee
Therefore, surface divergences have an illusory character.

For further discussion on surface divergences, see \sref{sec:surfdiv}.
\section{Casimir Effect Between Real Materials}
\label{secreal}
\subsection{The Lifshitz Formula Revisited}
\label{sec:lifshitz}
As a prolegomena to the derivation of the Lifshitz formula for the
Casimir force between parallel dielectric slabs, let us note that the
results in the previous section may be easily derived geometrically, in
terms of multiple reflections.  Suppose we have translational invariance
in the $y$ and $z$ directions, so in terms of reduced Green's functions,
everything is one-dimensional.  Suppose at $x=0$ and $x=a$ we have 
discontinuities giving rise to reflection and transmission coefficients.
That is, if we only had the $x=0$ interface, the reduced Green's function
would have the form
\numparts
\be
g(x,x')=\frac1{2\kappa}\left(\rme^{-\kappa|x-x'|}+r\rme^{-\kappa(x+x')}\right),
\label{oneinterface}
\ee
for $x,x'>0$, while for $x'>0>x$,
\be
g(x,x')=\frac1{2\kappa}t \rme^{-\kappa(x'-x)}.
\ee
\endnumparts
Similarly, if we only had the interface at $x=a$, we would have similarly
defined reflection and transmission coefficients $r'$ and $t'$.
Transmission and reflection coefficients defined for a wave incident from the 
left instead of the right will be denoted with tildes.
If both interfaces are present, we can calculate the Green's function
in the region to the right of the rightmost interface $x,x'>a$ in the form
\numparts
\be
g(x,x')=\frac1{2\kappa}\left(\rme^{-\kappa|x-x'|}+R\rme^{-\kappa(x+x'-2a)}
\right),
\ee
where $R$ may be easily computed by summing multiple reflections:
\bea
R=r'+t'\rme^{-\kappa a}r\rme^{-\kappa a}\tilde t'
+t'\rme^{-\kappa a}r\rme^{-\kappa a}\tilde r'
\rme^{-\kappa a}r\rme^{-\kappa a}\tilde t'
+\dots\nonumber\\
=r'+\frac{rt'\tilde t'}{\rme^{2\kappa a}-r\tilde r'}.
\label{R}
\eea
\endnumparts
For the TE $\delta$-function potential (\ref{oneplusonelag}), 
$r=\tilde r=-(1+2\kappa a/\lambda)^{-1}$,
and $t=\tilde t=1+r$, and we immediately recover the result (\ref{gout}).
But the same formula applies to electromagnetic modes in a dielectric
medium with two parallel interfaces, where the permittivity is
\be
\varepsilon(x)=\left\{\begin{array}{cc}
\varepsilon_1,&x<0,\\
\varepsilon_3,&0<x<a,\\
\varepsilon_2,&a<x.
\end{array}\right..
\ee
In that case \cite{ce}
\numparts
\be
r=\frac{\kappa_3-\kappa_1}{\kappa_3+\kappa_1},\quad r'=\frac{\kappa_2-\kappa_3}
{\kappa_2+\kappa_3},\quad \tilde r'=-r',
\label{reflcoef}
\ee
and
\be
t'=1+r',\quad \tilde t'=1-r',
\ee
\endnumparts
where $\kappa_i^2=k^2-\omega^2\epsilon_i$.
Substituting these expressions into (\ref{R}) we obtain
\be 
R=\frac{\kappa_2-\kappa_3}{\kappa_2+\kappa_3}+\frac{4\kappa_2\kappa_3}
{\kappa_3^2-\kappa_2^2}\frac1{\frac{\kappa_3+\kappa_1}{\kappa_3-\kappa_1}
\frac{\kappa_3+\kappa_2}{\kappa_3-\kappa_2}\rme^{2\kappa_3 a}-1},
\ee
which coincides with the formula (3.16) given in Ref.~\cite{miltonbook}.

However, to calculate most readily the force between the slabs, we need
the corresponding formula for the reduced Green's function between the
interfaces.  This may also be readily derived by multiple reflections:
\bea
\fl g(x,x')=\frac1{2\kappa}\bigg[\rme^{-\kappa|x-x'|}+\tilde r'
\rme^{-\kappa(2a-x-x')}
+r\tilde r'\rme^{-\kappa(2a-x'+x)}
+r\tilde r^{\prime2}\rme^{-\kappa(4a-x-x')}\nonumber\\
\mbox{}+r^2\tilde r^{\prime2}\rme^{-\kappa(4a+x-x')}+\dots\nonumber\\
\mbox{}+r\rme^{-\kappa(x'+x)}+r\tilde r'\rme^{-\kappa(2a+x'-x)}
+r^2\tilde r'\rme^{-\kappa(2a+x'+x)}+r^2\tilde r^{\prime 2}\rme^{-\kappa(
4a+x'-x)}+\dots\bigg]\nonumber\\
\fl=\frac1{2\kappa}\bigg\{\rme^{-\kappa|x-x'|}+\frac1{\rme^{2\kappa a}-r\tilde r'}
\bigg[2r\tilde r'\cosh\kappa(x-x')
+\tilde r'\rme^{\kappa(x+x')}+r\rme^{-\kappa(x+x'-2a)}\bigg]\bigg\}.\nonumber\\
\label{mfin}
\eea
Indeed, this reduces to (\ref{gin}) when the appropriate reflection 
coefficients are inserted.  The pressure on the planes may be computed
from the discontinuity in the stress tensor, or
\be
t_{xx}\bigg|_{x=a-}-t_{xx}\bigg|_{x=a+}=\frac1{2\rmi}(-\kappa^2+\partial_x
\partial_{x'})g(x,x')\bigg|_{x=x'=a+}^{x=x'=a-}
=\frac{\rmi\kappa}{\frac1r\frac1{\tilde r'}
\rme^{2\kappa a}-1},
\ee
from which the $\delta$-potential results (\ref{stressin}) and 
(\ref{stressout}) follow immediately.
For the case of parallel dielectric slabs the TE modes therefore contribute
the following expression for the pressure\footnote{For the case of
dielectric slabs, the propagation constant $\kappa$ is different on the two
sides; we omit the term corresponding to the free propagator, however.
In the energy, the omitted terms are proportional to the volume of each
slab, and therefore correspond to the volume or bulk energy of the material.}:
\be
P^{\rm TE}=\int_{-\infty}^\infty\frac{\rmd \omega}{2\pi}\int\frac{(\rmd
\bi{k})}{(2\pi)^2}\frac{\rmi \kappa_3}{\frac{\kappa_3+\kappa_2}{\kappa_3-
\kappa_2}\frac{\kappa_3+\kappa_1}{\kappa_3-\kappa_1}\rme^{2\kappa_3a}-1}.
\ee
The contribution from the TM modes are obtained  by the replacement
\be
\kappa\to\kappa'=\frac\kappa\varepsilon,
\ee
except in the exponentials \cite{ce}.
This gives for the force per unit area at zero temperature\index{Lifshitz' 
formula}
\begin{eqnarray}
P^{T=0}_{\rm Casimir}=-{1\over4\pi^2}\int_0^\infty \rmd\zeta \int_0^\infty
\rmd k^2\,\kappa_3 \left(d^{-1}+d^{\prime-1}
\right),
\label{dielectricforce}
\end{eqnarray}
with the denominators here being [$\kappa_i=\sqrt{k^2+\zeta^2\varepsilon_i(\rmi
\zeta)}$]
\begin{equation}
d={\kappa_3+\kappa_1\over\kappa_3-\kappa_1}
{\kappa_3+\kappa_2\over\kappa_3-\kappa_2}\rme^{2\kappa_3 a}-1,\qquad
\quad d'={\kappa'_3+\kappa'_1\over\kappa'_3-\kappa'_1}
{\kappa'_3+\kappa'_2\over\kappa'_3-\kappa'_2}\rme^{2\kappa_3 a}-1,
\end{equation}
which correspond to the TE and TM Green's functions, respectively.
This is the celebrated Lifshitz formula 
\cite{lifshitz,dzyaloshinskii0,dzyaloshinskii,landauandlifshitz}, 
which we shall discuss further in the following subsections.
We merely note here that if we take the limit $\varepsilon_{1,2}\to
\infty$, and set $\varepsilon_3=1$, we recover Casimir's result for
the attractive force between parallel, perfectly conducting plates
(\ref{casresult}).

Henkel \etal \cite{henkel} have computed the
Casimir force at short distances ($\sim 1$ nm) from interactions between
polaritons.  Their result agrees with the Lifshitz formula with the plasma
formula (\ref{plasmad}) employed, see Ref.~\cite{lambrecht,genetx}.
\subsection{The Relation to van der Waals Forces}

\label{vanderwaalsforces}
Now suppose the central slab consists of a tenuous medium and the surrounding
medium is vacuum, so that the dielectric constant in the slab
differs only slightly from unity,
\begin{equation}
\epsilon-1\ll1.
\end{equation}
Then, with a simple change of variable,
\begin{equation}
\kappa=\zeta p,
\end{equation}
we can recast the Lifshitz formula (\ref{dielectricforce}) into the
form\index{Lifshitz' formula}
\begin{equation}
P\approx-{1\over32\pi^2}\int_0^\infty \rmd\zeta\,\zeta^3
[\varepsilon(\zeta)-1]^2\int_1^\infty
{\rmd p\over p^2}[(2p^2-1)^2+1]\rme^{-2\zeta pa}.
\label{weaklif}
\end{equation}
If the separation of the surfaces is large compared to the
wavelength characterizing $\varepsilon$, $a\zeta_c\gg1$, we can
disregard
the frequency dependence of the dielectric constant,
and we find
\begin{equation}
P\approx-{23(\varepsilon-1)^2\over640\pi^2a^4}.
\label{longdist}
\end{equation}
For short distances, $a\zeta_c\ll1$, the approximation is
\begin{equation}
P\approx-{1\over32\pi^2}{1\over a^3}\int_0^\infty
\rmd\zeta(\varepsilon(\zeta)-1)^2.
\label{shortdist}
\end{equation}
These formulas are identical with the well-known forces
 found for the complementary geometry in Ref.~\cite{schdermil}.
 \index{Casimir force!between tenuous dielectric slabs}

Now we wish to obtain these results from the sum of van der Waals
forces,\index{van der Waals forces}
derivable from a potential of the form
\begin{equation}
V=-{B\over r^\gamma}.
\label{vbgamma}
\end{equation}
We do this by computing the energy (${\cal N}= $ density of molecules)
\begin{equation}
E=-{1\over2}B {\cal N}^2\int_0^a \rmd z\int_0^a \rmd z'\int(\rmd{\bf r_\perp})
(\rmd{\bf r'_\perp})
{1\over[({\bf r_\perp-r'_\perp})^2+(z-z')^2]^{\gamma/2}}.
\end{equation}
If we disregard the infinite self-interaction terms (analogous to
dropping the volume energy terms in the Casimir calculation), we
get \cite{schdermil,sonokm2}
\begin{equation}
P=-{\partial\over\partial a}{E\over A}=-{2\pi B
{\cal N}^2\over(2-\gamma)(3-\gamma)}
{1\over a^{\gamma-3}}.
\end{equation}
So then, upon comparison with (\ref{longdist}), we set $\gamma=7$
and in terms of the polarizability,\index{Polarizability}
\begin{equation}
\alpha={\varepsilon-1\over4\pi {\cal N}},
\label{eq:polarepsilon}
\end{equation}
we find
\begin{equation}
B={23\over4\pi}\alpha^2,
\label{bee}
\end{equation}
or, equivalently, we recover the retarded dispersion potential\index{Retarded
dispersion forces}
of Casimir and Polder \cite{casimirandpolder},\index{Casimir-Polder forces}
\begin{equation}
V=-{23\over4\pi}{\alpha^2\over r^7},
\label{caspol}
\end{equation}
whereas for short distances we recover from (\ref{shortdist})
the London potential \cite{london},\index{London interaction}
\begin{equation}
V=-{3\over\pi}{1\over r^6}\int_0^\infty \rmd\zeta\,\alpha(\zeta)^2.
\end{equation}

Recent, nonperturbative approaches to Casimir-Polder forces include that
of Buhmann \etal \cite{buhmann}.  

\subsubsection{Force Between a Molecule and a Plate}
One can also calculate the force between a polarizable molecule, with
electric polarizability $\alpha(\omega)$, and a dielectric slab.  A simple,
gauge-invariant way of doing this starts from the variational form
\cite{schdermil,miltonbook}
\be
\delta W=-\int_{-\infty}^\infty \rmd t\,\delta E=-\frac\rmi2\int(\rmd x)\delta
\varepsilon(x)\Gamma_{kk}(x,x),
\ee
where $\delta\varepsilon(\bi{r})=4\pi\alpha(\omega)\delta(\bi{r-R})$,
$\bi{R}$ denoting the position of the molecule. 
Here $\bGamma$ is the electromagnetic Green's dyadic, defined by
\be
\bGamma(\bi{r},\bi{r'})=\rmi \langle \bi{E}(\bi{r})\bi{E}(\bi{r'})
\rangle.
\ee
 In terms of the
reduced Green's function, defined by (\ref{redgreen}),  then
\be
\delta E=\frac\rmi 2 4\pi\int\frac{\rmd \omega}{2\pi}\frac{\rmd^2k}{(2\pi)^2}
\alpha(\omega)g_{kk}(x,x;\omega,\bi{k}).
\ee
It is easily seen how the trace of the reduced Green's function can be
expressed in terms of the reduced TE and TM Green's functions,
\be
g_{kk}=\left(\omega^2g^{\rm TE}+\frac{k^2}{\varepsilon\varepsilon'}g^{\rm TM}
+\frac1\varepsilon\frac\partial{\partial x}\frac1{\varepsilon'}\frac\partial
{\partial x'}g^{\rm TM}\right)\Bigg|_{x=x'}.
\ee
For a single interface, the Green's functions to the right of a dielectric
slab situated in the half-space $x<0$ are given by (\ref{oneinterface})
with the reflection coefficients in the vacuum
\be
r^{\rm TE}=\frac{\kappa-\kappa_1}{\kappa+\kappa_1},\qquad
r^{\rm TM}=\frac{\kappa-\kappa_1/\varepsilon_1}{\kappa+\kappa_1/\varepsilon_1},
\ee
where $\kappa^2=k^2+\zeta^2$ and $\kappa_1^2=k^2+\zeta^2\varepsilon_1$.
In this way, we immediately obtain the energy between a dielectric slab 
(permittivity $\varepsilon_1$) and a polarizable molecule a distance $Z$
from it:
\bea
\fl E_{\rm slab,mol}=-\frac1{16\pi^2}\int_0^\infty \rmd\zeta 4\pi\alpha(\zeta)
\int_0^\infty \rmd k^2\frac1\kappa\rme^{-2\kappa Z}\left[-\zeta^2\frac{\kappa
-\kappa_1}{\kappa+\kappa_1}+(2k^2+\zeta^2)\frac{\varepsilon_1\kappa-\kappa_1}
{\varepsilon_1\kappa+\kappa_1}\right].\nonumber\\
\eea
If the separation between the plate and the molecule is large, we expect that
we may neglect the frequency dependence of the polarizability, $\alpha(\zeta)
\to\alpha(0)$.  There are then two simple limits.  If we take $\varepsilon_1
\to \infty$ we are describing a perfectly conducting plane, in which case we
immediately obtain the result first given by Casimir and Polder 
\cite{casimirandpolder}
\be
E_{\rm metal,mol}=-\frac{3\alpha(0)}{8\pi Z^4}.
\label{cpmetal}
\ee
On the other hand, we could consider a tenuous medium, $(\varepsilon_1-1)\ll1$,
in which case
\be
E_{\rm dilute,mol}=-\frac{23}{160\pi}\frac{\alpha(0)(\varepsilon_1-1)}{Z^4}.
\label{dilslabmol}
\ee

The latter should be, as in the previous subsection, interpretable as
the sum of pairwise van der Waals interactions between the external molecule
and the molecules which make up the slab, given by the Casimir-Polder
interaction (\ref{caspol}).  The net energy then is
\be
-\frac{23}{4\pi}\alpha\mathcal{N}
\int_{Z}^\infty \rmd z\int_0^\infty \rmd \rho\,\rho\int_0^{2\pi}\rmd\phi
\frac{\alpha(0)}{(z^2+\rho^2)^{7/2}}=-\frac{23}{4\pi}\alpha\frac{\mathcal{N}2\pi}{20}
\frac{\alpha(0)}{Z^4},
\ee
which coincides with (\ref{dilslabmol}) when (\ref{eq:polarepsilon}) is used.

The force between a molecule and a plate has been measured by Sukenik et 
al.~\cite{sukenik}, who actually verified the force between a molecule
and two plates \cite{bartonatomplates} at the roughly 10\% level.  Recently,
this result has been questioned (at about the same level of accuracy) by
Bordag \cite{Bordag:2004dn}, who argued that a subtle error involving the
quantization of gauge fields in the presence of boundaries was made by
Casimir and Polder \cite{casimirandpolder} and subsequent workers.  The
fact that the result can be given an unambiguous gauge-invariant derivation,
and that it is closely related to the Lifshitz formula and the retarded
dispersion van der Waals force suggests that this critique is invalid.
(Bordag now concedes that the usual result is valid for ``thick'' plates,
where the normal component of $\bi{E}$ is given by the surface charge
density.)  

For a recent rederivation of (\ref{cpmetal}) see Hu \etal \cite{hu04}.
A very recent paper by Babb, Klimchitskaya, and Mostepanenko \cite{babb04}
gives a rederivation of the Casimir-Polder energy (\ref{cpmetal}) in the
retarded limit, and finds no support for Bordag's modification.  They then go
on to discuss the dynamical polarizability and thermal corrections for
real materials, and find substantial (35\%) corrections at short distances
$\sim 100$ nm.

In this connection we might also
mention the work of Noguez and Rom\'an-Vel\'azquez \cite{noguez04}, who
calculate the force between a sphere and a plate made of dissimilar
materials in the non-retarded limit (see also van Kampen \cite{kampen}
and Gerlach \cite{gerlach}) in terms of multipolar interactions.
They find significant deviations from the proximity approximation
(\sref{sec:prox}), which says that there is no difference between
the force between a sphere made of material A and a plate made of material
B and the reversed situation, when the separation is comparable or
large compared to the radius of the sphere, and that under the above-%
mentioned A-B interchange the forces change by up to 6\%.  See also 
Ref.~\cite{noguezqfext,noguez68}.

Ford and Sopova \cite{fordqfext, fordpra}
consider Casimir forces between small metal spheres and dielectric (and
conducting) plates, modeled by a plasma dispersion relation
\be
\varepsilon(\omega)=1-\frac{\omega_p^2}{\omega^2}.
\label{plasmad}
\ee
The electric dipole approximation used requires $a\omega_p\ll1$, that is,
the radius of the sphere $a$ must be in the 10--100 nm range.  The force
is oscillatory, being alternatively attractive and repulsive as a function
of the height $Z$ of the atom above the plate.  Thus levitation in the
earth's gravitational field might be possible, for $Z\sim 1$ $\mu$m.

\subsection{Roughness and Conductivity Corrections}

\subsubsection{Roughness Corrections}
No real material surface is completely smooth.  Even beyond the atomic
level, there will be regions of higher and lower elevations. Insofar as
these are plateaus large compared to the separation between the disjoint
surfaces, the corrections can be easily incorporated by use of the 
proximity approximation (see \sref{sec:prox} below).  This is nothing other than the
naively obvious statement that if $P(a)$ is the force per unit area
between two parallel plates separated by a distance $a$, the average
force per area between rough surfaces made up of large plateaus and
valleys, with the perpendicular distance between two
adjacent points on the two surfaces
in terms of transverse coordinates
$(x,y)$ being $a(x,y)$, is
\be
P=\frac1{A}\int \rmd x\,\rmd y \, P(a(x,y)).
\ee
In Ref.~\cite{chen04}, for example, an equivalent
 expression is used directly with
data obtained by topography of the surfaces using an
atomic force microscope.  Traditionally, a stochastic estimate has
been used.   Let the separations $a$ be distributed around the mean
$a_0$ according to
a Gaussian, with the probability of finding separation $a$ being given by
\be
p(a)=\frac1{\sqrt{\pi}\delta a}\rme^{-(a-a_0)^2/(\delta a)^2}.
\ee
We will assume $\delta a\ll a_0$.  Then, $\langle a\rangle = a_0$,
$\langle (a-a_0)^2\rangle=\frac12(\delta a)^2$,
and in general
\bea
\fl\langle a^\alpha\rangle=\int_0^\infty \rmd a\,a^\alpha p(a)=
\frac1{\sqrt{\pi}\delta a}\int_{-\infty}^\infty \rmd a \,\rme^{-a^2/(\delta a)^2}
(a+a_0)^\alpha\nonumber\\
\lo=a_0^\alpha\left[1+\frac{\alpha(\alpha-1)}2\frac12\frac{(\delta a)^2}{a_0^2}
+\frac{\alpha(\alpha-1)(\alpha-2)(\alpha-3)}{4!}\frac34\frac{(\delta a)^4}
{a_0^4}+\dots\right],
\eea
The force between a sphere and a plate depends on the closest distance $d$
between them like $d^{-3}$, see (\ref{eq:proxthm}) below, so the stochastic
estimate for the roughness correction in that case, in terms of the
mean-square fluctuation amplitude $A=\delta a/\sqrt{2}$, is
\be
F_{\rm sph-pl, rough}=F_{\rm sph-pl}\left[1+6\left(\frac{A}{d}\right)^2
+45\left(\frac{A}{d}\right)^4+\dots\right].
\ee
A much more detailed discussion may be found in Ref.~\cite{newreview}.
It must be appreciated that the approximate treatment based on the
proximity approximation is invalid for short wavelength deformations
\cite{reynaudepl}.

\subsubsection{Finite Conductivity}

Another interesting result, important for the recent experiments 
\cite{mohideen,mohideen2,mohideen2a,decca03},
is the correction for an imperfect conductor,\index{Conductor!imperfect}
 where for frequencies above
the infrared, an adequate representation for the dielectric constant is
\cite{ce} that given by the plasma model (\ref{plasmad})
where the plasma frequency is, in Gaussian units
\begin{equation}
\omega_p^2={4\pi e^2 N\over m},
\label{plasma}
\end{equation}
where $e$ and $m$ are the charge and mass of the electron, and $N$ is the
number density of free electrons in the conductor.  A simple calculation
shows, at zero temperature \cite{hargreaves,schdermil},\index{Casimir
force!imperfect conductors}
\begin{equation}
P\approx-{\pi^2\over240 a^4}\left[1-{8\over3\sqrt{\pi}}{1\over ea}\left(
m\over N\right)^{1/2}\right].
\end{equation}
If we define a penetration parameter, or skin depth,\index{Skin depth} 
by $\delta=1/\omega_p$,
we can write the force per area for parallel plates out to fourth order as 
\cite{mostturn,mostbook,bezerra00,newreview}
\begin{equation}
\fl P\approx-{\pi^2\over240 a^4}\left[1-{16\over3}{\delta\over a}
+24{\delta^2\over a^2}-\frac{640}7\left(1-\frac{\pi^2}{210}\right)
\frac{\delta^3}{a^3}+\frac{2800}{9}\left(1-\frac{163\pi^2}{7350}\right)
\frac{\delta^4}{a^4}\right],
\end{equation}\index{Casimir effect!imperfect conductor}\index{Imperfect
conductor}
while using the proximity force theorem (see \sref{sec:prox}),
to convert pressures between parallel plates to forces between a lens of
radius $R$ and a plate,
\be
\mathcal{F}_{n-1}=\frac{2\pi R}{n-1}aP_n,
\ee
for a term in the pressure going like $P_n\propto a^{-n}$,
the force between a spherical surface and a plate separated by a
distance $d$ is
\be
\fl \mathcal{F}\approx-\frac{\pi^3 R}{360 d^3}\left[
1-4{\delta\over d}
+\frac{72}5{\delta^2\over d^2}-\frac{320}7\left(1-\frac{\pi^2}{210}\right)
\frac{\delta^3}{d^3}+\frac{400}{3}\left(1-\frac{163\pi^2}{7350}\right)
\frac{\delta^4}{d^4}\right].
\end{equation}

Lambrecht, Jaekel, and Reynaud \cite{lambrecht2} analyzed the
Casimir force between mirrors with arbitrary frequency-dependent
reflectivity,\index{Reflectivity!frequency-dependent}
 and found that it is always smaller than that between perfect
reflectors.\index{Casimir force!mirrors of variable reflectivity}

We might also mention here the interesting suggestion that repulsive
Casimir forces might exist \cite{Kenneth:2002ij} between parallel
plates.  This harks back to an old suggestion of Boyer \cite{boyerunusual},
that repulsion will occur between two plates, one of which is a perfect
electrical conductor, $\varepsilon\to\infty$, and the other a perfect magnetic
conductor, $\mu\to\infty$,
\be
P=\frac78\frac{\pi^2}{240}\frac1{a^4}.
\ee
 However, it appears
that it will prove very difficult to observe such effects in the laboratory
\cite{commentonrepulsion}.
Klich \cite{klichqfext} now seems to agree with this assessment.

\subsection{Thermal Corrections}
\label{sec:temp}
The discussion in this subsection is adapted from that in 
Refs.~\cite{Hoye:2002at,Brevik:2003rg}.
We begin by reviewing how temperature effects are incorporated into the
expression for the force between parallel dielectric (or conducting)
plates separated by a distance $a$.
To obtain the finite temperature Casimir force from the zero-temperature
expression, one conventionally makes the following substitution in the
imaginary frequency,
\numparts
\label{AB}
\begin{equation}
\zeta\to\zeta_m=\frac{2\pi m}{\beta},\label{A}
\end{equation}
and replaces the integral over frequencies by a sum,
\begin{equation}
\int_{-\infty}^\infty \frac{\rmd\zeta}{2\pi}\to\frac1\beta\sum_{m=-\infty}^\infty.
\label{B}
\end{equation}
\endnumparts
This reflects the requirement that thermal Green's functions be periodic
in imaginary time with period $\beta$ \cite{ms}.
  Suppose we write the finite-temperature pressure as
  [for the explicit form, see (\ref{dielectricforce}) and (\ref{3}) below]
\begin{equation}
P^T=\sum_{m=0}^\infty {}'f_m,
\label{primedsum}
\end{equation}
where the prime on the summation sign means that the $m=0$ term is counted
with half weight.
To get the low temperature limit, one can  use the Euler-Maclaurin
(EM) sum  formula,
\begin{equation}
\sum_{k=0}^\infty f(k)=\int_0^\infty f(k)\,\rmd k+\frac{1}{2}f(0)-\sum_{q=
1}^\infty \frac{B_{2q}}{(2q)!} f^{(2q-1)}(0),
\label{22}
\end{equation}
where $B_n$ is the $n$th Bernoulli number.
This means here, with half-weight for the $m=0$ term,
\begin{equation}
P^T=\int_0^\infty f(m)\, \rmd m-\sum_{k=1}^\infty
\frac{B_{2k}}{(2k)!}f^{(2k-1)}(0).
\label{em}
\end{equation}
It is noteworthy that the terms involving $f(0)$ cancel in (\ref{em}).
The reason for this is that the EM formula equates an integral
to its trapezoidal-rule approximation plus a series of corrections;
thus the $1/2$ for $m=0$ in (\ref{primedsum}) is built in automatically.
For perfectly conducting plates separated by vacuum
 [see the $\lambda\to\infty$ limit of (\ref{31pressure})
or (\ref{ptm}), or the $\varepsilon_{1,2}\to\infty$ limit of 
(\ref{dielectricforce}) with $\varepsilon_3=1$]
\begin{equation}
f(x)=-\frac2{\pi\beta}\int_{2\pi x/\beta}^\infty \kappa^2\,\rmd 
\kappa\frac1{\rme^{2\kappa a}-1}.
\end{equation}
Of course, the integral in (\ref{em})
 is just the inverse of the finite-temperature
prescription (\ref{B}),
and gives the zero-temperature result.   The only nonzero odd
derivative occurring is 
\begin{equation}
f'''(0)=-\frac{16\pi^2}{\beta^4},
\label{stefan}
\end{equation}
which gives a Stefan's law type of term, seen in (\ref{linterm}) below.

The problem is that the EM formula only applies if $f(m)$ is continuous.
If we follow the  argument of 
Ref.~\cite{bostrom,bostrom2,Brevik:2002bi,brevik02},
 and take the $\epsilon_{1,2}\to\infty$ limit of (\ref{dielectricforce})
at the end\footnote{This is contrary to the ``Schwinger'' prescription
advocated in Refs.~\cite{schdermil,miltonbook}, in which the perfect-conductor
limit is taken before the zero-mode is extracted.}
 ($\epsilon_{1,2}$ are the permittivities of the
two parallel dielectric slabs), this is not the case, and for the TE mode
\numparts
\begin{eqnarray}
f_0&=&0,\\
f_m&=&-\frac{\zeta(3)}{4\pi\beta a^3},\quad 0<\frac{2\pi a m}{\beta}\ll1.
\end{eqnarray}
\endnumparts
Then we have to modify the argument as follows:
\begin{eqnarray}
P^T=\sum_{m=0}^\infty{}'f_m=\sum_{m=1}^\infty f_m
=\sum_{m=0}^\infty{}'\tilde f_m-\frac12\tilde f_0,\label{2.8}
\end{eqnarray}
where $\tilde f_m$ is defined by continuity,
\begin{equation}
\tilde f_m=\left\{\begin{array}{cc}
f_m,&m>0,\\
\lim_{m\to0}f_m,&m=0.\end{array}\right.
\end{equation}
Then by using the EM formula,
\begin{eqnarray}
P^T&=&\frac{\beta}{2\pi}\int_0^\infty \rmd\zeta\,f(\zeta)+\frac{\zeta(3)}
{8\pi\beta a^3}-\frac{\pi^2}{45}\frac1{\beta^4}\nonumber\\
&=&-\frac{\pi^2}{240 a^4}\left[1+\frac{16}{3}\left(\frac{a}{\beta}
\right)^4\right]+\frac{\zeta(3)}{8\pi a^3}T, \quad aT\ll1.
\label{linterm}
\end{eqnarray}
The same result for the low-temperature
limit is extracted through use of the Poisson sum formula, as, for example,
discussed in Ref.~\cite{miltonbook}.  Let us refer to these results,
with the TE zero mode excluded, as the modified ideal metal model (MIM).
The conventional result for an ideal metal (IM), obtained first by 
Lifshitz \cite{lifshitz,dzyaloshinskii} and by Sauer \cite{sauer} and
Mehra \cite{mehra}, is given by (\ref{linterm}) with the linear term in $T$ 
omitted.

Exclusion of the TE zero mode 
 will reduce the linear dependence at high temperature by a factor of
two, 
\be
P^T_{\rm IM}\sim -\frac{\zeta(3)}{4\pi a^3}T,\quad
P^T_{\rm MIM}\sim -\frac{\zeta(3)}{8\pi a^3}T,\quad aT\gg1,
\label{hitemp}
\ee
but this is not observable by present experiments.  
The observable consequence, however, is that it adds a linear
term at low temperature, which is given in (\ref{linterm}), 
up to exponentially small corrections \cite{miltonbook}.

 There are apparently two serious problems with the result (\ref{linterm}):
\begin{itemize}
\item  It would seem to be ruled out by experiment.  The ratio of the
linear term to the $T=0$ term is
\numparts
\begin{equation}
\Delta=\frac{30\zeta(3)}{\pi^3}aT=1.16 aT,
\end{equation}
or putting in the numbers (300 K $ = (38.7)^{-1}$ eV, $\hbar c=197$ MeV fm)
\begin{equation}
\Delta=0.15\left(\frac{T}{300\mbox{ K }}\right)\left(\frac{a}{1 \mu\mbox{m}}
\right),
\end{equation}
\endnumparts
or as Klimchitskaya observed \cite{klim02}, 
there is a 15\% effect at room temperature at
a separation of one micron. One would have expected
 this to have been been seen by Lamoreaux
\cite{lamoreaux}; his experiment was reported to be in agreement with the 
conventional theoretical prediction at the level of 5\%. (Lamoreaux
\cite{lamoreauxqfext} is now proposing a new experiment to resolve this
issue.)
\item Another  serious problem is the apparent thermodynamic 
inconsistency.  A linear term
in the force implies a linear term in the free energy (per unit area),
\begin{equation}
F=F_0+\frac{\zeta(3)}{16\pi a^2}T,\quad aT\ll1,
\label{linearfe}
\end{equation}
which implies a nonzero contribution to the entropy/area at zero temperature:
\begin{equation}
S=-\left(\frac{\partial F}{\partial T}\right)_V=-\frac{\zeta(3)}{16\pi a^2}.
\label{2.17}
\end{equation}
\end{itemize}
Taken at face value, this statement appears to be incorrect. 
We will discuss this problem more closely in \sref{entropy},
 and will find that although a linear temperature dependence will
occur in the free energy
 at room temperature, the entropy will go to zero as the temperature
goes to zero. The point is that the free energy $F$ for a finite 
$\varepsilon$ always will have a zero slope at $T=0$, 
thus ensuring that $S=0$ at $T=0$. The apparent conflict with 
(\ref{2.17}) or (\ref{linterm}) is due to the fact that the 
curvature of $F(T)$ near $T=0$ becomes infinite when 
$\varepsilon \rightarrow \infty$.
So (\ref{linearfe}) and (\ref{2.17}), corresponding to the modified
ideal metal model, describe real metals approximately only for low, but
not zero temperature -- See the following.

\subsubsection{Lifshitz formula at nonzero temperature}
The Casimir surface pressure at finite temperature
 $P^T$ between two dielectric plates
separated by a distance $a$ can be obtained from the Lifshitz formula
(\ref{dielectricforce}) by the prescription (\ref{B})\footnote{A 
rederivation of the Casimir force between dissipative metallic mirrors
at nonzero temperature has been given by Reynaud, Lambrecht, and Genet
\cite{reynaudqfext}.  They obtain formulas, generalizing those at
zero temperature \cite{genet03}, for the force valid even if the smoothness
condition necessary for the derivation of the Lifshitz formula is not
satisfied due to the failure of the Poisson summation formula.} 
\begin{equation}
P^T=-\frac{1}{\pi \beta}\sum_{m=0}^\infty {}^\prime
\int_{\zeta_m}^\infty \kappa^2\rmd \kappa
\left[\left(A^{-1}_m\rme^{2\kappa a}-1\right)^{-1}+\left(B^{-1}_m
\rme^{2\kappa a}
-1\right)^{-1}\right].
\label{3}
\end{equation}
 The relation between
$\kappa$ and the transverse wave vector $\bi{k}_\perp$ is $\kappa^2=k_\perp
^2+\zeta_m^2$, where  $\zeta_m=2\pi m/\beta$. 
Furthermore, the squared reflection coefficients are
\numparts
\begin{eqnarray}
 A_m&=&\left(\frac{\varepsilon p-s}{\varepsilon p+s}\right)^2,\quad
B_m=\left( \frac{s-p}{s+p}\right)^2, \label{4a}\\
s^2&=&\varepsilon -1+p^2, \quad p=\frac{\kappa}{\zeta_m},
\label{4b}
\end{eqnarray}
\endnumparts
with $\varepsilon(\rmi\zeta_m)$ being the permittivity.
Here, the first term in the square brackets in (\ref{3}) corresponds to
TM modes, the second to TE modes.
Note that whenever $\varepsilon$ is constant,
$A_m$ and $B_m$ depend on $m$ and $\kappa$ only in the combination
$p$,
\begin{equation}
A_m(\kappa)=A(p),\quad B_m(\kappa)=B(p).
\end{equation}

The free energy $F$ per unit area can be obtained from (\ref{3}) by
integration with respect to $a$ since $P^T=-\partial F/\partial a$.
We get \cite{aarseth01}
\numparts
\begin{equation}
\beta F=\frac{1}{2\pi}{\sum_{m=0}^\infty }{}^\prime
\int_{\zeta_m}^\infty\kappa\,\rmd\kappa\,
 [\ln(1-\lambda^{\rm TM})+\ln(1-\lambda^{\rm TE})],
\label{5a}
\end{equation}
where
\begin{equation}
\lambda^{\rm TM}=A_m\rme^{-2\kappa a},\quad \lambda^{\rm TE}=
B_m\rme^{-2\kappa a}.
\label{5b}
\end{equation}
\endnumparts

{}From thermodynamics the entropy $S$ and internal energy $U$ (both per
unit area) are related to $F$ by $F=U-TS$, implying
\begin{equation}
 S=-\frac{\partial F}{\partial T},\quad \textrm{and thus}\quad 
U=\frac{\partial(\beta F)}{\partial \beta}. 
\label{star}
\end{equation}
As mentioned above the behaviour of $S$ as $T\rightarrow 0$ has been
disputed, especially for metals where $ \varepsilon \rightarrow \infty$.
We now see the mathematical root of the problem: The quantities
$A_m=B_m\rightarrow 1$ in the  $\varepsilon\to\infty$ 
 limit except that $B_0=0$ for any
finite $\varepsilon$. So the question has been whether $B_0=0$ or
$B_0=1$ or something in between should be used in this limit as
results will differ for finite $T$, producing,
 as we saw above, a difference in the force
linear in
$T$. The corresponding difference in entropy will thus be  nonzero.
Such a difference would lead to a violation of the third law
of thermodynamics, which states that the entropy of a 
system with a nondegenerate ground state should be zero at $T=0$. 
Inclusion of the interaction between the plates at different separations
cannot change this general
property.
 We will show that this discrepancy vanishes when the limit
$\varepsilon \rightarrow \infty$ is considered carefully.

\subsubsection{Gold as a numerical example}
\label{V.3}
Let us go back to (\ref{3}) for the surface pressure, making use of 
the best available experimental results for $\varepsilon(\rmi\zeta)$ as 
input when calculating the coefficients $A_m$ and $B_m$. We choose gold 
as an example. Useful information about the real and imaginary parts, 
$n'$ and $n''$, of the complex permittivity $n=n'+\rmi n''$, versus the 
real frequency $\omega$,  is given in Palik's book \cite{palik} 
and similar sources. The range of photon energies given in Ref.~\cite{palik} 
is from 0.1 eV to $10^4$\, eV. (The conversion factor 
\begin{equation}
1 \,\textrm{eV}=1.519\times 10^{15}\; \textrm{rad/s} 
\label{87}
\end{equation}
is useful to have in mind.) When $n'$ and $n''$ are known  the permittivity 
$\varepsilon(\rmi\zeta)$ along the positive imaginary frequency axis, which is 
a real quantity, can be calculated by means of the Kramers-Kronig relations.
\begin{figure}
\centering
\includegraphics[height=10cm]{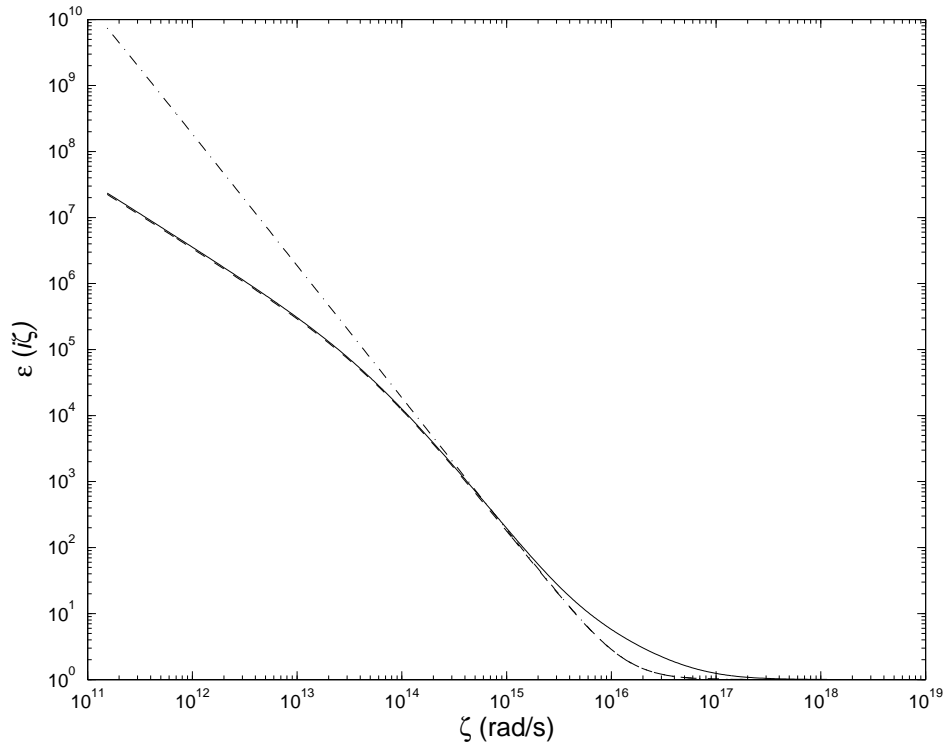}
\caption{Solid line: 
Permittivity $\varepsilon (\rmi\zeta)$ as function of imaginary 
frequency $\zeta$ for gold. The curve is calculated on the basis of 
experimental data.  Courtesy of Astrid 
Lambrecht and Serge Reynaud. Dashed lines: $\varepsilon(\rmi\zeta)$ versus 
$\zeta$ with $T$ as parameter, based upon the temperature dependent Drude
model; cf.~Appendix D of Ref.~\cite{Hoye:2002at}.
The upper curve is for $T=10$ K;
the lower is for $T=300$ K, which for energies below 1 eV ($1.5 \times 10^{15}$ rad/s)
nicely fits the experimental data.  Both curves are below the
experimental one for $\zeta>2\times 10^{15}$ rad/s.}
\label{fig1t}
\end{figure}

\begin{figure}
\centering
\includegraphics[height=10cm]{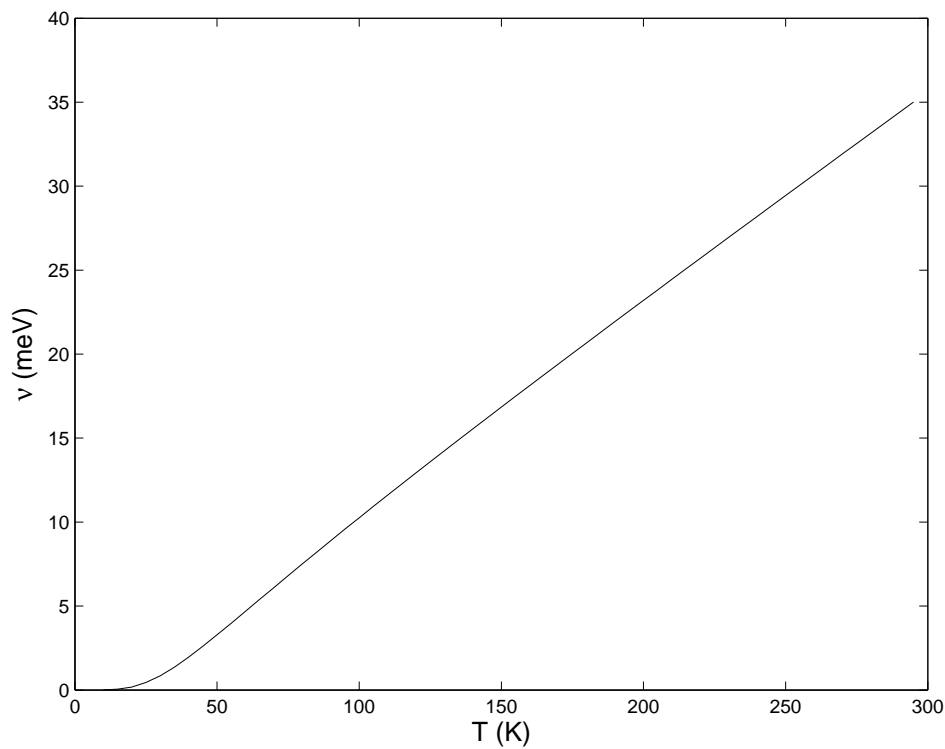}
\caption{Temperature dependence of the relaxation frequency for gold
based on the Bloch-Gr\"uneisen formula \cite{handbook67}.}
\label{fig6}
\end{figure}

Figure \ref{fig1t} 
shows how $\varepsilon(\rmi\zeta)$ varies with $\zeta$ over seven 
decades, $\zeta \in [10^{11}, 10^{18}]$ rad/s. The curve was given in 
Refs.~\cite{lambrecht,lambrecht3}, and is reproduced here for convenience. 
(We are grateful to A. Lambrecht and S. Reynaud for having given us the 
results of their accurate calculations.)  At low photon
energies, below about 1 eV, 
the data are well described by the Drude model, 
\begin{equation}
\varepsilon(\rmi\zeta)=1+\frac{\omega_p^2}{\zeta (\zeta+\nu)},
\label{1}
\end{equation}
where $\omega_p$ is the plasma frequency (\ref{plasma}) 
and $\nu$ the relaxation
frequency. (Usually, $\nu$ is taken to be a constant, equal to its 
room-temperature value, but see below.) 
The values appropriate for gold at room temperature are
 \cite{lambrecht,lambrecht3}
\begin{equation}
\omega_p=9.0 \; \textrm{eV}, \quad \nu=35\; \textrm{ meV}.
\label{88}
\end{equation}
The curve in Fig.~\ref{fig1t} shows a 
monotonic decrease of $\varepsilon(\rmi\zeta)$ 
with increasing  $\zeta$, as any permittivity as a function of imaginary
frequency has to follow according to thermodynamical requirements. 
The two dashed curves in the figure show, for comparison, how $\varepsilon 
(\rmi\zeta, T)$ varies
 with frequency if we accept the Drude model for all frequencies, and include 
 the temperature dependence of the relaxation frequency with $T$ as a 
 parameter.  (The latter is given in Fig.~\ref{fig6}, according to the
 Bloch-Gr\"uneisen formula \cite{handbook67}, which, however, does not
 take into account the physical fact that because of impurities, no actual
 conductor has zero resistivity at zero temperature \cite{resist}.
 See Appendix D of Ref.~\cite{Hoye:2002at}.)
For $T=300$ K, the Drude curve is seen to be good 
 for all frequencies up to $\zeta \sim 2\times 10^{15}$ rad/s; for higher 
 $\zeta$ it gives too low values of $\varepsilon$. Both Drude curves, for 
 $T=10$ K and $T=300$ K, are seen to give the same values when $\zeta \ge 
 3\times 10^{14}$ rad/s.

\begin{figure}
\centering
\includegraphics[height=10cm]{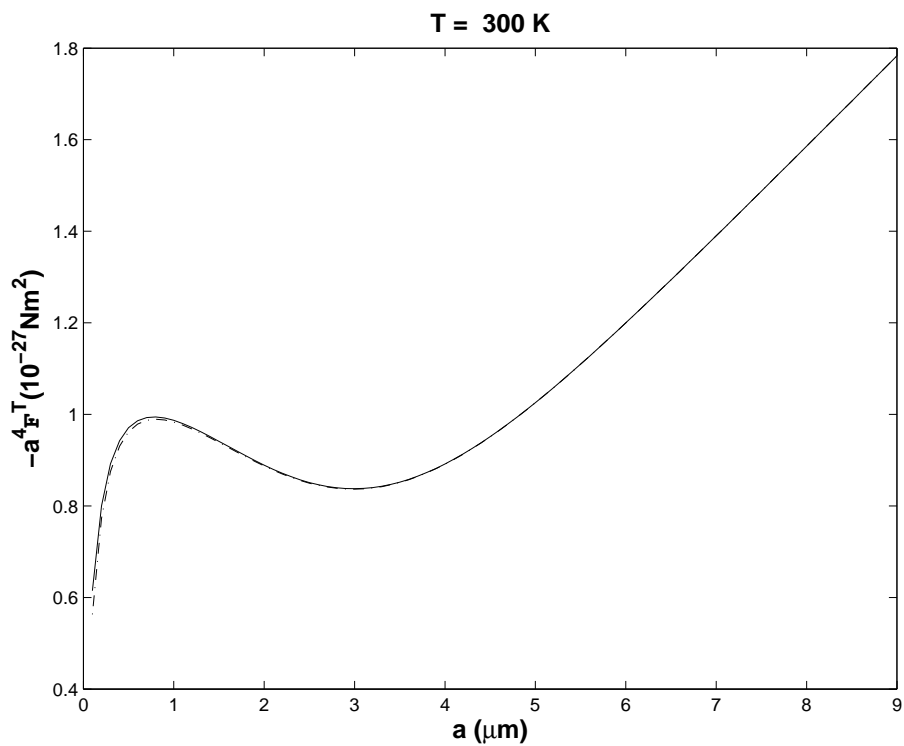}
\caption{ Surface pressure for gold, multiplied with $a^4$, versus $a$ 
when $T=300$~K. Input data for $\varepsilon(\rmi\zeta)$ are taken from 
Fig.~\ref{fig1t}.}
\label{fig4}
\end{figure}

As experiments are usually made at room temperature for various gap widths, we 
show in Fig.~\ref{fig4} how the surface force density for gold varies 
with $a$, at $T=300$~K. 
 The linear slope seen for $a\ge 4 \mu$m is nearly that predicted by
(\ref{hitemp})
for high temperatures when the TE zero-mode is excluded (modified ideal
metal), which gives a slope of $2.0\times 10^{-28}$ Nm$^2$/$\mu$m.
(This is in spite of the fact that $aT=0.5$ at $a=4$ $\mu$m.)
The linear region between 1 and 2 $\mu$m corresponds roughly to that in 
(\ref{linterm})  (intermediate temperatures).
Also shown is the prediction of the temperature dependent Drude model,
when $T=300$ K. The differences are seen to be very small.  Since the Drude 
values for the permittivity are lower than the empirical ones at high 
frequencies, as seen in 
Fig.~\ref{fig1t}, we expect the predicted Drude forces to be slightly 
weaker than those based upon the empirical permittivities. This expectation is 
borne out in Fig.~\ref{fig4}; the differences being large enough to be slightly visible 
at short distances, as we would expect since the plasma nature of the material 
becomes more pronounced for small distances.
Note that the temperature dependence of the permittivity is irrelevant here
because the temperature is fixed.

\begin{figure}
\centering
\includegraphics[height=10cm]{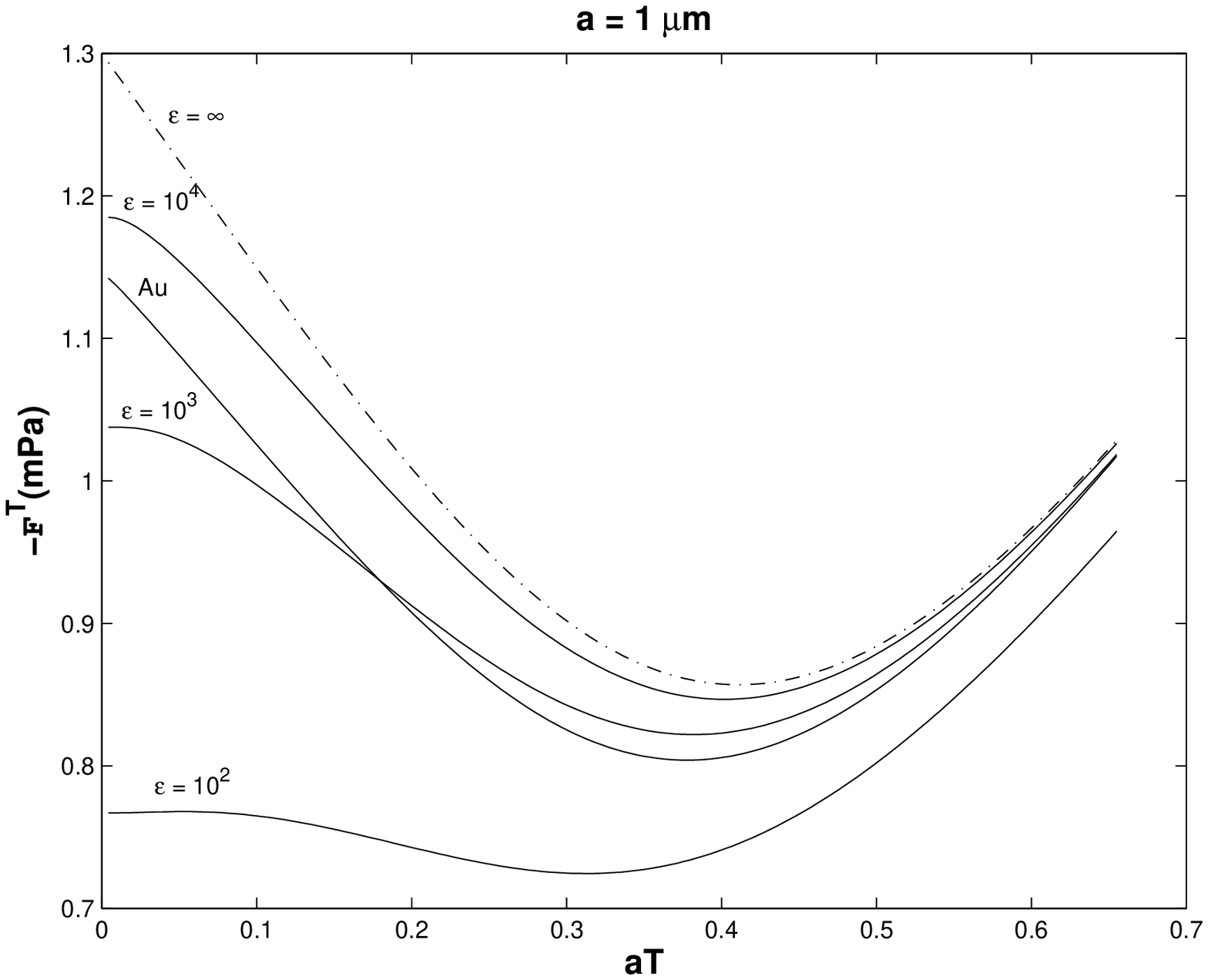}
\caption{Nondispersive theory: Surface pressure
 for $\varepsilon \in\{100, 1000, 10000, \infty \}$. 
For low values of $aT$ the latter coincides with the expression 
(\ref{linterm}).
Also shown for comparison is the dispersive result for gold, where experimental
  input data for $\varepsilon(\rmi\zeta)$ are taken from 
 Fig.~\ref{fig1t}. Gap width is $a=1\, \mu$m. The constraint $a=1\,\mu$m 
 applies only to the dispersive case, since otherwise $a^4P^T$
 is a function of $aT$ only.  Note that room temperature (300 K) corresponds
to $aT=0.13$.}
 \label{fig5}
 \end{figure}

It is of interest to check the magnitude of the dispersive effect in these 
cases. We have therefore made a separate calculation of the expression 
(\ref{3})
when $\varepsilon$ is taken to be  constant. Figure \ref{fig5} 
shows how the force 
varies with $aT$ in cases when $\varepsilon\in\{100, 1000, 10000, \infty \}$ are 
inserted in the expressions for $A_m$ and $B_m$ in (\ref{4a}).  

 It is seen from the figure that the first three curves asymptotically 
approach the $\varepsilon =\infty$ curve,  when 
$\varepsilon$ increases, as we would expect. Again, we emphasize that the 
dispersive curve for gold is calculated using the available room-temperature 
data for $\varepsilon(i\zeta)$ from Fig.~\ref{fig1t}. In the nondispersive 
case, 
there is of course no permittivity temperature problem since $\varepsilon$ is 
taken to be the same for all $T$.

There are several points worth noticing from Fig.~\ref{fig5}: 
\begin{enumerate}
\item
 The curves have 
a horizontal slope at $T=0$. For finite $\varepsilon$ this property is 
clearly visible on the curves. This has to be so on physical grounds: 
If the pressure had a linear dependence on $T$
for small $T$ so would the free energy $F$, in contradiction with the
requirement that 
the entropy $S=-\partial F/\partial T$ has to go to zero as 
$T \rightarrow 0$. For the gold data the initial horizontal slope is
not resolvable on the scale of this graph, but see the discussion in
\sref{entropy}.
\item The curves show that the magnitude of the
force \textit{diminishes} 
with increasing $T$ (for a fixed $a$), in a certain temperature interval up 
to $aT \simeq 0.3$. This perhaps counterintuitive effect is thus clear from 
the nondispersive curves.  This is qualitatively similar to the behavior
seen in Fig.~\ref{fig4} for fixed $T$, where the minimum occurs for $aT\sim
0.4$.
\item It is seen that the curve for 
 $\varepsilon = \mbox{const.} = 1000$ gives a reasonably good approximation to 
 the real dispersive curve for gold when $a=1~\mu$m; the deviations are less 
  than about 5\% except for the lowest values of $aT$ ($aT <0.1$). 
This fact makes our neglect of the temperature dependence of 
$\varepsilon(i\zeta)$ appear physically reasonable; the various curves turn 
out to be rather insensitive with respect to variations in the input values of 
$\varepsilon(i\zeta)$.
 \item Also, it can be remarked that $B_0=0$ is required when 
  $\varepsilon$ is finite. Otherwise the curves in Fig.~\ref{fig5}, and thus 
  the free energy, would have a finite slope at $T=0$ which again would 
  imply a finite entropy contribution at $T=0$ in violation with the third 
 law of thermodynamics.
\end{enumerate}
 
 \subsubsection{Behavior of the Free Energy at Low Temperature}
 \label{entropy}
\begin{figure}[t]
\begin{center}
\epsfxsize=20pc 
\begin{turn}{270}
\epsfbox{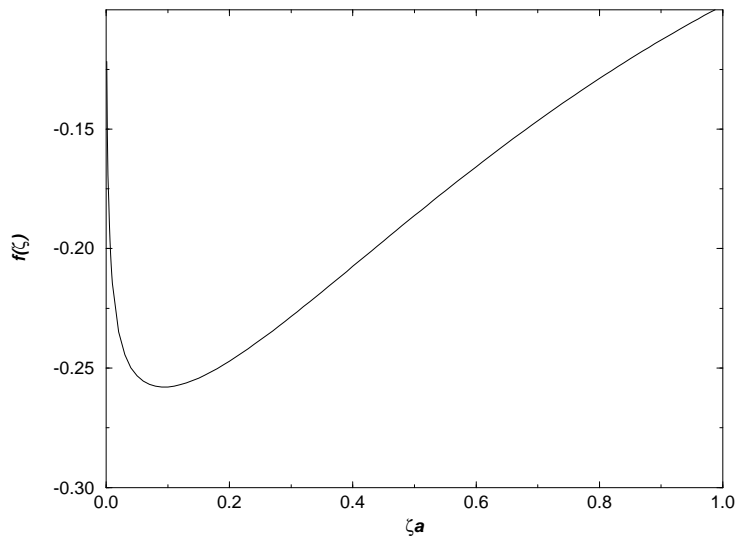} 
\end{turn}
\end{center}
\caption{The behavior of the free energy for low frequencies, in the Drude
model, with parameters suitable for gold, and a plate separation of
$a=1$ $\mu$m.  Here 
$ F^{\rm TE}=\frac{T}{2\pi a^2}\sum_{m=0}^\infty{}'f(\zeta_m)$.
Here, we have used the room temperature value of the relaxation
parameter.\label{fig:5}}
\end{figure}

The low temperature correction is dominated by low frequencies,\footnote{This
statement is in the context of using of the Euler-Maclaurin summation formula
to evaluate (\ref{3}), for
example.} where the Drude formula is extremely accurate.  Using this fact, we
have performed analytic and numerical calculations which
show that the free energy has a quadratic low-temperature dependence,
independent of the plate separation:
\begin{equation}
\fl F(T)=F_0+T^2\frac{\omega_p^2}{48\nu}(2\ln2-1)=F_0+T^2(19\mbox{ eV}),
\qquad T\ll\frac\nu{\omega_p^2 a^2}\approx 20 \mbox{ mK },
\label{quad}
\end{equation}
where we have put in the numbers for gold, (\ref{88}), (the temperature restriction
refers to a 1 $\mu$m plate separation)
rather than the naive extrapolation (\ref{linearfe})
\begin{equation}
F=F_0+T\frac{\zeta(3)}{16\pi a^2}=F_0+\frac{T}{4\pi a^2}0.30.
\label{naiveext}
\end{equation}
We see from Fig.~\ref{fig:5} that the value 
in (\ref{naiveext}) indeed results if one extrapolates
the approximately linear curve there for $\zeta a>0.25$ to zero, following
the argument given in (\ref{2.8}).  However, we see
that the free energy smoothly changes to the quadratic behavior exhibited
in (\ref{quad}).  Of course, the turn-over will be much sharper if we
replace the room-temperature relaxation frequency $\nu(300\mbox{ K})$ by
the positive value at zero temperature, due to elastic scattering from
defects or impurities.

Results consistent with these have been reported by Sernelius and 
Bostr\"om \cite{sernelius03}. In particular they show that one cannot ignore
the constant value $\nu(0 \mbox{ K})$, so there is no relevant temperature
dependence of the relaxation parameter.  Although there is a region of
negative entropy, the Nernst heat theorem is not violated,
but rather $S\to0$ as $T\to0$
if one goes to sufficiently low temperature, in contradiction to 
Refs.~\cite{bezerra65,bezerra66}.

\subsubsection{Surface impedance form of reflection coefficient}
It has been proposed that the resolution to the temperature problem
for the Casimir effect is that the surface impedance form of the reflection
coefficients should be used in the Lifshitz formula 
\cite{geyer03,mostqfext,klimqfext,bezerra04},
rather than that based on the bulk permittivity.  Here we show that the
two approaches are in fact equivalent, and that the former must include
transverse momentum dependence.

For the TE modes, the reflection coefficient is given by (\ref{reflcoef}) 
\cite{ce} 
\begin{equation}
r^{\rm TE}=-\frac{k_{1z}-k_{2z}}{k_{1z}+k_{2z}},
\label{rte}
\end{equation}
where 
\begin{equation}
k_{az}=\sqrt{\omega^2\varepsilon-k_\perp^2}\to \rmi\sqrt{\zeta^2[
\varepsilon(\rmi\zeta)-1]
+\kappa^2}=i\kappa_a,\label{disp}
\end{equation}
with $\kappa^2=\kappa_2^2=k_\perp^2+\zeta^2$,
and the subscripts 1 and 2 refer to the metal and the
vacuum regions, respectively.  Now from Maxwell's equations outside sources
we easily derive  just inside
the metal (the tangential components, designated by $\perp$,
 of $\bi{E}$ and $\bi{B}$ are continuous across the interface)
\numparts
\begin{eqnarray}
-\rmi k_{1z}\bi{k_\perp\cdot B_\perp}-\rmi\omega \varepsilon\left(1-\frac{k_\perp^2}
{\omega^2\varepsilon}\right)\bi{k_\perp\cdot( n\times E_\perp)}=0,
\label{me11}
\\
-\rmi k_{1z}\bi{k_\perp\cdot(n\times E_\perp)}-\rmi\omega\bi{k_\perp\cdot B_\perp}
=0.\label{me22}
\end{eqnarray}
\endnumparts
Here $\bi{n}$ is the normal to the interface. Now the surface impedance
is defined by
\begin{equation}
\bi{E_\perp}=Z(\omega,\bi{k_\perp})\bi{B_\perp\times n},\quad
\bi{n\times E_\perp}=Z(\omega,\bi{k_\perp}) \bi{B_\perp}.
\end{equation}
So eliminating $\bi{B}_\perp$ using this definition
we find two equations:
\begin{eqnarray}
k_{1z}=-\frac{\omega}Z,\label{kZ}\\
k_{1z}^2=\omega^2\varepsilon-k_\perp^2,
\end{eqnarray}
the latter being the expected dispersion relation (\ref{disp}).
Substituting this into the expression for the reflection coefficient (\ref{rte})
we find
\begin{equation}
r^{\rm TE}=-\frac{\zeta+Z\kappa}{\zeta-Z\kappa}
=-\frac{1+Zp}{1-Zp},\quad p=\frac{\kappa}{\zeta},
\end{equation}
which apart from (relative) signs (presumably just a different convention choice)
coincides with that given in Geyer \etal \cite{geyer03} or Bezerra 
\etal~\cite{bezerra03}. See also Refs.~\cite{bezerra02,svetovoy03}.
The first discussion of the Lifshitz formula in this approach was given
in Ref.~\cite{mostturn}.  


However, it is crucial to note that the ``surface impedance'' so defined
depends on the transverse momentum,
\begin{equation}
Z=-\frac\zeta{\sqrt{\zeta^2[\varepsilon(\rmi\zeta)-1]+\kappa^2}},
\end{equation}
and so $r^{\rm TE}\to 0$ as $\zeta\to 0$ just as in the dielectric constant
formulation.  Of course, we have exactly the same result 
for the energy as before, since
this is nothing but a slight change of notation, as noted in 
Ref.~\cite{esquivel02,bostrom2}.

It is therefore incorrect to assume 
that $Z$ is only a function of frequency, not of transverse momentum, and
to use the normal and anomalous skin effect formulas derived for real waves
impinging on imperfect conductors.\footnote{Of course, in general, the
permittivity will be a function both of the frequency and the transverse
momentum, $\varepsilon(\omega,\bi{k}_\perp)$, but we believe the latter
dependence
is not significant for separations larger than $\hbar c/\omega_p=0.02$ $\mu$m.}
In the above-cited references, this necessary dependence was not included. 
(For further comments on the insufficiency of the argument in Ref.~\cite{geyer03}
see Ref.~\cite{svet174}.

 How does the usual argument go?   The normal component of the
wavevector in a conductor is given by
\begin{equation}
k_z=\left[\omega^2\left(\varepsilon+\rmi\frac{4\pi\sigma}\omega\right)-k_\perp^2
\right]^{1/2}\to\sqrt{\rmi4\pi\omega\sigma},\qquad \omega\to0,
\label{nse}
\end{equation}
from which the usual normal skin effect formula follows immediately,
\begin{equation}
Z(\omega)=-(1-\rmi)\sqrt{\frac\omega{8\pi\sigma}}.
\end{equation}
However, the limit in (\ref{nse}) here consists in omitting two ``small''
terms: $\omega^2\varepsilon$ 
(which is legitimate) and $k_\perp^2\le\omega^2$.  
Here this last is
not valid because in going to finite temperature we have severed the
connection between $\omega\to \rmi\zeta$ and $k_\perp$; the latter is in no
sense ignorable as we take $\zeta\to0$ to determine the low temperature
dependence. This is the same error
to which we refer in Ref.~\cite{Hoye:2002at}.  (This $\bi{k}_\perp$
dependence still seems to be ignored in a recent reanalysis by
Torgerson and Lamoreaux \cite{torgerson1} (see also Ref.~\cite{torgerson}) 
who argue that
low frequencies of order of the inverse transverse size of the plates
dominate the low temperature behavior so that a linear term in the temperature
does not appear.  This seems unlikely since the zero-temperature
dependence is extracted by an analytic continuation procedure.)

Not only do Mostepanenko,
Klimchitskaya, \etal \cite{bezerra02,geyer03} ignore transverse
momentum dependence, but they apparently do not use the correct values of
the frequency in their evaluation of the surface impedance.  They use
the impedance appropriate to the domain of infrared optics, thereby
extrapolating the surface impedance at what they consider a characteristic
frequency $\sim 1/2a$ rather than using the actual zero frequency
value \cite{klimqfext}.  This seems to be a completely {\it ad hoc\/} 
prescription,
as opposed to the procedure advocated in Brevik \etal \cite{Hoye:2002at}, 
which uses the actual electrical properties of the materials. 

A beginning of a general discussion of nonlocal effects, including the
anomalous skin effect, in Casimir phenomena has recently been given
by Esquivel and Svetovoy \cite{esqsvet}.  There they argue that 
the Leontovich approach \cite{lifandpit,abrikosov} advocated by 
\cite{bezerra02,geyer03} 
only applies to normal incidence, which is why the surface impedances
only depend on frequency.  In fact, this is incorrect in general, and if
only local functions are used for the permittivity, that is $\varepsilon=
\varepsilon(\omega)$, the dependence for the TE surface impedance given
above is reproduced.  For propagating waves the Leontovich approximation
is appropriate, but not for the evanescent fields relevant to the Casimir
effect, where $k_\perp/\omega>1$ occur. 
They do not calculate temperature effects; the nonlocal anomalous skin
effect for $\omega<\omega_p$ that they compute gives a correction to
the Casimir force of order 0.5\%, but other nonlocal effects, such as
plasmon excitations, could be more significant \cite{esquivel,esquivelqfext}.

\subsection{Beyond the Proximity Approximation}
\label{sec:prox}
As we will discuss in the next section, to avoid problems of parallelism,
most recent experiments to measure the force between conductors have
not been made between parallel plates, but between a plate and a spherical
surface, or between crossed cylinders.  The Lifshitz and Casimir formulas
do not apply to these situations.  However, in the 1930s, it was recognized
that if the separation between the sphere and the plate is very small compared
to the radius of curvature of the sphere, the latter force may be derived
from the force for the parallel plate configuration.  This result is usually
called the Proximity Force Theorem \cite{blocki}, which here says that the 
attractive force $\mathcal{F}$
between a sphere of radius $R$ and a flat surface is simply the 
circumference of the sphere times the energy per unit area for parallel plates,
or, from (\ref{1/2casimir}),\index{Casimir force!between sphere and plate}
\begin{equation}
\mathcal{F}=2\pi R \, {\cal E}(d)=-{\pi^3\over360}{R\over d}{\hbar c\over d^2},
\quad R\gg d,
\label{eq:proxthm}
\end{equation}
where $d$ is the\index{Proximity theorem}
distance between the plate and the sphere at the point of closest
approach, and $R$ is the radius of curvature of the sphere at that 
point.  (The exact shape of the ``sphere'' is not relevant in the
strict approximation $R\gg d$.) 
 The proof of (\ref{eq:proxthm}) is quite simple.  If $R\gg d$,
each element of the sphere may be regarded as parallel to the plane,
so the potential energy of the sphere is
\begin{equation}
\fl V(d)=\int_{0}^\pi 2\pi R\sin\theta R\,\rmd\theta 
\,{\cal E}(d+R(1-\cos\theta))=2\pi R\int_{-R}^R
\rmd x\,{\cal E}(d+R-x).
\end{equation}
To obtain the force between the sphere and the plate, we differentiate
with respect to $d$:
\begin{eqnarray}
\mathcal{F}&=&
-{\partial V\over\partial d}=2\pi R\int_{-R}^R\rmd x\,{\partial\over 
\partial x}{\cal E}(d+R-x)\nonumber\\
&=&2\pi R[{\cal E}(d)-{\cal E}(d+2R)]\approx 2\pi R\, {\cal E}(d),\quad d\ll R,
\end{eqnarray}
provided that ${\cal E}(a)$ falls off with $a$.
This result was already given in Refs.~\cite{abrikosova,deriagin,derjaguin}.  
The proximity theorem
itself dates back to a paper by Derjaguin in 1934 \cite{derpt,derpt2}.  

Let us apply this theorem to the MIM model (\ref{linterm}) for the force
between parallel plates at low temperature.  The corresponding free energy is
\be
F=-\frac{\pi^2}{720 a^3}+\frac{\pi^2}{45}a T^4-\frac{\zeta(3)}{2\pi} T^3
+\frac{\zeta(3)}{16\pi a^2}T,
\ee
where the term constant in $a$ is determined by the high-temperature limit
(\ref{hitemp}) -- see Ref.~\cite{miltonbook}, p.~56.  This free energy is to
be used in the proximity force theorem, with the result for the force between
a sphere and a plate \cite{klim2,lamoreaux,lamoreaux2,newreview}
\be
\mathcal{F}=-\frac{\pi^3}{360}\frac{R}{d^3}\left[1-16(Td)^4+\frac{360\zeta(3)}
{\pi^3}(Td)^3-\frac{45\zeta(3)}{\pi^3}Td\right].
\ee
The terms linear in $T$ would not be present in the IM model.  At room 
temperature, $300$ K,  and at 1 $\mu$m separation, 
the successive terms correspond to corrections of $-0.46$\%,
$+3.1$\%, and $-23$\%, respectively.  This model, of course, does not begin
to reflect the true temperature dependence, discussed for parallel plates
above.  A full discussion of the temperature dependence for the force between
a spherical lens and a plate will appear elsewhere.

Emig has recently presented exact results  for Casimir forces between
periodically deformed surfaces \cite{Emig:2002xz,emigqfext,Buscher:2004tb}.
In the latest paper, the authors calculate the force between a flat plate
and one with a rectangular (square) corrugation, of amplitude $\Delta a$.
This was probed experimentally by Roy and Mohideen \cite{mohideen3},
with clear deviations from the proximity approximation. 
(See also Refs.~\cite{mohideenlat,mohideenlat2} for measurements of the
so-called ``lateral Casimir effect.'')
 For short wavelength corrugations for either TE or TM modes one gets
\be
P=-\frac{\pi^2}{480}\frac1{(a-\Delta a)^4}\approx-\frac{\pi^2}{480 a^4}
\left(1+\frac{4\Delta a}{a}\right),
\ee
while for long wavelength corrugations
\be
P=-\frac{\pi^2}{480}\frac12
\left(\frac1{(a-\Delta a)^4}+\frac1{(a+\Delta a)^4}\right),
\ee
which is as expected from the proximity approximation.
For intermediate wavelength corrugations numerical results are given.
The force approaches that given by the proximity approximation for
large $\lambda$ like $\Delta a/\lambda$, as compared to $(\Delta a/\lambda)^2$
for sinusoidal corrugations, due to the sharp edges.  These behaviors can be
understood from the ray optics approach of Jaffe and Scardicchio 
\cite{Jaffe:2003mb} discussed in the following subsection.
The relative contributions of the TE and TM modes vary with the wavelength
and the shape of the corrugation, the ratio of the modes approaching unity
as $a/\Delta a$ tends to 1 or $\infty$. 
Insofar as first approximations to these interactions were extracted
through use of the proximity force theorem, these results shed valuable
light on how to move beyond that approximation.

\subsubsection{Optical Paths}
A very interesting strategy for moving beyond the proximity approximation
has been suggested by Jaffe and Scardicchio \cite{Jaffe:2003mb}.
This is related to the semiclassical closed orbit approach advocated
by Schaden and Spruch \cite{schaden,schaden1,schaden2} and earlier by
Gutzweiler \cite{gutzweiler,gutzbook}, and also to that of Balian and Bloch
\cite{balianandbloch3,balianandbloch4,balianandbloch5}.  Fulling has also
recently proposed similar ideas \cite{fullingpo,fullingqfext}.

In the simplest context, that of parallel plates, the approach is, of
course, exact, and is precisely what we wrote down in (\ref{mfin}).
We simply compute the energy using (\ref{casenergy}) with
\be
\mathcal{G}(\bi{r,r})=\int\frac{(\rmd\bi{k}_\perp)}{(2\pi)^2}g(x,x),
\ee
where $g(x,x')$ is given by (\ref{mfin}).  Rather than carry out the sum
as given there, let us sum the terms with even and odd numbers of reflections
separately.  The former give, when the zero reflection term is omitted,
\be g_{\rm even}(x,x)=
\frac1{2\kappa}2\left[\tilde r'r\rme^{-2\kappa a}+(\tilde r'r)^2
\rme^{-4\kappa a}+\dots\right]
=\frac1{2\kappa}(\coth\kappa a-1),
\ee
where in the last step we have inserted the values for the reflection
amplitudes appropriate to Dirichlet boundaries, $r=\tilde r'=-1$.
When this is inserted into the expression for the energy we obtain
rather immediately the usual result for the Casimir energy between Dirichlet
plates:
\be
\mathcal{E}=-\frac1{96\pi^2a^3}\int_0^\infty \rmd u\frac{u^3}{\rme^u-1}
=-\frac{\pi^2}{1440a^3}.
\label{dirichletce}
\ee
Keeping only the first term in the sum (2 reflections) gives
\be
\mathcal{E}^{(2)}=-\frac1{16\pi^2a^3},
\ee
which is in magnitude only 7.6\% low, while keeping 2 plus 4 reflections
give an error of 1.8\%:
\be
\mathcal{E}^{(2)}+\mathcal{E}^{(4)}=-\frac1{16\pi^2a^3}\left(1+\frac1{16}\right).
\ee
The odd reflections give a term in $g(x,x)$ which depends on $x$:
\be
g_{\rm odd}(x,x)=-\frac1{2\kappa}\left(\rme^{-2\kappa x}+\rme^{2\kappa(x-a)}
\right)\frac1{1-\rme^{-2\kappa a}}.
\ee
when this is integrated over $x$, the $a$ dependence of this term disappears,
so this gives rise to an irrelevant constant in the energy.  Keeping it 
and the zero-reflection term gives
the expression for the total energy as obtained directly from (\ref{mfin})
\be
\mathcal{E}_0+\mathcal{E}_{\rm even}+\mathcal{E}_{\rm odd}
=-\frac1{12\pi^2}\frac1{a^3}
\int_0^\infty \rmd y\,y^3\left(\coth y-\frac1y\right).
\ee

Jaffe and Scardicchio \cite{Jaffe:2003mb}
use this method to estimate the force between a sphere
an a plate.  The results disagree with the proximity approximation when
$d/R$ is bigger than a few percent, but agrees with an exact numerical
calculation \cite{gies}, described in the following subsection,
 up to $d/R\approx0.1$, where the proximity theorem
fails badly.
 
\subsubsection{Worldline Approach to the Casimir Energy}

Gies, Moyaerts, and Langfeld \cite{gies,moyaerts} have developed a numerical
technique for extracting Casimir energies in nontrivial geometries, such as
between a sphere and a plate.  It is based on the string-inspired worldline
approach.  They consider, like Graham \etal 
\cite{graham,Graham:2002fw,Graham:2003ib} 
a scalar field in a smooth background
potential like (\ref{bkgdpot}).  The worldline representation of the effective
action is obtained by introducing a proper time representation of the
functional logarithm with ultraviolet regularization, doing the trace
in configuration space, and interpreting the matrix element there as a
Feynman path integral over all worldlines $x(\tau)$.  Field theoretic
divergences can thus be handled.  Other divergences arise from the potential
itself, when it approaches some idealized limit, which may  not be removed
in a physically meaningful way and may 
or may not contribute to physical observables.
The expectation value is evaluated by the ``loop-cloud'' method, using
techniques from statistical mechanics.  Although in the ``sharp'' and 
``strong'' limits in the sense of Graham \etal
\cite{graham,Graham:2002fw,Graham:2003ib} divergences occur in the theory,
a finite \textit{force\/} between rigid bodies can be obtained.  The general
result for $\delta$-function planes, discussed in \sref{Sec:2.5}, is
reproduced numerically, and then the sphere-plate system is considered.
The numerical results, for $d/R$ from $10^{-3}$ to 10, agree closely
with the \textit{geometric mean}\footnote{The geometric mean version of
the proximity force approximation, which coincides with the semiclassical
periodic orbit method of Schaden and Spruch \cite{schaden,schaden1,schaden2},
 has been found to be the most accurate
also for concentric cylindrical shells, the Casimir energy for which was
calculated by Mazzitelli \etal \cite{mazzitelli,mazzitelliqfext}.}
of the plate-based and the sphere-based proximity force approximation
(deviation from either becomes sizable for $d/R>0.02$).  Note that
electromagnetic fluctuations (e.g., TM modes) have not been considered
in this approach.

\subsection{Status of the Experimental Measurements on Casimir Forces}

\label{sec:exp}
\index{Experimental verification of the Casimir effect|(}
Attempts to measure the Casimir effect between solid bodies date back to the
middle 1950s.  The early measurements were, not surprisingly, somewhat
inconclusive \cite{deriagin,derjaguin,kitchener,sparnaay,%
black,silfhout,tabor0,tabor,winterton,israelachivili}.
The Lifshitz theory (\ref{dielectricforce}), 
for zero temperature, was, however, confirmed accurately
in the experiment of Sabisky and Anderson in 1973 \cite{sabisky}.  So there
could be no serious doubt of the reality of zero-point fluctuation forces.
For a review of the earlier experiments, see 
Refs.~\cite{sparnaayrev,israelrev}.

New technological developments allowed for dramatic improvements in 
experimental
techniques in recent years, and thereby permitted nearly direct confirmation of
the Casimir force between parallel conductors.  First, in 1997 Lamoreaux used
a electromechanical system based on a torsion pendulum to measure the force
between a conducting plate and a sphere \cite{lamoreaux,lamoreaux2},
as given by the proximity force theorem (\ref{eq:proxthm}).
Lamoreaux \cite{lamoreaux,lamoreaux2}
claimed agreement with this theoretical value at the 5\% level,
although it seems that finite conductivity was not included correctly,
nor were roughness corrections incorporated \cite{mohcom}.
Further, Lambrecht and Reynaud \cite{lambrecht} analyzed the effect of
conductivity and found discrepancies with Lamoreaux \cite{lamoreaux3}, and
therefore stated that it was too early to claim agreement between theory and
experiment.\index{Effects of conductivity}  See also 
Refs.~\cite{lambrecht3,lamreply}.

An improved experimental measurement was reported in 1998 by Mohideen and
Roy \cite{mohideen}, based on the use of an atomic force microscope.
They included finite conductivity, roughness,\index{Roughness corrections}
 and conventional temperature corrections,
\index{Temperature effects!parallel plates}\index{Roughness corrections}
although no evidence for latter has been claimed.
  Spectacular agreement
with theory at the 1\% level was attained.  Improvements were subsequently
reported \cite{mohideen2,mohideen2a}. 
(The nontrivial effects of corrugations in the
surface were examined in Ref.~\cite{mohideen3,mohideenlat,mohideenlat2}.)  
Erdeth \cite{ederth}
measured the Casimir forces between
crossed cylinders at separations of 20--100 nm.  The highest precision
was achieved with very smooth, gold-plated surfaces.
Rather complete analyses of the roughness,
conductivity, and temperature corrections to the Lamoreaux and Mohideen
experiments have been published \cite{klim4,klim,klim2}. 

More recently, a new measurement of the Casimir force (\ref{eq:proxthm})
was presented by a group at Bell Labs \cite{belllabs,belllabs2}, using
a micromachined torsional device, a micro-electromechanical system or MEMS,
by which they measured the attraction
between a polysilicon plate and a spherical metallic surface.  Both
surfaces were plated with a 200~nm film of gold.  The authors included
finite conductivity \cite{lambrecht,klim5}
 and surface roughness corrections \cite{maradudin,bezerra}, and obtained
 agreement with theory at better than 0.5\% at the smallest separations
of about 75~nm.  However, potential corrections of greater than 1\% exist,
so that limits the level of verification of the theory.  Their experimental
 work, which now continues at Harvard,
suggests novel nanoelectromechanical applications.\index{Nanoelectromechanical
applications}\index{Practical applications of the Casimir effect}

There is only one experiment with a parallel-plate geometry \cite{bressi},
which is of limited accuracy ($\sim15$\%) due to the difficulty of maintaining 
parallelism.  It is, however, of considerable interest because the
interpretation does not depend on the proximity theorem, corrections to
which are problematic \cite{emigqfext,Jaffe:2003mb}; see \sref{sec:prox}.
The importance of improving the accuracy of the parallel-plate configuration
has been emphasized by Onofrio \cite{onofrioqfext}.

The most precise experiment to date, using a MEMS, makes use of 
both static and  dynamical
procedures and yields a claimed accuracy of about 0.25\% \cite{decca03,decca03a},
but this accuracy has been disputed \cite{iannuzzi}, due to difficulty
in controlling roughness and the concomittant uncertainty in the ability
to determine the separation distance.  It has been asserted \cite{decca03a}
that this experiment rules out the temperature dependence claimed in
Ref.~\cite{Hoye:2002at} (see \sref{V.3}), but this is problematic at this 
point, especially as comparison is only made with the MIM model
(\ref{linterm}), rather than with the detailed calculation given there.

Very recently, the Harvard group has performed a very interesting Casimir
force measurement between a gold-covered plate and a sphere coated with
a hydrogen-switchable mirror \cite{harvard}.  Although the mirror becomes
transparent in the visible upon hydrogenation, no effect was observed
on the Casimir force when the mirror was switched on and off.  This
shows that, in contradiction to the claims of 
Mostepanenko \etal~, for example in Ref.~\cite{bezerra04},
the Casimir force is responsive to a very wide range of frequencies, in
accordance with the Lifshitz formula and the general dispersion relation
for the permittivity.\footnote{Iannuzzi quotes Klimchitskaya as now agreeing
with this statement.  This, however, is hard to reconcile with statements
made in Ref.~\cite{klimqfext} that one should use the extrapolated
surface impedance value at $\omega_c=1/2a$ rather than the actual 
zero-frequency value.} (See also Ref.~\cite{bostrom99,sernelius00}.)  
In particular, their results show that wavelengths much larger than the
separation between the surfaces play a crucial role.

Because all the recent experiments measure forces between relatively
thin films, rather than between bulk metals, significant deviations
from the Lifshitz formula ($\sim2\%$) may be expected \cite{svetovoyqfext}.
This may also be relevant to the claimed accuracy of the first Mohideen
experiment \cite{mohideen}, which uses a thin metallic coating, regarded as
completely transparent.

This may be an appropriate point to comment on the recent paper of
Chen \etal~\cite{chen04}.  This is based on a reanalysis of experimental
data obtained four years ago in Ref.~\cite{mohideen2a}.  Experimental precision
of 1.75\% and theoretical accuracy of 1.69\% is claimed at the shortest
distances, 62 nm.  However, their analysis seems flawed.  They obtain
average experimental forces by averaging many measurements, which is
only permissible if the averaging is carried out at exactly the same
separation between the surfaces.  Of course they have no way of knowing
this.  Furthermore, they apparently use the mean separation parameter
$d_0$ as a free variable in their fit, which essentially negates the
possibility of testing the theory, which is most sensitive at the shortest
separations \cite{ederth}.  Iannuzzi asserts that at distance of order
100 nm, errors of a few \AA ngstroms preclude a 1\% measurement.  
Therefore this analysis cannot be used as a serious constraint
for either new forces or for setting limits on temperature 
corrections.\footnote{I thank Davide Iannuzzi for discussion of these points.}

A difference force experiment has been proposed by Mohideen and
collaborators \cite{chenqfext,chen03}. The idea is to measure the
difference in the force between a lens and a plate at room temperature,
before
and after both surfaces have been heated 50 K by a laser pulse.  The 
measurements are not yet good enough to distinguish between the plasma
and the Drude modes of the permittivity, or between the simplified
impedance model versus the measured bulk permittivity approach, as
discussed in \sref{sec:temp}.

A proposal has been made to measure the force between eccentric
cylinders, in which the axes are parallel but slightly offset
\cite{mazzitelliqfext}.  The net force on the inner cylinder
is zero, of course, when the cylinders are concentric, but this 
equilibrium point is unstable.  The idea is to look for a shift in the
mechanical resonant frequency of the outer cylinder due to the Casimir
force exerted by the inner one.  The chief difficulty may be in maintaining
parallelism.

Another active area of experimental effort involving Casimir measurements
is the search for new forces at the submicron level.  These are based on
looking for a discrepancy between the measured and predicted Casimir forces.
The most recent limits are given in Krause, Decca, \etal \cite{krauseqfext,decca03}.
Unfortunately, the limits, for an assumed potential of the form
\be
V(r)=-\frac{Gm_1m_2}{r}\left(1+\alpha\rme^{-r/\lambda}\right),
\ee
for $\lambda<10^{-7}$ m are only for absurdly large strengths, $\alpha\le
10^{14}$, and as $\lambda$ decreases the upper limit on $\alpha$
increases.  The Purdue group has also proposed iso-electronic experiments
to look at the force between a sphere and two different plates, composed
of material with similar electronic properties (and hence similar Casimir
forces) but different nuclear properties (and hence presumably different
new forces). See Ref.~\cite{deccaqfext} for a brief description of
their experiment and the detection of a small, but probably not significant,
residual force.  

Very recently, there has been a report \cite{baessler}
of an experiment \cite{nesvizhevsky} 
of dropping ultracold neutrons onto a surface. 
They are trapped between the mirror and the earth's gravitational
potential. These gravitational bound states would be
modified by any deviation from Newtonian gravity. No such deviations from
Newton's law is found down to the 1--10 nm range.  See also
Nesvizhevsky and Protasov \cite{Nesvizhevsky:2004qb} who obtain limits
on non-Newtonian forces inferior to those of Casimir measurements, that
is, $\alpha<10^{21}$ at $\lambda=10^{-7}$ m, although  it is relatively
better that the Casimir limits in the nanometer range, but the limits
are extremely weak there, $\alpha<10^{26}$. 

It is clear that as micro engineering comes into its own, Casimir forces
will have to be taken into account and utilized.  A recent interesting
paper by Chumak, Milonni, and Berman \cite{chumak} suggests that the noncontact
friction observed by Stipe \etal \cite{stipe} on a cantilever near a surface
is due in major part to Casimir forces.  The Casimir force is responsible
for the frequency shift observed of about 4.5\% for a gold sample at a
separation of 2 nm.

For another example along these lines, Lin \etal \cite{lin} have shown that 
Casimir-Polder forces between atoms and the surface can provide
fundamental limitations on stability of a Bose-Einstein condensate near a
microfabricated silicon chip, a system which holds great promise
for technological applications.

The recent intense experimental activity is very encouraging to
the development of the field.  Coming years, therefore, promise ever 
increasing experimental input into a field that has been dominated by
theory for five decades.

\setcounter{footnote}{1}
\section{Self-Stress}
\label{sec:selfstress}
\subsection{Surface and Volume Divergences}
\label{sec:surfdiv}

It is well known  that in general the Casimir energy
density diverges in the neighborhood of a surface.  For flat surfaces
and conformal theories (such as the conformal scalar theory considered
in Ref.~\cite{Milton:2002vm}, or electromagnetism) those divergences are not 
present.\footnote{In general, this need not be the case.  For example,
Romeo and Saharian \cite{Romeo:2000wt} show that with mixed boundary conditions
the surface divergences need not vanish for parallel plates.  
For additional work on local effects with mixed (Robin)
boundary conditions, applied to spheres and cylinders, and corresponding
global effects, see Refs.~\cite{saharian2,saharian3,Fulling:2003zx}.
See also \sref{sec:2.4} and Ref.~\cite{Graham:2002yr,Olum:2002ra}.}  
We saw hints of this
in \sref{sec:2.4}.  In particular, Brown and Maclay \cite{brown}
calculated the local stress tensor for two ideal plates separated by a distance
$a$ along the $z$ axis, with the result for a conformal scalar
\be
\langle T^{\mu\nu}\rangle
=-\frac{\pi^2}{1440 a^4}[4\hat z^\mu\hat z^\nu-g^{\mu\nu}].
\ee
This result was given recent rederivations in
\cite{Actor:1996zj,Milton:2002vm}.
Dowker and Kennedy \cite{dowkerandkennedy} and Deutsch and
Candelas \cite{deutsch} considered the local
stress tensor between planes inclined at an angle $\alpha$,
with the result, in cylindrical coordinates $(t,r,\theta,z)$,
\be
\langle T^{\mu\nu}\rangle=-\frac{f(\alpha)}{720 \pi^2r^4}\left(
\begin{array}{cccc}
1&0&0&0\\
0&-1&0&0\\
0&0&3&0\\
0&0&0&-1
\end{array}\right),
\ee
where for a conformal scalar, with Dirichlet boundary conditions,
\be
f(\alpha)=\frac{\pi^2}{2\alpha^2}\left(\frac{\pi^2}{\alpha^2}-\frac{\alpha^2}{\pi^2}
\right),
\ee
and for electromagnetism, with perfect conductor boundary conditions,
\be
f(\alpha)=\left(\frac{\pi^2}{\alpha^2}+11\right)\left(\frac{\pi^2}{\alpha^2}-1
\right).
\ee
For $\alpha\to0$ we recover the pressures and energies for parallel plates,
(\ref{strongte}), (\ref{casresult}) and (\ref{dirichletce}).
(These results were later discussed in Ref.~\cite{brevly}.)

Although for perfectly conducting
flat surfaces, the energy density is finite, for electromagnetism
the individual electric and magnetic fields have divergent RMS values,
\be
\langle E^2\rangle\sim-\langle B^2\rangle\sim\frac1{\epsilon^4},
\qquad\epsilon\to0,
\ee
a distance $\epsilon$ above a conducting surface.  However, if the surface
is a dielectric, characterized by a plasma dispersion relation (\ref{plasmad}),
 these divergences are softened
\be
\langle E^2\rangle\sim\frac1{\epsilon^3},\qquad
-\langle B^2\rangle\sim\frac1{\epsilon^2},\qquad\epsilon\to0,
\ee
so that the energy density also diverges \cite{sopovaqfext}
\be
\langle T^{00}\rangle\sim\frac1{\epsilon^3},\qquad\epsilon\to0.
\ee
The null energy condition ($n_\mu n^\mu=0$)
\be
T^{\mu\nu}n_\mu n_\nu\ge0
\label{null}
\ee
is satisfied, so that gravity still focuses light.

Graham \cite{grahamqfext} examined the general relativistic energy conditions
required by causality.  In the neighborhood of a smooth domain wall,
given by a hyperbolic tangent, the energy is always negative
at large enough distances.  Thus the weak energy condition is violated, as is
the null energy condition (\ref{null}).  However, when (\ref{null}) is 
integrated over a complete geodesic, positivity is satisfied.  It is not clear
if this last condition, the Averaged Null Energy Condition, is always obeyed
in flat space.  Certainly it is violated in curved space, but the effects 
always seem small, so that exotic effects such as time travel are prohibited.

However, as Deutsch and Candelas \cite{deutsch} showed many years ago,
in the neighborhood of a curved surface
for conformally invariant theories, $\langle T_{\mu\nu}\rangle$ diverges
as $\epsilon^{-3}$, where $\epsilon$ is the distance from the surface,
with a coefficient proportional to the sum of the principal curvatures of
the surface.  In particular they obtain the result, in the vicinity of
the surface,\index{Stress tensor!vacuum expectation value!conformal}
\begin{equation}
\langle T_{\mu\nu}\rangle\sim\epsilon^{-3}T^{(3)}_{\mu\nu}+
\epsilon^{-2}T^{(2)}_{\mu\nu}+\epsilon^{-1}T^{(1)}_{\mu\nu},
\end{equation}
and obtain explicit expressions for the coefficient tensors $T^{(3)}_{\mu\nu}$
and $T^{(2)}_{\mu\nu}$ in terms of the extrinsic curvature of the boundary.
\index{Divergences!curvature}

For example, for the case of a sphere, the leading surface divergence has the
form, for conformal fields, for $r=a+\epsilon$, $\epsilon\to0$\index{Casimir
stress!local!sphere}\index{Local Casimir effect!sphere}
\begin{equation}
\langle T_{\mu\nu}\rangle={A\over\epsilon^3}\left(\begin{array}{cccc}
2/a&\,0\,&\,0\,&0\\
0&0&0&0\\
0&0&1&0\\
0&0&0&\sin\theta\end{array}\right),
\label{candelas}
\end{equation}
in spherical polar coordinates, where the constant is 
$A=1/1440\pi^2$ for a scalar, or $A=1/120\pi^2$
for the electromagnetic field.
Note that (\ref{candelas}) is properly traceless.  The cubic divergence
in the energy density 
near the surface translates into the quadratic divergence
in the energy found for a conducting ball \cite{miltonballs}.
The corresponding quadratic divergence in the stress corresponds to
the absence of the cubic divergence in $\langle T_{rr}\rangle$.

This is all completely sensible.  However, in their paper Deutsch and Candelas
\cite{deutsch} expressed a certain skepticism about the validity of the result 
of Ref.~\cite{mildersch} for the spherical shell case (described in part in
\sref{Sec3.3})
where the divergences cancel.  That skepticism was reinforced in a later
paper by Candelas \cite{candelas}, who criticized the authors of 
Ref.~\cite{mildersch} for omitting $\delta$ function terms, and constants
in the energy.  These objections seem utterly without merit.  
In a later critical paper by the same author
\cite{candelas2}, it was asserted that errors were made, rather than
a conscious removal of unphysical divergences.

Of course, surface curvature divergences are present. 
As Candelas noted \cite{candelas,candelas2}, they have the form
\begin{eqnarray}
\fl E=E^S\int \rmd S+E^C\int \rmd S\,(\kappa_1+\kappa_2)
+E_I^C\int \rmd S\,(\kappa_1-\kappa_2)^2+E_{II}^C\int \rmd S\kappa_1\kappa_2+
\dots,
\label{divstructure}
\end{eqnarray}
where $\kappa_1$ and $\kappa_2$ are the principal curvatures of the surface.
\index{Principal curvatures} The question is
to what extent are they observable.  After all, as has been shown
in Ref.~\cite{miltonbook,Milton:2002vm} and in \sref{sec:2.4},
 we can drastically change the local structure
of the vacuum expectation value of the energy-momentum tensor
in the neighborhood of flat plates by merely
exploiting the ambiguity in the definition of that tensor, yet each
yields the same finite, observable (and observed!) energy of interaction
between the plates.  For curved boundaries, much the same is true.
{\it A priori}, we do not know which energy-momentum tensor to employ,
and the local vacuum-fluctuation energy density is to a large extent
meaningless.  It is the global energy, or the force between distinct bodies,
that has an unambiguous value.  It is the belief of the author that divergences
in the energy which go like a power of the cutoff are probably unobservable,
being subsumed in the properties of matter. Moreover, the
coefficients of the divergent terms depend on the regularization scheme.
 Logarithmic divergences, of course, are of another class \cite{bkv}.
\index{Divergences!surface|)}\index{Regularization}
\index{Casimir effect!local|)}
\index{Local effects|)}

Dramatic cancellations of these curvature terms can occur.  It might
be thought that the reason a finite result was found for the Casimir
energy of a perfectly conducting spherical shell \cite{boyersphere,balian,
mildersch} is that the term involving the squared difference of curvatures
in (\ref{divstructure}) is zero  only in that case.  However, for reasons not
yet apparent to the present author, it has been shown that at least for
the case of electromagnetism the corresponding term 
is not present (or has a vanishing coefficient) for an arbitrary smooth
cavity \cite{Bernasconi}, and so the Casimir energy for a perfectly conducting
ellipsoid of revolution, for example, is finite.  This finiteness of the
Casimir energy (usually referred to as the vanishing of the second
heat-kernel coefficient \cite{newreview}) for an ideal smooth closed surface
was anticipated already in Ref.~\cite{balian},
but contradicted by Ref.~\cite{deutsch}. More specifically, although
odd curvature terms cancel inside and outside for any thin shell, it would
be anticipated that the squared-curvature term, which is present as a
surface divergence in the energy density, would be reflected as an
unremovable divergence in the energy.  For a closed surface the last term in
(\ref{divstructure}) 
is a topological invariant, so gives an irrelevant constant,
while no term of the type of the penultimate term can appear due to the
structure of the traced cylinder expansion \cite{Fulling:2003zx}.
 It would be extraordinarily
interesting if this Casimir energy could be computed for an ellipsoidal
boundary, but the calculation appears extremely difficult because the 
Helmholtz equation is not separable in the exterior region.

\subsection{Casimir Forces on Spheres via $\delta$-Function Potentials}
\label{Sec3}
This section is an adaptation and an extension of calculations presented
in Ref.~\cite{Milton:2004vy}.  This investigation was carried out in
response to the program of the MIT group \cite{graham,graham2,
Graham:2002fw,Weigel:2003tp,Graham:2003ib}.
They rediscovered irremovable divergences in the Casimir energy 
for a circle in $2+1$ dimensions first 
discovered by Sen \cite{sen,sen2}, but then found divergences in the case of
a spherical surface, thereby casting doubt on the validity of the
Boyer calculation \cite{boyersphere}.  Some of their results, as we shall
see, are spurious, and the rest are well known \cite{bkv}.  However, their
work has been valuable in sparking new investigations of the problems of
surface energies and divergences.

We now carry out the calculation we presented in \sref{Sec1}
in three spatial dimensions,
with a radially symmetric background
\be
\mathcal{L}_{\rm int}=-\frac12\frac{\lambda}a\delta(r-a)\phi^2(x),
\ee
which would correspond to a Dirichlet shell in the limit $\lambda\to\infty$.
The time-Fourier transformed Green's function satisfies the equation
($\kappa^2=-\omega^2$)
\be
\left[-\nabla^2+\kappa^2+\frac{\lambda}a\delta(r-a)\right]\mathcal{G}(\bi{r,r'})=
\delta(\bi{r-r'}).
\ee
We write $\mathcal{G}$ in terms of a reduced Green's function
\be
\mathcal{G}(\bi{r,r'})=\sum_{lm}g_l(r,r')Y_{lm}(\Omega)Y^*_{lm}(\Omega'),
\ee
where $g_l$ satisfies
\be
\left[-\frac1{r^2}\frac{\rmd}{\rmd r}r^2\frac{\rmd}{\rmd r}+
\frac{l(l+1)}{r^2}+\kappa^2
+\frac{\lambda}a\delta(r-a)\right]g_l(r,r')=\frac1{r^2}\delta(r-r').
\label{redgf}
\ee
We solve this in terms of modified Bessel functions, $I_\nu(x)$, $K_\nu(x)$,
where $\nu=l+1/2$, which satisfy the Wronskian condition
\be
I'_\nu(x)K_\nu(x)-K'_\nu(x)I_\nu(x)=\frac1x.
\ee
The solution to (\ref{redgf}) is obtained
 by requiring continuity of $g_l$ at each
singularity, $r'$ and $a$, and the appropriate discontinuity of the
derivative.  Inside the sphere we then find ($0<r,r'<a$)
\be
\fl g_l(r,r')=\frac1{\kappa r r'}\left[e_l(\kappa r_>)s_l(\kappa r_<)
-\frac{\lambda}{\kappa a}s_l(\kappa r)s_l(\kappa r')\frac{e_l^2(\kappa a)}{1
+\frac{\lambda}{\kappa a}s_l(\kappa a)e_l(\kappa a)}\right].
\label{insphgf}
\ee
Here we have introduced the modified Riccati-Bessel functions,
\be
s_l(x)=\sqrt{\frac{\pi x}2}I_{l+1/2}(x),\quad
e_l(x)=\sqrt{\frac{2 x}\pi}K_{l+1/2}(x).
\ee
Note that (\ref{insphgf}) 
reduces to the expected result, vanishing as $r\to a$,
in the limit of strong coupling:
\be
\lim_{\lambda\to\infty} g_l(r,r')=\frac1{\kappa r r'}\left[e_l(\kappa r_>)
s_l(\kappa r_<)-\frac{e_l(\kappa a)}{s_l(\kappa a)}s_l(\kappa r)s_l
(\kappa r')\right].
\label{halfcasimir}
\ee
When both points are outside the sphere, $r,r'>a$, we obtain a similar
result:
\be
\fl g_l(r,r')=\frac1{\kappa r r'}\left[e_l(\kappa r_>)s_l(\kappa r_<)
-\frac{\lambda}{\kappa a}e_l(\kappa r)e_l(\kappa r')\frac{s_l^2(\kappa a)}{1
+\frac{\lambda}{\kappa a}s_l(\kappa a)e_l(\kappa a)}\right].
\label{outsphgf}
\ee
which similarly reduces to the expected result as $\lambda\to\infty$.

Now we want to get the radial-radial component of the stress tensor
to extract the pressure on the sphere, which is obtained by applying the
operator
\be
\partial_r\partial_{r'}-\frac12(-\partial^0\partial^{\prime0}+\bnabla
\cdot\bnabla')\to\frac12\left[\partial_r\partial_{r'}-\kappa^2
-\frac{l(l+1)}{r^2}\right]
\label{radop}
\ee to the Green's function, where in the last term we have 
averaged over the surface of the sphere.
In this way we find, from the discontinuity of
$\langle T_{rr}\rangle$ across the $r=a$ surface, the net stress
\be
\mathcal{S}
=\frac{\lambda}{2\pi a^2}\sum_{l=0}^\infty  (2l+1)\int_0^\infty \rmd x\,
\frac{\left(e_l(x)s_l(x)\right)'-\frac{2e_l(x)s_l(x)}x}{1+\frac{\lambda
e_l(x)s_l(x)}x}.
\label{teforce}
\ee

The same result can be deduced by computing the total energy (\ref{casenergy}).
The free Green's function, the first term in (\ref{insphgf}) or 
(\ref{outsphgf}), evidently makes no significant contribution to the energy,
for it gives a term independent of the radius of the sphere, $a$, so we
omit it.  The remaining radial integrals are simply
\numparts
\label{radints}
\bea
\int_0^x \rmd y\,s_l^2(y)&=\frac1{2x}\left[\left(x^2+l(l+1)\right)s_l^2
+xs_ls_l'-x^2s_l^{\prime2}\right],\label{radinta}\\
\int_x^\infty \rmd y\,e_l^2(y)&=
-\frac1{2x}\left[\left(x^2+l(l+1)\right)e_l^2
+xe_le_l'-x^2e_l^{\prime2}\right],\label{radintb}
\eea
\endnumparts
where all the Bessel functions on the right-hand-sides of these equations
are evaluated at $x$.  Then using the Wronskian, we find that
the Casimir energy is
\be
E=-\frac{1}{2\pi a}\sum_{l=0}^\infty  (2l+1)\int_0^\infty \rmd x\,x\,
\frac{\rmd}{\rmd x}\ln\left[1+\lambda I_\nu(x)K_\nu(x)\right].
\label{teenergy}
\ee
If we differentiate with respect to $a$, with $\lambda/a$ fixed, we
immediately recover the force (\ref{teforce}).  This expression, upon
integration by parts, coincides with that given by Barton \cite{barton03},
and was first analyzed in detail by Scandurra \cite{Scandurra:1998xa}.
For strong coupling,
it reduces to the well-known expression for the Casimir energy
of a massless scalar field  inside and
outside a sphere upon which Dirichlet boundary conditions are imposed,
that is, that the field must vanish at $r=a$:
\be
\lim_{\lambda\to\infty}E=-\frac{1}{2\pi a}\sum_{l=0}^\infty (2l+1)\int_0^\infty
\rmd x\,x\,\frac{\rmd}{\rmd x}\ln\left[I_\nu(x)K_\nu(x)\right],\label{dsph}
\ee
because multiplying the argument of the logarithm by a power of $x$ is
without effect, corresponding to a contact term.  Details of the evaluation
of Eq.~(\ref{dsph}) are given in Ref.~\cite{Milton:2002vm}, and will be
considered in  \sref{Sec3.3} below.  (See also 
Refs.~\cite{benmil,lesed1,lesed2}.)

The opposite limit is of interest here.  The expansion of the logarithm
is immediate for small $\lambda$.  The first term, of order $\lambda$, 
is evidently
divergent, but irrelevant, since that may be removed by renormalization
of the tadpole graph.  In contradistinction to the claim of 
Refs.~\cite{Graham:2002fw,graham2,Graham:2003ib,Weigel:2003tp}, 
the order $\lambda^2$ term is finite,
as established in Ref.~\cite{Milton:2002vm}.  That term is
\be
E^{(\lambda^2)}=\frac{\lambda^2}{4\pi a}
\sum_{l=0}^\infty(2l+1)\int_0^\infty \rmd x\,x
\frac{\rmd}{\rmd x}[I_{l+1/2}(x)K_{l+1/2}(x)]^2.\label{og}
\ee
The sum on $l$ can be carried out using a trick due to Klich \cite{klich}:
The sum rule
\be
\sum_{l=0}^\infty (2l+1)e_l(x)s_l(y)P_l(\cos\theta)=\frac{xy}\rho \rme^{-\rho},
\ee
where $\rho=\sqrt{x^2+y^2-2xy\cos\theta}$, is squared, and then integrated
over $\theta$, according to
\be
\int_{-1}^1 \rmd(\cos\theta) 
P_l(\cos\theta)P_{l'}(\cos\theta)=\delta_{ll'}\frac2{2l+1}.
\ee
In this way we learn that
\be
\sum_{l=0}^\infty (2l+1)e_l^2(x)s_l^2(x)=\frac{x^2}2\int_0^{4x}\frac{\rmd w}w
\rme^{-w}.
\ee
Although this integral is divergent, because we did not integrate by parts
in (\ref{og}), that divergence does not contribute:
\be
E^{(\lambda^2)}=\frac{\lambda^2}{4\pi a}\int_0^\infty \rmd x\,
\frac12 x \,\frac{\rmd}{\rmd x}\int_0^{4x}
\frac{\rmd w}w \rme^{-w}=\frac{\lambda^2}{32\pi a},
\label{4.25}
\ee
which is exactly the result (4.25) of Ref.~\cite{Milton:2002vm}, 
which also follows from (\ref{Ed}) here.

However, before we wax too euphoric, we recognize that the order $\lambda^3$ 
term
appears logarithmically divergent, just as Refs.~\cite{Graham:2003ib} and 
\cite{Weigel:2003tp} 
claim.  This does not signal a breakdown in perturbation
theory, as the divergence (\ref{d=1div})
in the $D=1$ calculation did.  Suppose we
subtract off the two leading terms,
\be
\fl E=-\frac1{2\pi a}\sum_{l=0}^\infty(2l+1)\int_0^\infty \rmd x\,x\,\frac{\rmd}
{\rmd x}
\left[\ln\left(1+\lambda I_\nu K_\nu\right)-\lambda I_\nu K_\nu+
\frac{\lambda^2}2(I_\nu K_\nu)^2
\right]+\frac{\lambda^2}{32\pi a}.\label{fulle}
\ee
To study the behavior of the sum for large values of $l$, we can use the
uniform asymptotic expansion (Debye expansion),
\be
\nu\gg1:\quad I_\nu(x)K_\nu(x)\sim\frac{t}{2\nu}\left[1+\frac{A(t)}{\nu^2}
+\frac{B(t)}{\nu^4}+\dots\right].
\label{uae}
\ee
Here $x=\nu z$, and $t=1/\sqrt{1+z^2}$.  The functions $A$ and $B$, etc., are
polynomials in $t$.  We now insert this into (\ref{fulle}) and expand
not in $\lambda$ but in $\nu$; the leading term is
\be
E^{(\lambda^3)}\sim
\frac{\lambda^3}{24\pi a}\sum_{l=0}^\infty\frac1\nu\int_0^\infty
\frac{\rmd z}{(1+z^2)^{3/2}}=\frac{\lambda^3}{24\pi a}\zeta(1).
\ee
Although the frequency integral is finite, the angular momentum sum is
divergent.  The appearance here of the divergent $\zeta(1)$ seems to
signal an insuperable barrier to extraction of a finite Casimir energy
for finite $\lambda$.  The situation is different in the limit
$\lambda\to\infty$ -- See \sref{Sec3.3}.

This divergence has been known for many years, and was first calculated
explicitly in 1998 by Bordag \etal~\cite{bkv}, where the second heat kernel
coefficient gave an equivalent result,
\be
E\sim \frac{\lambda^3}{48\pi a}\frac1s,\quad s\to0.
\ee
A possible way of dealing with this divergence was advocated in 
Ref.~\cite{Scandurra:1998xa}. Very recently, Bordag and Vassilevich
\cite{Bordag:2004rx}  have
reanalyzed such problems from the heat kernel approach.  They show that this
$\Or(\lambda^3)$ divergence corresponds to a surface tension counterterm,
an idea proposed by me in 1980 \cite{miltonbag,miltonfermion} in connection
with the zero-point energy contribution to the bag model.  Such a surface
term corresponds to $\lambda/a$ fixed, which then necessarily implies
a divergence of order $\lambda^3$.  Bordag argues that it is perfectly
appropriate to insert a surface tension counterterm so that this divergence
may be rendered finite by renormalization.

\subsection{TM Spherical Potential}
\label{Sec3.2}

Of course, the scalar model considered in the previous subsection is merely
a toy model, and something analogous to electrodynamics is of far more
physical relevance.  There are good reasons for believing that cancellations
occur in general between TE (Dirichlet) and TM (Robin) modes.  Certainly
they do occur in the classic Boyer energy of a perfectly conducting spherical
shell \cite{boyersphere, balian, mildersch}, and the indications
are that such cancellations occur even with imperfect boundary conditions
\cite{barton03}.  Following the latter reference, let us consider the
potential
\be
\mathcal{L}_{\rm int}=\frac12\lambda a\frac1r\frac\partial{\partial r}
\delta(r-a)\phi^2(x).
\label{tmlag}
\ee
In the limit $\lambda\to\infty$ this corresponds to TM boundary conditions.
The reduced Green's function is thus taken to satisfy
\be
\left[-\frac1{r^2}\frac{\rmd}{\rmd r}r^2\frac{\rmd}{\rmd r}+\frac{l(l+1)}{r^2}
+\kappa^2-\frac{\lambda a}r\frac\partial{\partial r}\delta(r-a)\right]g_l(r,r')
=\frac1{r^2}\delta(r-r').
\ee
At $r=r'$ we have the usual boundary conditions, that $g_l$ be continuous, but
that its derivative be discontinuous,
\be
r^2\frac{\partial}{\partial r}g_l\bigg|_{r=r'-}^{r=r'+}=-1,
\ee
while at the surface of the sphere the derivative is continuous,
\numparts
\be
\frac\partial{\partial r}rg_l\bigg|_{r=a-}^{r=a+}=0,
\label{dcont}
\ee
while the function is discontinuous,
\be
g_l\bigg|_{r=a-}^{r=a+}=-\lambda \frac\partial{\partial r}rg_l\bigg|_{r=a}.
\label{fndiscont}
\ee
\endnumparts
Equations (\ref{dcont}) and (\ref{fndiscont}) are the analogues of the
boundary conditions (\ref{tmbc1}), (\ref{tmbc2}) treated in \sref{sec:tm11}.
 
 It is then easy to find the Green's function.  When both points are
inside the sphere,
\numparts
\be
\fl r,r'<a:\quad g_l(r,r')=\frac1{\kappa rr'}\left[s_l(\kappa r_<)e_l(\kappa r_>)
-\frac{\lambda\kappa a[e_l'(\kappa a)]^2s_l(\kappa r)s_l(\kappa r')}
{1+\lambda\kappa a e_l'(\kappa a)s_l'(\kappa a)}\right],
\ee
and when both points are outside the sphere,
\be
\fl r,r'>a:\quad g_l(r,r')=\frac1{\kappa rr'}\left[s_l(\kappa r_<)e_l(\kappa r_>)
-\frac{\lambda\kappa a[s_l'(\kappa a)]^2e_l(\kappa r)e_l(\kappa r')}
{1+\lambda\kappa a e_l'(\kappa a)s_l'(\kappa a)}\right].
\ee
\endnumparts
It is immediate that these supply the appropriate Robin boundary conditions
in the $\lambda\to\infty$ limit:
\be
\lim_{\lambda\to0}\frac\partial{\partial r}rg_l\bigg|_{r=a}=0.
\ee

The Casimir energy may be readily obtained from (\ref{casenergy}),
and we find, using the integrals (\ref{radinta}), (\ref{radintb})
\be
E=-\frac1{2\pi a}\sum_{l=0}^\infty (2l+1)\int_0^\infty \rmd x\,x\frac{\rmd}
{\rmd x}
\ln\left[1+\lambda x e_l'(x)s_l'(x)\right].
\label{tmenergy}
\ee
The stress may be obtained from this by applying $-\partial/\partial a$,
and regarding $\lambda a$ as constant [see (\ref{tmlag})], 
or directly, from the Green's function by applying the operator,
\be
t_{rr} =\frac1{2\rmi}\left[\nabla_r\nabla_{r'}-\kappa^2-\frac
{l(l+1)}{r^2}\right]g_l\bigg|_{r'=r},
\ee
which is the same as that in (\ref{radop}), except that
\be
\nabla_r=\frac1r\partial_r r,
\ee
appropriate to TM boundary conditions (see Ref.~\cite{mildim}, for example).
Either way, the total stress on the sphere is
\be
\mathcal{S}=
-\frac\lambda{2\pi a^2}\sum_{l=0}^\infty(2l+1)\int_0^\infty dx\,x^2\frac{[e_l'(x)
s_l'(x)]'}{1+\lambda x e_l'(x)s_l'(x)}.
\ee
The result for the energy (\ref{tmenergy}) is similar, but not identical, to
that given by Barton \cite{barton03}.

Suppose we now combine the TE and TM Casimir energies, (\ref{teenergy}) 
and (\ref{tmenergy}):
\be
\fl E^{\rm TE}+E^{\rm TM}=-\frac1{2\pi a}\sum_{l=0}^\infty (2l+1)\int_0^\infty 
\rmd x\,x\frac \rmd{\rmd x}
\ln\left[\left(1+\lambda\frac{e_ls_l}x\right)\left(1+\lambda x
e_l's_l'\right)\right].
\label{combenergy}
\ee
In the limit $\lambda\to\infty$ this reduces to the familiar expression
for the perfectly conducting spherical shell \cite{mildersch}:
\be
\lim_{\lambda\to\infty}E=-\frac1{2\pi a}\sum_{l=1}^\infty(2l+1)\int_0^\infty
\rmd x\,x\left(\frac{e_l'}{e_l}+\frac{e_l''}{e_l'}+\frac{s_l'}{s_l}
+\frac{s_l''}{s_l'}\right).
\label{spherece}
\ee
Here we have, as appropriate to the electrodynamic situation, omitted the
$l=0$ mode. This expression
 yields a finite Casimir energy, as we will see in \sref{Sec3.3}. 
What about finite $\lambda$?  In general,
it appears that there is no chance that the divergence found in the previous
section in order $\lambda^3$ can be cancelled.  But suppose the coupling
for the TE and TM modes are different.  If $\lambda^{\rm TE}\lambda^{\rm 
TM}=4$, a cancellation appears possible.

Let us illustrate this by retaining only the leading terms in the uniform
 asymptotic expansions: ($x=\nu z$)
\be
\frac{e_l(x)s_l(x)}x\sim\frac{t}{2\nu},\qquad x e_l'(x)s_l'(x)\sim 
-\frac\nu{2t},\quad\nu\to\infty.
\ee
Then the logarithm appearing in the integral for the energy (\ref{combenergy}) is
approximately
\be
\ln\sim\ln\left(-\frac{\lambda^{\rm TM}\nu}{2t}\right)+\ln\left(1+\frac
{\lambda^{\rm TE}t}{2\nu}\right)+\ln\left(1-\frac{2t}{\lambda^{\rm TM}\nu}
\right).
\ee
The first term here presumably gives no contribution to the energy,
because it is independent of $\lambda$ upon differentiation, and further
we may interpret $\sum_{l=0}^\infty\nu^2=0$ [see (\ref{sumzeta})].  
Now if we make the above identification of
the couplings, 
\be
\hat\lambda=\frac{\lambda^{\rm TE}}2=\frac2{\lambda^{\rm TM}},
\label{hatl}
\ee
all the odd powers of $\nu$ cancel out, and
\be
E\sim -\frac1{2\pi a}\sum_{l=0}^\infty(2l+1)\int_0^\infty \rmd x\,x 
\frac{\rmd}{\rmd x}
\ln\left(1-\frac{{\hat\lambda}^2 t^2}{\nu^2}\right).
\label{leadingtetm}
\ee
The divergence encountered for the TE mode is thus removed, and the power
series is simply twice the sum of the even terms there.  This will be
finite.  Presumably, the same is true if the subleading terms in the uniform
asymptotic expansion are retained.

It is interesting to approximately evaluate (\ref{leadingtetm}).
The integral over $z$ may be easily evaluated as a contour integral,
leaving
\be
E\sim-\frac1{a}\sum_{l=0}^\infty \nu^2
\left(1-\sqrt{1-\frac{{\hat\lambda}^2}{\nu^2}}\right).
\ee
This $l$ sum appears to be divergent, an artifact of the asymptotic
expansion, since we know the $\lambda^2$ term is finite.  However, if we expand
the square root for small ${\hat\lambda}^2/\nu^2$, we see that the
$\Or({\hat\lambda}^2)$ term vanishes if we interpret the sum as
\be
\sum_{l=0}^\infty \nu^{-s}=(2^s-1)\zeta(s),
\label{sumzeta}
\ee
in terms of the Riemann zeta function.  The leading term is 
$\Or({\hat\lambda}^4)$:
\be
E\sim-\frac{{\hat\lambda}^4}{8a}\sum_{l=0}^\infty \frac1{\nu^2}=
-\frac{{\hat\lambda}^4\pi^2}
{16a}.
\ee
To recover the correct leading $\lambda$ behavior in (\ref{4.25}) requires
the inclusion of the subleading $\nu^{-2n}$ terms displayed in (\ref{uae}).

Much faster convergence is achieved if we consider the results with the
$l=0$ term removed, as appropriate for electromagnetic modes.  Let's illustrate
this for the order $\lambda^2$ TE mode (now, for simplicity, write $\lambda
=\lambda^{\rm TE}$)  Then, in place of the energy (\ref{4.25}) we have
\be
\fl \tilde E^{(\lambda^2)}=\frac{\lambda^2}{32\pi a}+\frac{\lambda^2}{4\pi a}
\int_0^\infty \frac{\rmd x}{x^2}\sinh^2x \,e^{-2x}=\frac{\lambda^2}a\left(
\frac1{32\pi}+\frac{\ln2}{4\pi}\right)=\frac{\lambda^2}a(0.065\,106\,1).
\label{exact}
\ee
Now the leading term in the uniform asymptotic expansion is no longer zero:
\bea
E^{(0)}&=-\frac1{2\pi a}\sum_{l=1}^\infty (2l+1)\int_0^\infty \rmd x \, x
\frac{\rmd}{\rmd x}\left(-\frac{\lambda^2 t^2}{8\nu^2}\right)\nonumber\\
&=\frac{\lambda^2}{8\pi a}\sum_{l=1}^\infty \nu^0\left(-\frac\pi2\right)
=\frac{\lambda^2}{16a}=\frac{\lambda^2}a(0.0625),
\eea
which is 4\% lower than the exact answer (\ref{exact}).  The next term
in the uniform asymptotic expansion is
\bea
E^{(2)}&=-\frac{\lambda^2}{4\pi a}[3\zeta(2)-4]\int_0^\infty \rmd z\,t^2\frac
{t^2-6t^4+5t^6}8\nonumber\\
&=\frac{\lambda^2}a\left(\frac{3\pi^2}{2048}-\frac{3}{256}
\right)=\frac{\lambda^2}a(0.002\,736\,8),
\eea
which reduces the estimate to
\be
E^{(0)}+E^{(2)}=\frac{\lambda^2}a(0.065\,236\,8), 
\ee
which is now 0.2\% high.  Going out one more term gives
\bea
\fl E^{(4)}=-\frac{\lambda^2}{8\pi a}\left[15\zeta(4)-16\right]\int_0^\infty
\rmd z\,t^2\frac{t^4}{16}(7-148t^2+554t^4-708t^6+295t^8)\nonumber\\
\lo =-\frac{\lambda^2}a\left(\frac{59\pi^4}{524288}-\frac{177}
{16328}\right)=-\frac{\lambda^2}a(0.000\,158\,570),
\eea
and the estimate for the energy is now only 0.04\% low:
\be
E^{(0)}+E^{(2)}+E^{(4)}=\frac{\lambda^2}a(0.065\,078\,23).
\ee

We could also make similar remarks about the TM contributions.  

\subsection{Perfectly Conducting Spherical Shell}
\label{Sec3.3}
  Now we consider a massless scalar in three space
dimensions, with a spherical boundary on which the field vanishes.
This corresponds to the TE modes for the electrodynamic situation first
solved by Boyer \cite{boyersphere,balian,mildersch}.   The purpose of this 
section (adapted from Ref.~\cite{Milton:2002vm}) is to
emphasize anew that, contrary to the implication of Ref.~\cite{graham2,
Graham:2002fw,Graham:2003ib,Weigel:2003tp},
the corresponding Casimir energy is also finite for this configuration.

 The general calculation in $D$ spatial
dimensions was given in Ref.~\cite{benmil}; the pressure is
given by the formula
\be
\fl P=-\sum_{l=0}^\infty\frac{(2l+D-2)\Gamma(l+D-2)}{l!2^D\pi^{(D+1)/2}
\Gamma(\frac{D-1}2)a^{D+1}}\int_0^\infty \rmd x\,x\frac{\rmd}{\rmd x}\ln\left[I_\nu(x)
K_\nu(x)x^{2-D}\right].\label{3.1}\ee
Here $\nu=l-1+D/2$.  For $D=3$ this expression reduces to
\be
P=-\frac1{8\pi^2a^4}\sum_{l=0}^\infty(2l+1)\int_0^\infty\rmd x\,x
\frac{\rmd}
{\rmd x}\ln\left[I_{l+1/2}(x)K_{l+1/2}(x)/x\right].
\label{fsphere}
\ee
This precisely corresponds to the strong limit $\lambda\to\infty$ given
in (\ref{dsph}), if we recall the comment made about contact terms there.
In Ref.~\cite{benmil} we evaluated  expression (\ref{3.1}) 
by continuing in $D$ from
a region where both the sum and integrals existed.  In that way, a completely
finite result was found for all positive $D$ not equal to an even integer.

Here we will adopt a perhaps more physical approach, that of allowing the 
time-coordinates in the underlying Green's function to approach each other,
as described in Ref.~\cite{mildersch}.  That is, we recognize that the $x$
integration above is actually a (dimensionless) imaginary 
frequency integral, and therefore we should replace
\be
\int_0^\infty \rmd x\,f(x)=\frac12\int_{-\infty}^\infty \rmd y\,\rme^{\rmi
y\delta}f(|y|),
\label{timesplit}
\ee
where at the end we are to take $\delta\to0$.  Immediately, we can
replace the $x^{-1}$ inside the logarithm in (\ref{fsphere})
 by $x$, which makes the integrals
converge, because the difference is proportional to a delta function in
the time separation, a contact term without physical significance.

To proceed, we use the uniform asymptotic expansions for the modified
Bessel functions, (\ref{uae}).  This
is an expansion in inverse powers of $\nu=l+1/2$, low terms in which turn
out to be remarkably accurate even for modest $l$.  The leading terms
in this expansion are
\be
\ln\left[x I_{l+1/2}(x)K_{l+1/2}(x)\right]
\sim\ln\frac{zt}2+\frac1{\nu^2}g(t)+\frac1{\nu^4}
h(t)+\dots,
\label{uae2}
\ee
where $x=\nu z$ and $t=(1+z^2)^{-1/2}$.
Here
\numparts
\bea
g(t)&=&\frac18(t^2-6t^4+5t^6),\\
h(t)&=&\frac1{64}(13t^4-284t^6+1062t^8-1356t^{10}+565t^{12}).
\eea
\endnumparts
The leading term in the pressure is therefore
\bea
\fl P_0=
-\frac1{8\pi^2a^4}\sum_{l=0}^\infty(2l+1)\nu\int_0^\infty \rmd z\, t^2
=-\frac1{8\pi a^4}\sum_{l=0}^\infty\nu^2=\frac3{32\pi a^4}\zeta(-2)=0.
\eea
where in the last step we have used the formal zeta function 
evaluation (\ref{sumzeta}).\footnote{Note that the corresponding TE contribution for
the electromagnetic Casimir pressure would not be zero, for there the sum
starts from $l=1$.}  Here the rigorous way to argue is to recall the
presence of the point-splitting factor $\rme^{\rmi \nu z\delta}$ and to carry out
the sum on $l$ using
\be
\sum_{l=0}^\infty \rme^{\rmi\nu z\delta}=-\frac1{2\rmi}\frac1{\sin z\delta/2},
\label{lsum}
\ee
so
\bea
\sum_{l=0}^\infty \nu^2\rme^{\rmi\nu z\delta}=-\frac{\rmd^2}{\rmd(z\delta)^2}
\frac{\rmi}
{2\sin z\delta/2}
=\frac{\rmi}8\left(-\frac2{\sin^3z\delta/2}+\frac1{\sin z\delta/2}\right).
\eea
Then $P_0$ is given by the divergent expression
\be
P_0=\frac{\rmi}{\pi^2 a^4\delta^3}\int_{-\infty}^\infty \frac{\rmd z}{z^3}
\frac1{1+z^2},
\ee
which we argue is zero because the integrand is odd, as justified by averaging
over contours passing above and below the pole at $z=0$.

The next term in the uniform asymptotic expansion (\ref{uae2}), 
that involving $g$, 
likewise gives zero pressure, as intimated by the formal zeta function identity
(\ref{sumzeta}),
which vanishes at $s=0$.  The same conclusion follows from point splitting,
using (\ref{lsum}) and arguing that the resulting integrand
$\sim z^2t^3 g'(t)/z\delta$ is odd in $z$.
  Again, this cancellation does not occur in the electromagnetic case
because there the sum starts at $l=1$.

So here the leading term which survives is that of order $\nu^{-4}$
in (\ref{uae2}),
namely
\be
P_2=\frac1{4\pi^2a^4}\sum_{l=0}^\infty \frac1{\nu^2}\int_0^\infty
\rmd z \,h(t),
\ee
where we have now dropped the point-splitting factor because this expression
is completely convergent.  The integral over $z$ is
\be
\int_0^\infty \rmd z \, h(t)=\frac{35\pi}{32768}
\ee and the sum over $l$ is $3\zeta(2)=\pi^2/2$, so the leading contribution
to the stress on the sphere is
\be
{\cal S}_2=4\pi a^2P_2=\frac{35\pi^2}{65536a^2}=\frac{0.00527094}{a^2}.
\ee
Numerically this is a terrible approximation.

What we must do now is return to the full expression and add and subtract
the leading asymptotic terms.  This gives
\be
{\cal S}={\cal S}_2-\frac1{2\pi a^2}\sum_{l=0}^\infty(2l+1)R_l,
\ee
where 
\be
R_l=Q_l+\int_0^\infty \rmd x\left[\ln zt+\frac1{\nu^2}g(t)+\frac1{\nu^4}h(t)\right],
\label{remainder}
\ee
where the integral
\be
Q_l=-\int_0^\infty \rmd x\ln[2xI_\nu(x)K_{\nu}(x)]
\ee
was given the asymptotic form  in Ref.~\cite{benmil,miltonbook} ($l\gg1$):
\bea
\fl Q_l\sim\frac{\nu\pi}2+\frac\pi{128\nu}-\frac{35\pi}{32768\nu^3}
+\frac{565\pi}{1048577\nu^5}
-\frac{1208767\pi}{2147483648\nu^7}
+\frac{138008357\pi}{137438953472\nu^9}.
\label{ql}
\eea
The first two terms in (\ref{ql}) cancel the second and third terms in 
(\ref{remainder}), of course.
The third term in (\ref{ql}) corresponds to $h(t)$, so the last three terms 
displayed in (\ref{ql}) give the asymptotic behavior of the remainder,
which we call $w(\nu)$.  Then we have, approximately,
\be
{\cal S}\approx {\cal  S}_2-\frac1{\pi a^2}\sum_{l=0}^n\nu R_l-\frac1{\pi a^2}
\sum_{l=n+1}^\infty \nu w(\nu).
\ee
For $n=1$ this gives ${\cal S}\approx0.002\,852\,78/a^2$, and for larger $n$
this rapidly approaches the value first  given in Ref.~\cite{benmil},
and rederived in \cite{lesed1,lesed2,lesedflux}
\be
{\cal S}^{\rm TE}=0.002817/a^2,
\ee
a value much smaller than the famous electromagnetic result \cite{boyersphere,
davis,mildersch,balian},
\be
{\cal S}^{\rm EM}=\frac{0.04618}{a^2},
\label{boyerresult}
\ee
because of the cancellation of the leading terms noted above.
Indeed, the TM contribution was calculated separately in Ref.~\cite{mildim},
with the result
\be
\mathcal{S}^{\rm TM}=-0.02204\frac1{a^2},
\ee
and then subtracting the $l=0$ modes from both contributions we obtain
(\ref{boyerresult})
\be
{\cal S}^{\rm EM}=\mathcal{S}^{\rm TE}+\mathcal{S}^{\rm TM}+\frac{\pi}{48a^2}
=\frac{0.0462}{a^2}.
\ee

\subsection{Dielectric Spheres}

The Casimir self-stress on a uniform dielectric sphere was first worked
out in 1979 \cite{miltonballs}.  It was generalized to the case when
both electric permittivity and magnetic permeability are present in 1997
\cite{sonokm}.  Since this calculation is summarized in my monograph  
\cite{miltonbook}, we content ourselves here with simply stating the result
for the pressure, ($x=\sqrt{\varepsilon\mu}|y|$, $x'=\sqrt{\varepsilon'\mu'}|y|$
where $\varepsilon',\mu'$ are the interior, and $\varepsilon, \mu$ are the
exterior, values of the permittivity and the permeability)
\begin{eqnarray}
P=-{1\over2a^4}\int_{-\infty}^\infty{\rmd y\over2\pi}\rme^{\rmi y\delta}
\sum_{l=1}^\infty{2l+1\over4\pi}
\Bigg\{x{\rmd\over \rmd x}\ln D_l\nonumber\\
\mbox{}+2x'[ s_l'(x') e_l'(x')- e_l(x') s_l''(x')]
-2x[ s_l'(x) e_l'(x)- e_l(x) s_l''(x)]\Bigg\}\;,
\label{stress}
\end{eqnarray}
where the ``bulk'' pressure has been subtracted, and
\begin{equation}
D_l=[s_l(x') e_l'(x)-
 s_l'(x')e_l(x)]^2-\xi^2[s_l(x') e_l'(x)+
s_l'(x') e_l(x)]^2,
\end{equation}
with the parameter $\xi$ being
\begin{equation}
\xi=\frac{\sqrt{\frac{\varepsilon'}{\epsilon}\frac{\mu}{\mu'}}-1}{
\sqrt{\frac{\epsilon'}{\epsilon}\frac{\mu}{\mu'}}+1},
\end{equation}
and $\delta$ is the temporal regulator introduced in (\ref{timesplit}).
This result is obtained either by computing the radial-radial component
of the stress tensor, or from the total energy.

In general, this result is divergent.  However, consider
the special case
$\sqrt{\epsilon\mu}=\sqrt{\epsilon'\mu'}$, that is, when the speed of light
is the same in both media.
Then $x=x'$ and the Casimir energy 
derived from (\ref{stress}) reduces to
\begin{equation}
E=4\pi a^3P=-\frac{1}{4\pi a}\int_{-\infty}^\infty \rmd y\,
\rme^{\rmi y\delta}\sum_{l=1}^\infty
(2l+1)x\frac{\rmd}{\rmd x}\ln[1-\xi^2((s_le_l)')^2],
\label{special}
\end{equation}
where
\begin{equation}
\xi=\frac{\mu-\mu'}{\mu+\mu'}=-\frac{\varepsilon-\varepsilon'}
{\varepsilon+\varepsilon'}.
\label{emu}
\end{equation}
If $\xi=1$ we recover the case of a perfectly conducting spherical
shell, treated in \sref{Sec3.3} [cf.\ (\ref{spherece})], 
for which $E$ is finite.  In fact (\ref{special})
is finite for all $\xi$.

Of particular interest is the dilute limit, where \cite{klich}
\begin{equation}
E\approx\frac{5\xi^2}{32\pi a}=\frac{0.099\,4718\xi^2}{2a}, \quad \xi\ll1.
\label{smallxisphere}
\end{equation}
[This evaluation is carried out in the same manner as that of (\ref{og}).]
It is  remarkable that
the value for a spherical conducting shell
(\ref{boyerresult}), for which $\xi=1$, is only 7\% lower, which as
Klich remarks, is accounted for nearly entirely by the next term in the
small $\xi$ expansion.

There is another dilute limit which is also quite surprising.  For a purely
dielectric sphere ($\mu=1$) the leading term in an expansion in powers
of $\varepsilon-1$ is finite \cite{bmm,barton,bkv,hb}:
\be
E=\frac{23}{1536\pi}\frac{(\varepsilon-1)^2}a=(\varepsilon-1)^2\frac{0.004\,767}
{a}.
\label{dilutesph}
\ee
This result coincides with the sum of van der Waals energies of the
material making up the ball \cite{sonokm2}.
The term of order $(\varepsilon-1)^3$ is divergent \cite{bkv}.
The establishment of the result (\ref{dilutesph}) 
was the death knell for the Casimir
energy explanation of sonoluminescence \cite{sonorev02} -- See \sref{sec:dce}.

The temperature correction to this result was first worked out
by Nesterenko, Lambiase, and Scapetta \cite{scar,scar2}.  See also
Ref.~\cite{vvn}.

\subsection{Cylinders}
It is much more difficult to carry out Casimir calculations for
cylindrical geometries.  We restrict our attention here to cylinders
of circular cross section and infinite length.  
Although calculations have been carried out
for parallelopiped geometries 
\cite{lukosz,lukosz1,lukosz2,ruggiero,ruggiero2,ambjorn,caruso2,caruso,
actor3,actor,li,queiroz}, 
the effects included refer only to the
interior modes of oscillation.  This is because the wave equation is
not separable outside a cube or a rectangular solid.  As a result,
divergences occur which cannot be legitimately removed, which nevertheless
are artificially removed by zeta-function methods.  It is the view of the
author that such finite results are without meaning.

But even though circular-cylinder calculations are possible, they are
considerably more complex than the corresponding spherical calculations.
This is not merely because spherical Bessel functions are simpler than
cylinder functions.  The fundamental difficulty in these geometries is
that there is in general no decoupling between TE and TM modes \cite{stratton}.
Progress in understanding has therefore been much slower in this regime.
It was only in 1981 that it was found that the electromagnetic Casimir
energy of a perfectly conducting cylinder was attractive, the energy
per unit length being \cite{deraadcyl}
\be
\mathcal{E}_{\mathrm{em,cyl}}=-\frac{0.01356}{a^2},
\ee
for a circular cylinder of radius $a$.  The corresponding result for a
scalar field satisfying Dirichlet boundary conditions of the cylinder is
repulsive \cite{nestcyl},
\be
\mathcal{E}_{\mathrm{D,cyl}}=\frac{0.000606}{a^2}.
\ee

These ideal limits are finite, but, as with the spherical geometry, less
ideal configurations have unremovable divergences.  For example, a cylindrical
$\delta$-shell potential, as described earlier, has divergences (in third
order) \cite{scancyl}.  And it is expected that a dielectric cylinder
will have a divergent Casimir energy, although the coefficient of 
$(\varepsilon-1)^2$ will be finite for a dilute dielectric cylinder 
\cite{borpiro},
corresponding to a finite van der Waals energy between the molecules
that make up the material \cite{dicyl}.  Recent progress in understanding these
points will be described below.

\subsubsection{Dielectric cylinders}
The following calculation represents work in progress with Ines Cavero-Pelaez.
Although the calculation remains incomplete, we offer it here as a detailed 
example of how a complicated electromagnetic calculation is formulated in
the Green's function approach.
We start from the equations satisfied by the Green's dyadics for Maxwell's
equations in a medium characterized by a permittivity $\varepsilon$ and
a permeability $\mu$
(see Ref.~\cite{miltonballs}):
\numparts
\begin{eqnarray}
\bnabla\times\bGamma'-\rmi\omega\mu\bPhi=\frac{1}{\varepsilon}
\bnabla\times\mathbf{1},\label{me1}\\
-\bnabla\times\bPhi-\rmi\omega\varepsilon\bGamma'=\mathbf{0},\label{me2}
\end{eqnarray}
\endnumparts
where
\begin{equation}
\bGamma'(\bi{r,r'},\omega)=\bGamma(\bi{ r,r'};\omega)+\frac{\mathbf{1}}{
\varepsilon(\omega)},
\end{equation}
and where the unit dyadic
$\mathbf{1}$ includes a three-dimensional $\delta$ function,
\begin{equation}
\mathbf{1}=\mathbf{1}\delta(\bi{r-r'}).
\end{equation}
The two dyadics are solenoidal,
\numparts
\begin{eqnarray}
\bnabla\cdot \bPhi=\mathbf{0},\\
\bnabla\cdot \bGamma'=\mathbf{0}.
\end{eqnarray}
\endnumparts
The corresponding second-order equations are
\numparts
\begin{eqnarray}
(\nabla^2+\omega^2\varepsilon\mu)\bGamma'=-\frac{1}{\varepsilon}\bnabla
\times(\bnabla\times\mathbf{1}),\\
(\nabla^2+\omega^2\varepsilon\mu)\bPhi=\rmi\omega\bnabla\times\mathbf{1}.
\label{helmholtz}
\end{eqnarray}
\endnumparts

Classically, these Green's dyadic equations are equivalent to Maxwell's
equations, and give the solution thereto when a polarization source $\bi{P}$
is present,
\begin{equation}
\bi{E}(x)=\int (\rmd x')\bGamma(x,x')\cdot \bi{P}(x').
\end{equation}
Quantum mechanically, they give the one-loop vacuum expectation values of the
product of fields (at a given frequency $\omega$)
\numparts
\begin{eqnarray}
\langle\bi{E(r)E(r')}\rangle&=\frac{\hbar}{\rmi}\bGamma(\bi{r,r'}),
\label{evev}\\
\langle\bi{H(r)H(r')}\rangle&=-\frac{\hbar}{\rmi}\frac{1}{\omega^2\mu^2}
\bnabla\times\bGamma(\bi{r,r'})\times\overleftarrow
{\bnabla}'.\label{mvev}
\end{eqnarray}
\endnumparts
Thus, from knowledge of the classical Green's dyadics, we can calculate
the one-loop vacuum energy or stress.

We now introduce the appropriate partial wave decomposition for a cylinder,
a slight modification of that given for a conducting cylindrical shell
\cite{deraadcyl}\footnote{It might be thought that we could immediately
use the general waveguide decomposition of modes into those of TE and
TM type, for example as given in Ref.~\cite{jskmer}.  However, this is
here impossible because the TE and TM modes do not separate.  See 
Ref.~\cite{stratton}.}:
\numparts
\begin{eqnarray}
\bGamma'(\bi{r,r'};\omega)=\sum_{m=-\infty}^\infty\int_{-\infty}
^\infty\frac{\rmd k}{2\pi}\bigg\{(\bnabla\times\hat\bi{z})f_m(r;k,\omega)
\chi_{mk}(\theta,z)\nonumber\\
\mbox{}+\frac{\rmi}{\omega\varepsilon}\bnabla\times(\bnabla
\times\hat\bi{z})g_m(r;k,\omega)\chi_{mk}(\theta,z)\bigg\},\\
\bPhi(\bi{r,r'};\omega)=\sum_{m=-\infty}^\infty\int_{-\infty}
^\infty\frac{\rmd k}{2\pi}\bigg\{(\bnabla\times\hat\bi{z})\tilde
g_m(r;k,\omega)
\chi_{mk}(\theta,z)\nonumber\\
\mbox{}-\frac{\rmi\varepsilon}{\omega\mu}\bnabla\times(\bnabla
\times\hat\bi{z})\tilde f_m(r;k,\omega)\chi_{mk}(\theta,z)\bigg\},
\end{eqnarray}
\endnumparts
where the cylindrical harmonics are
\begin{equation}
\chi_{mk}(\theta,z)=\frac{1}{\sqrt{2\pi}}\rme^{\rmi m\theta}\rme^{\rmi kz},
\end{equation}
and the dependence of $f_m$ etc. on $\bi{r'}$ is implicit (they are further
vectors in the second tensor index).
Because of the presence of these harmonics, we have
\numparts
\begin{eqnarray}
\bnabla\times\hat\bi{z}\to \hat\bi{r}\frac{\rmi m}{r}-\boldsymbol{\hat\theta}
\frac{\partial}{\partial r}\equiv\boldsymbol{\mathcal{M}},\\
\bnabla\times(\bnabla\times\hat\bi{z})\to \hat\bi{r}\rmi k\frac{\partial}
{\partial r}-\boldsymbol{\hat\theta}
\frac{mk}{r}-\hat\bi{z} d_m\equiv\boldsymbol{\mathcal{N}},
\end{eqnarray}
\endnumparts
in terms of the cylinder operator
\begin{equation}
d_m=\frac{1}{r}\frac{\partial}{\partial r}r\frac{\partial}{\partial r}
-\frac{m^2}{r^2}.
\end{equation}
Now if we use the Maxwell equation (\ref{me2}) we conclude\footnote{The
ambiguity in solving for these equations is absorbed in the definition
of subsequent constants of integration.}
\numparts
\begin{eqnarray}
\tilde g_m=g_m,\\
(d_m-k^2)\tilde f_m=-\omega^2\mu f_m.
\end{eqnarray}
\endnumparts

From the other Maxwell equation (\ref{me1}) we deduce (we now make the second,
previously suppressed, position arguments explicit) (the prime on the
differential operator signifies action on the second, primed argument)
\numparts
\begin{eqnarray}
d_m\mathcal{D}_m\tilde \bi{f}_m(r;r',\theta',z')=\frac{\omega^2\mu}{\varepsilon}
\chi^*_{mk}(\theta',
z')\boldsymbol{\mathcal{M}}^{\ast\prime}\frac{1}{r}\delta(r-r'),\\
d_m\mathcal{D}_m \bi{g}_m(r;r',\theta',z')=-\rmi\omega
\chi^*_{mk}(\theta',z')\boldsymbol{\mathcal{N}}^{\ast\prime}\frac{1}{r}
\delta(r-r'),
\end{eqnarray}
\endnumparts
where the Bessel operator appears,
\begin{equation}
\mathcal{D}_m=d_m+\lambda^2,\qquad \lambda^2=\omega^2\varepsilon\mu-k^2.
\end{equation}
Now we do the separation of variables on the second argument,
\numparts
\begin{eqnarray}
\fl\tilde\bi{f}_m(r,\bi{r'})&=\left[\boldsymbol{\mathcal{M}}^{\ast\prime}
F_m(r,r';k,\omega)+\boldsymbol{\mathcal{N}}^{\ast\prime}
\tilde F_m(r,r';k,\omega)\right]
\chi_{mk}^*(\theta',z'),\\
\fl\bi{g}_m(r,\bi{r'})&=-\frac{\rmi}{\omega}
\left[\boldsymbol{\mathcal{N}}^{\ast\prime}G_m(r,r';k,\omega)+
\boldsymbol{\mathcal{M}}^{\ast\prime}\tilde G_m(r,r';k,\omega)\right]
\chi_{mk}^*(\theta',z'),
\end{eqnarray}
\endnumparts
where we have now introduced the two scalar Green's functions $F_m$, $G_m$,
which satisfy
\numparts
\begin{eqnarray}
d_m\mathcal{D}_mF_m(r,r')&=\frac{\omega^2\mu}{\varepsilon}
\frac{1}{r}\delta(r-r'),\label{defm}\\
d_m\mathcal{D}_mG_m(r,r')&=\omega^2\frac{1}{r}\delta(r-r'),
\end{eqnarray}
\endnumparts
while $\tilde F_m$ and $\tilde G_m$ are annihilated by the operator
$d_m\mathcal{D}_m$.

In the following we will apply these equations to a dielectric-diamagnetic
cylinder of radius $a$, where the interior of the cylinder is characterized
by a permittivity $\varepsilon$ and a permeability $\mu$, while the outside
is vacuum, so $\varepsilon=\mu=1$ there.  Let us compute the Green's dyadics
for the case that the source point is outside, $r'>a$.  If the field point 
is also outside, $r,r'>a$, the Green's dyadics have the form
($\mu=\varepsilon=1$)
\numparts
\bea
\fl\bGamma'=\sum_{m=-\infty}^\infty\int_{-\infty}
^\infty\frac{\rmd k}{2\pi}\bigg\{\boldsymbol{\mathcal{M}}\left(-\frac{d_m-k^2}
{\omega^2}\right)\left[\boldsymbol{\mathcal{M}}^{\ast\prime}
F_m(r,r';k,\omega)+\boldsymbol{\mathcal{N}}^{\ast\prime}
\tilde F_m(r,r';k,\omega)\right]\nonumber\\
\fl\qquad\mbox{}+\frac1{\omega^2}\boldsymbol{\mathcal{N}}
\left[\boldsymbol{\mathcal{N}}^{\ast\prime}G_m(r,r';k,\omega)+
\boldsymbol{\mathcal{M}}^{\ast\prime}\tilde G_m(r,r';k,\omega)\right]\bigg\}
\chi_{mk}(\theta,z)\chi_{mk}^*(\theta',z'),\\
\fl\bPhi=\sum_{m=-\infty}^\infty\int_{-\infty}
^\infty{\rmd k\over2\pi}\bigg\{\frac{\rm i}\omega
\boldsymbol{\mathcal{M}}\left[\boldsymbol{\mathcal{N}}^{\ast\prime}
G_m(r,r';k,\omega)+\boldsymbol{\mathcal{M}}^{\ast\prime}
\tilde G_m(r,r';k,\omega)\right]\nonumber\\
\fl\qquad\mbox{}-\frac\rmi{\omega}\boldsymbol{\mathcal{N}}
\left[\boldsymbol{\mathcal{M}}^{\ast\prime}F_m(r,r';k,\omega)+
\boldsymbol{\mathcal{N}}^{\ast\prime}\tilde F_m(r,r';k,\omega)\right]\bigg\}
\chi_{mk}(\theta,z)\chi_{mk}^*(\theta',z').
\eea
\endnumparts
From the differential equation (\ref{defm}) we see that the Green's function $F$
has the form ($m\ne0$)
\bea
\fl F_m=-\frac{\omega^2}{\lambda^2}\left[\frac1{2|m|}\left(\frac{r_<}{r_>}\right)^{|
m|}+\frac\pi{2\rmi}J_m(kr_<)H_m(kr_>)\right]\nonumber\\
\fl\qquad\mbox{}+ a_mH_m(\lambda r)H_m(\lambda r')+b_m r^{-|m|}H_m(\lambda r')
+c_m r^{\prime-|m|}H_m(\lambda r)+d_m r^{-|m|}r^{\prime-|m|},
\eea
while $G_m$ has the same form with the constants $a_m$, $b_m$, $c_m$, $d_m$
replaced by $a'_m$, $b'_m$, $c'_m$, $d'_m$, respectively.
The homogeneous functions have the form
\bea
\fl \tilde F_m=\tilde a_mH_m(\lambda r)H_m(\lambda r')+\tilde b_m r^{-|m|}H_m(\lambda r')
+\tilde c_m r^{\prime-|m|}H_m(\lambda r)+\tilde d_m r^{-|m|}r^{\prime-|m|},
\eea
and $\tilde G_m$ replaces $\tilde a\to \tilde a'$, etc.

When the source point is outside and the field point is inside, there are only
homogeneous solutions of the equations, so we may write for
$r<a, r'>a$
\be
\fl F_m=e_m r^{|m|}r^{\prime-|m|}+f_m r^{|m|}
H_m(\lambda r')+g_m J_m(\lambda' r)
r^{\prime -|m|}+h_m J_m(\lambda'r)H_m(\lambda r'),
\ee
and similarly for $G_m$, $\tilde F_m$, $\tilde G_m$, with the constants
denoted by $e_m'$, $\tilde e_m$, and $\tilde e_m'$, respectively.
Here the outside and inside forms of $\lambda$ are given by
\be
\lambda^2=\omega^2-k^2,\qquad \lambda^{\prime2}=\omega^2\mu\varepsilon-k^2.
\ee

The various constants are to be determined, as far as possible, by the
boundary conditions at $r=a$.
The boundary conditions at the surface of the dielectric cylinder are
the continuity of the tangential components of the electric field,
of the normal component of the electric displacement, of 
the normal component of the magnetic induction, and of the tangential
components of the magnetic field:
\begin{eqnarray}
\bi{E}_t\quad\mbox{is continuous},\qquad\varepsilon E_n\quad
\mbox{is continuous},\nonumber\\
\bi{H}_t\quad\mbox{is continuous},\qquad \mu H_n\quad\mbox{is continuous}.
\label{conteqns}
\end{eqnarray}
These conditions are redundant, but we will impose all of them as a check of
consistency.  In terms of the Green's dyadics, the conditions read
\numparts
\begin{eqnarray}
\boldsymbol{\hat\theta}\cdot \bGamma'\bigg|_{r=a}\quad\mbox{is continuous},
\label{bc1}\\
\hat\bi{z}\cdot \bGamma'\bigg|_{r=a}\quad\mbox{is continuous},
\label{bc2}\\
\hat\bi{r}\cdot\varepsilon \bGamma'\bigg|_{r=a}\quad\mbox{is continuous},
\label{bc3}\\
\hat\bi{r}\cdot \mu\bPhi\bigg|_{r=a}\quad\mbox{is continuous},
\label{bc4}\\
\boldsymbol{\hat\theta}\cdot \bPhi\bigg|_{r=a}\quad\mbox{is continuous},
\label{bc5}\\
\hat\bi{z}\cdot \bPhi\bigg|_{r=a}\quad\mbox{is continuous}.
\label{bc6}
\end{eqnarray}
\endnumparts
A fairly elaborate system of linear equations for the various constants
results.  However, they are not quite sufficient to determine all the
relevant physical combinations.  We also need to impose one of the Helmholtz
equations, say (\ref{helmholtz}).  From that equation we learn
\numparts
\bea
b'-k\,\mbox{sgn}m\, \tilde b=0,\label{bee1}\\
b-\frac{\mbox{sgn}m}k\tilde b'=0,\label{bee2}\\
\hat d+\hat d'=0,\label{dee}\\
f+\frac\mu\varepsilon\frac{\mbox{sgn}m}k\tilde f'=0,\label{eff1}\\
f'+\frac\varepsilon\mu k\,\mbox{sgn}m \,\tilde f=0,\label{eff2}\\
\hat e-\frac\mu\varepsilon \hat e'=0,\label{eee}
\eea
\endnumparts
where we have introduced the abbreviations for any constant $K$
\be
\hat K=K-k\,\mbox{sgn}m\,\tilde K,\qquad \hat K'=K'-\frac{\mbox{sgn}m}k \tilde K'.
\ee
Then from the boundary conditions we can solve for the remaining constants:
First,
\numparts
\bea
\hat c=\hat c'=0,\label{cee}\\
\hat g=\hat g'=0,\label{gee}
\eea
\endnumparts
and
\numparts
\bea
\tilde h_m'=-\frac{\varepsilon^2}\mu(1-\varepsilon\mu)\frac{\omega^2mk}{\lambda
\lambda' D}h_mH_m(\lambda a)J_m(\lambda'a),\\
\tilde a_m'=\frac{\lambda^{\prime2}}{\lambda^2}
\frac{\varepsilon}\mu(1-\varepsilon\mu)\frac{\omega^2mk}{\lambda
\lambda' D}h_mJ_m(\lambda'a)^2,\\
a_m=\frac{\omega^2}{\lambda^2}\frac\pi{2\rmi}\frac{J_m(\lambda a)}{H_m(\lambda a)}
+\frac{\lambda^{\prime 2}\varepsilon}{\lambda^2\mu}h_m\frac{J_m(\lambda' a)}
{H_m(\lambda a)},
\eea
all in terms of
\be
h_m=\frac\mu\varepsilon\omega^2\frac{\lambda\lambda' D}\Xi,
\ee
\endnumparts
where the denominators occurring here are
\numparts
\bea
D=\varepsilon \lambda a J_m'(\lambda'a)H_m(\lambda a)-\lambda'aJ_m(\lambda'a)
H_m'(\lambda a),\\
\tilde D=\mu \lambda a J_m'(\lambda'a)H_m(\lambda a)-\lambda'aJ_m(\lambda'a)
H_m'(\lambda a),\\
\Xi=(\lambda\lambda')^2D\tilde D-(\varepsilon\mu-1)^2k^2m^2\omega^2 
H_m^2(\lambda a)J_m^2(\lambda'a).
\eea
\endnumparts
The second set of constants is
\numparts
\bea
\tilde h_m=-\frac{\mu}{\varepsilon^2}
(1-\varepsilon\mu)\frac{mk}{\lambda
\lambda' \tilde D}h'_mH_m(\lambda a)J_m(\lambda'a),\\
\tilde a_m=-\frac{\lambda^{\prime2}}{\lambda^2}
\frac{(1-\varepsilon\mu)}\varepsilon\frac{mk}{\lambda
\lambda' \tilde D}h'_mJ_m(\lambda'a)^2,\\
a'_m=\frac{\omega^2}{\lambda^2}\frac\pi{2\rmi}\frac{J_m(\lambda a)}{H_m(\lambda a)}
+\frac{\lambda^{\prime 2}}{\lambda^2\varepsilon}h'_m\frac{J_m(\lambda' a)}
{H_m(\lambda a)},
\eea
in terms of
\be
h'_m=\varepsilon\omega^2\frac{\lambda\lambda' \tilde D}\Xi.
\ee
\endnumparts

It might be thought that $m=0$ is a special case, and indeed
\be
\frac1{2|m|}\left(\frac{r_<}{r_>}\right)^{|m|}\to\frac12\ln\frac{r_<}{r_>},
\ee
but just as the latter is correctly interpreted as the limit as $|m|\to0$,
so the coefficients in the Green's functions turn out to be just the $m=0$ 
limits of those given above, so the $m=0$ case is properly incorporated.

It is now easy to check that, as a result of the conditions (\ref{bee1}),
(\ref{bee2}), (\ref{dee}), (\ref{eff1}), (\ref{eff2}), (\ref{eee}), (\ref{cee})
and (\ref{gee}), the terms in the Green's functions that involve powers of
$r$ or $r'$ do not contribute to the electric or magnetic fields.
As we might well have anticipated, only the pure Bessel function terms
contribute.  (This observation was not noted in Ref.~\cite{deraadcyl}.)

We are now in a position to calculate the pressure on the surface of
the sphere from the radial-radial component of the stress tensor,
\be
T_{rr}=\frac12\left[\varepsilon(E_\theta^2+E_z^2-E_r^2)+\mu(H_\theta^2
+H_z^2-H_r^2)\right],
\ee
so as a result of the boundary conditions (\ref{conteqns}), the pressure
on the cylindrical walls are given by the expectation value of the 
squares of field components just outside the cylinder:
\bea
 T_{rr}\bigg|_{r=a-}-T_{rr}\Bigg|_{r=a+}=\frac{\varepsilon-1}2
\left(E_\theta^2+E_z^2+\frac1\varepsilon E_r^2\right)\Bigg|_{r=a+}\nonumber\\
\qquad\qquad\mbox{}+\frac{\mu-1}2\left(H_\theta^2+H_z^2
+\frac1\mu H_r^2\right)\Bigg|_{r=a+}.
\label{stressformula}
\eea
These expectation values are given by (\ref{evev}) and (\ref{mvev}),
where the latter may also be written as
\be
\langle \bi{H(r)H(r')}\rangle =-\frac1{\omega\mu}
\bPhi(\bi{r,r'})\times\overleftarrow\bnabla'.
\ee

It is quite straightforward to compute the vacuum expectation values in
terms of the coefficients given above.  Further details will be supplied
in a forthcoming publication.  The resulting expression for the pressure
may then, in a standard manner, be expressed after a Euclidean rotation,
\be
\omega\to\rmi\zeta,\qquad \lambda\to \rmi \kappa,
\ee
so that the Bessel functions are replaced by the modified Bessel functions,
\be
J_m(x')H_m(x)\to\frac2{\pi\rmi}I_m(y')K_m(y),
\ee
where $y=\kappa a$, $y'=\kappa'a$,
as
\bea
P=\frac{\varepsilon-1}{16\pi^3 a^4}\int \rmd \zeta a\,\rmd k a
\sum_{m=-\infty}^\infty\frac1{\tilde\Xi}
\Bigg\{\frac{k^2a^2-\zeta^2a^2\mu}{y^2}I_m(y')y'I_m'(y')[yK'_m(y)]^2\nonumber\\
\mbox{}-\frac\mu{y^{\prime2}}(k^2a^2-\zeta^2a^2\varepsilon)[y'I'_m(y')]^2yK'_m(y)
K_m(y)\nonumber\\
\mbox{}-\left[\mu y^2+\frac{m^2}{y^2}(\mu k^2a^2-\zeta^2a^2)\right]I_m(y')
y'I'_m(y')[K_m(y)]^2\nonumber\\
\mbox{}+\Bigg(-\frac{\varepsilon\mu-1}\varepsilon\frac{m^2k^2a^2\zeta^2a^2}{y^4}\left[
\frac{\varepsilon\mu-1}{y^{\prime2}}+2(\varepsilon+1)\right]\nonumber\\
\quad\mbox{}+y^{\prime2}
\left[1+\frac{m^2}{y^4}\left(k^2a^2-\frac{\zeta^2a^2}{\varepsilon}\right)
\right]\Bigg)[I_m(y')]^2yK_m'(y)K_m(y)\Bigg\},
\label{cylstress}
\eea
where 
\numparts
\bea
\tilde \Xi=\Delta\tilde \Delta+(\varepsilon\mu-1)^2\frac{m^2k^2a^2\zeta^2a^2
}{y^2y^{\prime2}}I^2_m(y')K^2_m(y),\label{tildexi}\\
\Delta=\varepsilon yI'_m(y')K_m(y)-y'K'_m(y)I_m(y'),\\
\tilde\Delta=\mu yI'_m(y')K_m(y)-y'K'_m(y)I_m(y').
\eea
\endnumparts

This result reduces to the well-known expression
for the Casimir pressure when the speed of light is the same inside and outside
the cylinder, that is, when $\varepsilon\mu=1$.  Then, it is easy to see
that the denominator reduces to
\be
\tilde \Xi=\Delta\tilde \Delta=\frac{(\varepsilon+1)^2}{4\varepsilon}
\left[1-\xi^2y^2[(I_mK_m)']^2\right],
\ee
where $\xi=(\varepsilon-1)/(\varepsilon+1)$.  In the numerator introduce
polar coordinates,
\be
y^2=k^2a^2+\zeta^2a^2,\qquad ka=y\sin\theta,\qquad \zeta a=y\cos\theta,
\label{polar}
\ee
and carry out the trivial integral over $\theta$.
The result is
\be
P=-\frac1{8\pi^2a^4}\int_0^\infty \rmd y\,y^2\sum_{m=-\infty}^\infty
\frac{\rmd}{\rmd y}\ln\left(1-\xi^2\left[y(K_mI_m)'\right]^2\right),
\label{equalspeedcyl}
\ee
which is exactly the finite result derived in Ref.~\cite{dicyl}, and 
analyzed in a number of papers \cite{brevny,gos,klichromeo}.  
For $\xi=1$ this is the formal result
for a perfectly conducting cylindrical shell first analyzed in 
Ref.~\cite{deraadcyl}.  On the other hand, if $\xi$ is regarded as
small, and (\ref{equalspeedcyl}) is expanded in powers of $\xi^2$, then
the term of order $\xi^2$ turns out to vanish, for reasons not yet understood
\cite{dicyl,klichromeo,miltonbook}.  Recall that the corresponding
coefficient for a dilute dielectric-diamagnetic sphere (\ref{smallxisphere})
is not zero.

\subsubsection{Bulk Casimir Stress}

The above expression is incomplete.  It contains an unobservable ``bulk'' 
energy
contribution, which the formalism would give if either medium, that of the
interior with dielectric constant $\varepsilon$ and permeability $\mu$, 
or that of the exterior with
dielectric constant and permeability unity, fills all space.
The corresponding stresses are computed from the
free Green's functions,
\begin{equation}
\fl F_m^{(0)}(r,r')=\frac\mu\varepsilon G_m^{(0)}(r,r')=
-\frac{\omega^2\mu}{\lambda^{\prime2}\varepsilon}\left[
{1\over2|m|}\left(\frac{r_<}{r_>}\right)^{|m|}+\frac{\pi}{2\rmi}
J_m(\lambda' r_<)H_m(\lambda'r_>)\right].
\label{bulkgf}
\end{equation}
 It should be noted that such a Green's function
does not satisfy the appropriate boundary conditions, and therefore we cannot
use (\ref{stressformula}), but rather one must compute the interior and
exterior stresses individually. Because the two scalar Green's functions 
differ only by a factor of $\mu/\varepsilon$ in this case, these are
\begin{eqnarray}
\fl T_{rr}^{(0)}(a-)=\frac1{2\pi\rmi}\sum_{m=-\infty}^\infty
\int_{-\infty}^\infty \frac{\rmd\omega}{2\pi}\int_{-\infty}^\infty
\frac{\rmd k}{2\pi}\frac{\lambda^{\prime2}}{\omega^2}
\left[\frac{\partial}{\partial r}\frac{\partial}{\partial r'}
G_m^{(0)}+\left(\lambda^{\prime2}-\frac{m^2}{a^2}\right)G_m^{(0)}
\right]\Bigg|_{r=r'=a-},\nonumber\\
\end{eqnarray}
while the outside bulk stress is given by the same expression with $\lambda'\to
\lambda$.
When we substitute the appropriate interior and exterior Green's functions
given in (\ref{bulkgf}), and perform the Euclidean rotation, $\omega\to\rmi
\zeta$
we obtain the following rather simple formula
for the bulk contribution to the pressure:
\begin{eqnarray}
P^b=T_{rr}^{(0)}(a-)-T_{rr}^{(0)}(a+)\nonumber\\
=\frac{1}{8\pi^3a^2}\sum_{m=-\infty}^\infty
\int_{-\infty}^\infty \rmd k \int_{-\infty}^\infty \rmd \zeta
\left[y^{\prime2}I'_m(y')K'_m(y')-(y^{\prime2}+m^2)I_m(y')K_m(y')
\right.\nonumber\\
\mbox{}-\left.y^2I'_m(y)K'_m(y)-(y^2+m^2)I_m(y)K_m(y)\right].
\label{bs}
\end{eqnarray}
This term must be \textit{subtracted\/} from the pressure given in
(\ref{cylstress}).  Note that this term is the direct analog of that
found in the case of a dielectric sphere in Ref.~\cite{miltonballs}
-- See (\ref{stress}).
Note also that $P^b=0$ in the special case $\varepsilon\mu=1$.

In the following, we will be interested in dilute dielectric media,
where $\mu=1$ and $\varepsilon-1\ll1$.  We easily find that when the
integrand in (\ref{bs}) is expanded in powers of $\varepsilon-1)$ the leading
terms yield
\bea
\fl P^b\approx-\frac{1}{4\pi^2a^4}\sum_{m=-\infty}^\infty
\int_0^\infty \rmd y\,y\int_0^{2\pi}\frac{\rmd \theta}{2\pi}
\Bigg[(\varepsilon-1)\zeta^2a^2I_m(y)K_m(y)\nonumber\\
\mbox{}+\frac14(\varepsilon-1)^2\frac{
(\zeta a)^4}y[I_m(y)K_m(y)]'+\Or(\varepsilon-1)^3\Bigg]\nonumber\\
\fl=-\frac{\varepsilon-1}{8\pi^2a^4}\sum_{m=-\infty}^\infty \int_0^\infty\rmd y
\,y^3\,\left[I_m(y)K_m(y)+\frac{3(\varepsilon-1)}{16}y[I_m(y)K_m(y)]'
+\Or\left((\varepsilon-1)^3\right)\right],\nonumber\\
\label{bs1}
\eea
where we have introduced polar coordinates as in (\ref{polar}).

\subsubsection{Dilute Dielectric Cylinder}

We now turn to the case of a dilute dielectric medium filling the cylinder,
that is, set $\mu=1$ and consider $\varepsilon-1$ as small.  The leading
term in the pressure, $\Or[(\varepsilon-1)^1]$, is obtained from
(\ref{cylstress}) by setting $\mu=\varepsilon=1$ everywhere in the integrand.
The denominator $\tilde\Xi$ is then unity, and we get
\be
P\approx -\frac{\varepsilon-1}{8\pi^2a^4}\sum_{m=-\infty}^\infty \int_0^\infty
\rmd y\,y^3\,I_m(y)K_m(y),
\ee
which is exactly what is obtained to leading order from the bulk stress
(\ref{bs1}):
\be
P-P^b=\Or[(\varepsilon-1)^2],
\ee
which is consistent with the interpretation of the Casimir energy as
arising from the pairwise interaction of dilutely distributed molecules.
In fact, from Ref.~\cite{dicyl,romeocomm}, 
we know that the van der Waals energy
vanishes even in order $(\varepsilon-1)^2$, so we expect the same to
occur with the Casimir energy, although the latter should diverge in
$\Or[(\varepsilon-1)^3]$ \cite{borpiro}.

We now obtain the expression for the $\Or[(\varepsilon-1)^2]$ term.
Because the general expression (\ref{cylstress}) is proportional to 
$\varepsilon-1$, we need only expand the integrand to first order in 
this quantity.  Let us write it as
\be
P=\frac{\varepsilon-1}{16\pi^3a^4}\int_{-\infty}^\infty \rmd \zeta a
\int_{-\infty}^\infty \rmd ka\sum_{m=-\infty}^\infty 
\frac{N}{\Delta\tilde\Delta},
\ee
where we have noted that the $(\varepsilon-1)^2$ in $\tilde \Xi$ 
(\ref{tildexi}) can be dropped. Then introducing polar coordinates
as in (\ref{polar}), and expanding the numerator and denominator according
to
\be
N=N^{(0)}+(\varepsilon-1)N^{(1)}+\dots,\qquad
\Delta\tilde\Delta=1+(\varepsilon-1)\Delta^{(1)}+\dots,
\ee
the second-order term in the unsubtracted Casimir pressure is given by
\be
P^{(2)}=\frac{(\varepsilon-1)^2}{16\pi^3a^4}
\int_0^\infty \rmd y\,y\int_0^{2\pi}
\rmd \theta\left(N^{(1)}-\Delta^{(1)}N^{(0)}\right).
\label{pee2}
\ee
Here the correction to the denominator is
\bea
\fl\Delta^{(1)}=yI'_m(y)K_m(y)-y\sin^2\theta[I_m(y)K_m(y)]'+\sin^2\theta
\left(m^2+y^2\right)I_m'(y)K_m'(y)\nonumber\\
\mbox{}-y^2\sin^2\theta I'_m(y)K'_m(y),
\eea and the first two term in the numerator expansion are
\numparts
\bea
\fl N^{(0)}=-\left[y^2+m^2(1-2\sin^2\theta)\right]I_m(y)K_m(y)-y^2(1-2\sin^2\theta)
I_m'(y)K_m'(y),\label{nzero}\\
\fl N^{(1)}=-\frac{1}2\left(m^2+y^2\right)\left[y^2+
m^2(1-2\sin^2\theta)\right]K^2_m(y)I^2_m(y)\nonumber\\
\fl\qquad\mbox{}+y\sin^2\theta\left[\left(
m^2+y^2\right)+m^2(1-2\sin^2\theta)
-4m^2\cos^2\theta\right]I^2_m(y)K_m(y)K_m'(y)
\nonumber\\
\fl\qquad\mbox{}+\frac12y^2\sin^2\theta(1-2\sin^2\theta)(m^2+y^2)I_m^2(y)K_m
^{\prime2}(y)\nonumber\\
\fl\qquad\mbox{}-\frac12 y^2\sin^2\theta\left[y^2+m^2(1-2\sin^2\theta)\right]
K_m^2(y)I_m^{\prime2}(y)\nonumber\\
\fl\qquad\mbox{}+y^4\left[\sin^2\theta-\sin^2\theta(1-2\sin^2\theta)\right]
I_m(y)I_m'(y)K_m(y)K_m'(y)\nonumber\\
\fl\qquad\mbox{}+y^3\left[\sin^2\theta+\sin^2\theta(1-2\sin^2\theta)
\right]I_m^{\prime2}(y)K_m(y)K_m'(y)\nonumber\\
\fl\qquad\mbox{}+\frac12 y^4\sin^2\theta(1-2\sin^2\theta)I_m^{\prime2}
K_m^{\prime2}.
\label{none}
\eea
\endnumparts
The angular integrals are trivially
\be
\int_0^{2\pi}\rmd\theta\sin^2\theta=\pi,
\qquad \int_0^{2\pi}\rmd\theta\sin^4\theta=\frac34\pi,
\ee
and then the straightforward reduction of (\ref{pee2}) is
\bea
\fl P^{(2)}=\frac{(\varepsilon-1)^2}{64\pi^2a^4}\sum_{m=-\infty}^\infty
\int_0^\infty \rmd y\bigg\{y(y^2+m^2)(2y^2-m^2)I_m^2(y)K_m^2(y)\nonumber\\
\fl\qquad\mbox{}+2y^2(2y^2+m^2)K_m^2(y)I_m(y)I_m'(y)\nonumber\\
\fl\qquad\mbox{}-y^3(y^2+m^2)I_m^2(y)K_m^{\prime2}(y)
-y^3(2y^2-m^2)K_m^2(y)I_m^{\prime2}(y)
\nonumber\\
\fl\qquad\mbox{}+4y^4K_m(y)K_m'(y)I_m^{\prime2}(y)
+2y^4I_m(y)I_m'(y)K_m^{\prime2}(y)+y^5\left[I_m'(y)K_m'(y)\right]^2\bigg\}.
\eea
Our challenge now is to evaluate this quantity.

\subsection{Perspective}  I have been working on this problem, on and off,
since 1998, when I learned of Romeo's proof \cite{romeocomm} that
the renormalized van der Waals energy for a dilute dielectric cylinder
was zero.  Unfortunately, I had labored under a misconception concerning
the form of the Green's dyadic, which was not in a sufficiently general
form until I started re-examining this problem with my graduate student
Ines Cavero-Pelaez this past year.  We now have a consistent formal result,
which only requires some delicate analysis to extract the answer.  The
results, and further details, will follow in a paper to appear later this
year.  This promises to add another bit of understanding to our knowledge
of Casimir forces, knowledge that seems to grow only incrementally based
on specific calculations, since a general understanding is still not
at hand.

\section{Casimir Effects for Solitons}
\label{sec:solitions}
Our discussion throughout this article so far has been confined to
idealized boundaries, although we alluded to a dynamical basis in the 
sections referring to the delta-function potentials.  Of course, from
the beginning of the subject, it has always been the goal to describe
the interactions due to real interfaces, be they constituted of atoms and
molecules, or due to solutions of the quantum field equations themselves.
The most natural thing to consider is a solitonic background, where the
soliton is a classical field configuration which minimizes the energy,
and then consider the effect of quantum fluctuations around this background
field.  Perhaps the first physical ideas along this direction were presented
in the context of the bag model 
\cite{chodos,chodos2,chodosthorn,degrand,johnson,donoghue2,schrock}. 
 The bag is supposed to represent, 
semiclassically, the notions of confinement, in which within the bag particles
carrying color charge (quarks and gluons)
are free to move subject to perturbative QCD
interactions, whereas outside the bag, no colored objects can exist.
Such a bag picture has never actually been derived from QCD, but it forms
an enormously fruitful phenomenological framework.  Similar pictures can
be derived from truncated models \cite{Adler:1978xi,Adler:1979we,Adler:1980ex,
Adler:1981as,Adler:1982pj,Adler:1982rk,savvidy}.

Casimir energies have been discussed in connection with the bag model since
1980 \cite{miltonbag,miltonfermion,johnsondpf}.  
(Actually, a zero-point energy parameter was put in the model from
the beginning.)  Unfortunately, no reliable result could be derived
because interior contributions alone are inherently divergent.  Efforts,
however, more or less successful, were made to extract finite parts
\cite{miltonfinite},
and a summary of some of the phenomenological results can be found in
Ref.~\cite{miltonbook}.  Progress toward understanding the divergences
promise to lead to more reliable predictions in the near future.

However, for kink and soliton solutions, reliable Casimir effects have
been found in a number of cases.  The reviews given at QFEXT03 \cite{QFEXT03}
by van Nieuwenhuizen, Bordag, and Jaffe are a useful starting point.  
For example, Refs.~\cite{Rebhan:2004vz,Goldhaber:2004kn,rebhan04} 
show that quantum corrections to the
mass and central charge of supersymmetric solitons are nonvanishing
even though zero-point energies of bosons and fermions seem to
cancel.  The Bogomolnyi bound is saturated because there is only one
fermionic zero mode.

Very interesting methods have been presented in the past few years
by the group led by Jaffe \cite{graham0,graham,quandtqfext}, based
on subtraction from the local spectral density (related to derivatives
of the phase shift) the first few Born terms, which correspond to low-order
Feynman diagrams, which may be renormalized in the standard manner.
For any smooth background a finite renormalized vacuum energy is obtained.
These methods have been used, as noted in \sref{Sec1} and \ref{sec:selfstress},
 to critically
discuss energies and self energies of idealized boundaries.  In solitonic
physics Fahri \etal \cite{fahri2} have used these methods to compute quantum
fluctuations around static classical solitons in Euclidean electroweak
theory, which are unstable, in an attempt to find stable quantum solitons.
(See also Ref.~\cite{khemaniqfext}.)  No solutions have yet been found, yet 
some promising nonspherical candidates exist.

Bordag \cite{bordagqfext} considers the vacuum energy of a fermion in the 
background of a Nielsen-Olesen vortex (string) \cite{nielsenolesen}.
The vacuum energy is defined by zeta-function regularization, and is expressed
in terms of the Jost function, evaluated by using the Abel-Plana formula.
The quantum correction found in this way, however, is very small.  It
may be that in other cases, such as electroweak strings, the quantum vacuum
energy might have more physical relevance, even leading to the stability of
the string.

We should also mention that Casimir energies play an important role in 
lattice simulations of QCD.  For its role in QCD string formation see, for
example, Juge, Kuti, and Morningstar \cite{juge,juge2} and L\"uscher and
Weisz \cite{Luscher:2002qv}.
\section{Dynamical Casimir Effects}
\label{sec:dce}

Everything discussed above referred to static configurations.  In such
a case the concept of energy is well-defined, but even then, as we have
seen, it is not easy or noncontroversial to extract a physically observable
effect.  When the boundaries are moving, the situation is far more difficult.

In one dimension, the problem seems tractable.  We can consider a point
undergoing harmonic oscillations, and ask what are the consequences for
a scalar field which must vanish at that point.  We expect that the result
is the production of real quanta of the field.  This is the dynamical
Casimir effect.  However the only reliable results seem to be for motions
which can be treated perturbatively, or in the opposite extreme, where the
adiabatic (instantaneous) approximation applies for very rapid changes.

In three dimensions, the situation is still more challenging.
Here we should mention the suggestion of Schwinger \cite{js,js2}, 
followed up by
Eberlein \cite{eberlein,eberlein2}, 
Chodos \cite{chodossono,chodossono2}, Carlson \cite{carlson,carlson2},
Visser \cite{cmp,liberati,liberati2,liberati3,liberati4,liberati4},  
and others, that the copious production
of light in sonoluminescence \cite{sonorev,sonorev02} was due to the dynamical
Casimir effect, due to the rapid expansion and contraction of a micron-sized
air bubbles in water.  The original estimation that there was sufficient
energy available for this mechanism was based on a naive use of the cutoff
value of $\frac12\sum\hbar\omega$.  An actual calculation showed that the
energy was insufficient by 10 orders of magnitude \cite{sonokm}.  
Dynamically, photons
indeed should be produced by QED by a rapidly oscillating bubble, but
to produce the requisite number ($10^6$ per flash) necessitated, if not
superluminal velocities at least macroscopic collapse time scales of order
$10^{-15}$ s, rather than the observed $10^{-11}$ s scale \cite{sonokm2}.

\subsection{Fulling/Unruh/Hawking Radiation}
One regime where definitive results exist for quantum particle production is in the
general relativistic context.  The Moore-Fulling-Davies Effect is the 
production of photons
by a mirror undergoing uniform acceleration
\cite{moore,fulldavies,daviesfull}.  The photon spectrum is thermal,
with the temperature proportional to the acceleration of the mirror.
The Unruh  effect is very similar \cite{unruh}.  If the free equations of
quantum field theory are examined in the frame of an accelerated observer,
with acceleration $a$,
it is found that such an observer sees a heat-bath of photons, again with
$T= a/2\pi$. (For a precise description of these phenomena see the classic
book by Birrell and Davies \cite{birrell}.)

These phenomena naturally are mirrored in gravitational phenomena.  The
celebrated Hawking radiation \cite{hawking}
 is the production of quanta by a black hole.
Energy is extracted from the black hole by particle-antiparticle production
outside the horizon.  One particle escapes, while the other falls into the
black hole.  The resulting thermal radiation has a temperature, 
in accordance with the expectation from the above, proportional to the
surface gravity of the black hole, or inversely proportional to its mass $M$:
\be
T=1.2\times 10^{26}\mbox{K}\left(\frac{1\,\mbox{gm}}{M}\right).
\ee
(Again, see Ref.~\cite{birrell}.)

Scully and collaborators \cite{scully03} have proposed an experiment to 
measure the Unruh effect by injecting atoms into a microwave or optical
cavity, which atoms then undergo acceleration.  Hu \etal \cite{hu04,hu04a}
persuasively argue that this experiment will not detect the Unruh effect,
because the latter does not refer to the radiation produced by the
accelerated detector (which is nil), Lorentz invariance, crucial to the
Unruh effect, is broken by the cavity, the thermal distribution of photons
in the cavity is not that of the Unruh effect, and finally that the injection
mechanism will produce cavity excitation so  that acceleration no longer
plays a crucial role.

For recent work on moving charges, detectors, and mirrors by Hu's group,
see Ref.~\cite{huqfext}.
\subsection{Terrestrial Applications}

Most of the calculations of the dynamical Casimir effect have considered
scalar fields.  For example, Crocce \etal \cite{crocce} consider a cavity
bifurcated by a semiconducting film whose properties can be changed in time
by laser pulses, modeled by a time-dependent potential
\be
\mathcal{L}_{\rm int}=-\frac12V(t)\delta(x)\phi^2.
\ee
The inhomogeneous wave equation is solved with the time-dependence given
as a first-order perturbation, with the result that if the film is driven
parametrically, that is, in resonance with one of the modes of the cavity,
particle (photon) production grows exponentially.  This is in line with the
expectations from the $1+1$ dimensional cavity, where if the length undergoes
periodic oscillations at a multiple of the fundamental frequency $\omega_0
=\pi/L$ of the cavity 
\be
L(t)=L+\Delta L\sin k\omega_0t,
\ee
for large times the energy produced is
\cite{wegrzyn,wegrzyn1,dodonov,dodonov1,newreview}
\be
E(t)=-\frac{k^2\omega_0}{24}+\frac{(k^2-1)\omega_0}{24}\cosh k\omega_0\frac
{\Delta L}{L}t,\qquad k=1,2,3,\dots.
\ee
The $k=0$ value is the Casimir energy corresponding to (\ref{eq:luscher}).
A numerical simulation method for calculating particle production
in both cosmological and terrestrial settings is given in Ref.~\cite{Antunes}.

One of the few treatments of the $3+1$ dimensional situation for 
electromagnetic fields is that of Uhlmann \etal \cite{uhlmann}, who
consider a rectangular cavity of length $L$ with perfectly conducting
walls, with a narrow dielectric slab of width $a$ at one end possessing
a time-dependent permittivity $\varepsilon(t)$.  The time dependence
is still treated perturbatively.  Only TM modes are effective in producing
photons, the number of which increase exponentially on resonance, just as
in Ref.~\cite{crocce}:
\be
\langle N\rangle (t)=\sinh^2\left(\frac{k_\perp^2}{\omega}\frac\chi
{\varepsilon_0}\frac{a}{L} t\right),
\ee
where $k_\perp^2$ is the square of the transverse wave vector, $\omega$ is the
resonant frequency, and $\chi$ is the amplitude of the sinusoidal time-varying
permittivity,
\be
\frac{\varepsilon_0}{\varepsilon(t)}=\mbox{constant}+\chi\sin\omega t.
\ee
The challenge will be to devise a practical experiment where this effect can
be observed in the microwave regime.

\section{Casimir Effect and the Cosmological Constant}
\label{sec:cc}

\subsection{Cosmological Constant Problem and Recent Observations}
It has been appreciated for many years that there is an apparently fundamental
conflict between quantum field theory and the smallness of the cosmological
constant \cite{weincc,weinccc,weincccc,zeldovich}.
  This is because the zero-point energy of the
quantum fields (including gravity) in the universe should give rise
to an observable cosmological vacuum energy density,
\begin{equation}
u_{\rm cosmo}\sim{1\over L_{\rm Pl}^4},
\end{equation}
where the Planck length is
\begin{equation}
L_{\rm Pl}=\sqrt{G_N}=1.6\times 10^{-33}\,\mbox{cm}.
\end{equation}
(We use natural units with $\hbar=c=1$.  The conversion factor is
$\hbar c \simeq 2 \times 10^{-14}\,\mbox{GeV\,cm}$.)  This means that the
cosmic vacuum energy density would be
\begin{equation}
u_{\rm cosmo}\sim 10^{118} \mbox{ GeV\,cm}^{-3},
\label{ccprob}
\end{equation}
which is 123 orders of magnitude larger than the critical mass density
required to close the universe:
\begin{equation}
\rho_c={3H_0^2\over8\pi G_N}=1.05\times 10^{-5}h_0^2 \,\mbox{GeV\,cm}^{-3},
\end{equation}
in terms of the dimensionless Hubble constant, 
$h_0=H_0/100 \; \mbox{km\,s}^{-1}\mbox{Mpc}^{-1}\approx0.7$. 
~From relativistic covariance 
the cosmological vacuum energy density must be the $00$ component 
of the expectation value of the energy-momentum tensor, 
which we can identify with the cosmological constant:
\begin{equation}
\langle T^{\mu\nu}\rangle =-u g^{\mu\nu}=-{\Lambda\over8\pi G}g^{\mu\nu}.
\end{equation}
[We use the metric with signature $(-1,1,1,1)$.]
Of course this is absurd with $u$ given by Eq.~(\ref{ccprob}), which would
have caused the universe to expand to zero density long ago.

For most of the past century, it was the prejudice of theoreticians that
the cosmological constant was exactly zero, although no one could give
a convincing argument.  In the last few years,
 however, with the new data gathered
on the brightness-redshift relation for very distant type Ia supernov\ae\
\cite{riess,perlmutter,sn,Tonry:2003zg,Riess:2004nr}, 
corroborated by observations of the
anisotropy in the cosmic microwave background \cite{wmap}, 
observations of large-scale structure \cite{Tegmark:2003ud},
and of the Sachs-Wolf effect \cite{Boughn:2004zm}.
Thus, it
seems clear that the cosmological constant is near the critical value, 
and in fact makes up the majority of the energy in the universe,
\begin{equation}
\Omega_\Lambda=\Lambda/8\pi G\rho_c \simeq 0.75.
\end{equation}
Dark matter makes up most of the rest.  Data are consistent with the
value for the ratio of pressure to energy predicted by the cosmological
constant interpretation, $w=p/\rho=-1$ \cite{Wang:2004gq,Wang:2004py}.
For reviews of the observational situation, see Ref.~\cite{pad,peebles}.
It is very hard to understand how the cosmological
constant can be nonzero but small.  (For a recent example of how
difficult this problem is to solve, see Dolgov \cite{dolgov}.)

\subsection{Quantum Fluctuations}
In Ref.~\cite{Milton:2001np,Milton:2002hx} we have presented 
a plausible scenario for understanding this puzzle.
It seems quite clear that vacuum fluctuations
in the gravitational and matter fields in flat Minkowski space give
a zero cosmological constant. 
On the other hand, since the work of Kaluza and Klein \cite{kaluza,klein,klein2}
it has been an exciting possibility that there exist extra dimensions
beyond those of Minkowski space-time.  
Why do we not experience those dimensions?  
The simplest possibility seems to be that those  extra dimensions
are curled up in a space $\cal S$ of size $a$, 
smaller than some observable limit.

Of course, in recent years, the idea of extra dimensions has become
much more compelling.  Superstring theory  requires at least 10 dimensions,
six of which must be compactified, and the putative M theory, supergravity,
is an 11 dimensional theory.
 Perhaps, if only gravity experiences the
extra dimensions, they could be of macroscopic size.
 Various scenarios
have been suggested \cite{add,add2,rsundrum}.

Macroscopic extra dimensions imply deviations from Newton's law at such
a scale.  Five years ago, millimeter scale deviations seemed plausible, and
many theorists hoped that the higher-dimensional world was on the brink
of discovery. Experiments were initiated \cite{long,long2}.
Recently,
the results of  definitive Cavendish-type experiments have appeared 
\cite{hoyle,Adelberger:2003zx,Long:2003ta,boulder},
which indicate
no deviation from Newton's law down to 100 $\mu$m.  
(The experimental constraints on non-Newtonian gravity discussed in
\sref{sec:exp} are so weak as to be useless in this connection.)

This poses an extremely 
serious constraint for model-builders.

Earlier we had proposed \cite{Milton:2001np}
that a very tight constraint indeed emerges if we recognize
that compact dimensions of size $a$ necessarily possess a quantum
vacuum or Casimir energy of order $u(z)\sim a^{-4}$.  
  These can be calculated in simple
cases.  Appelquist and Chodos \cite{appel,appel2}
found that the Casimir energy
for the case of scalar field on a circle, ${\cal S}=S^1$, was
\begin{equation}
u_C=-{3\zeta(5)\over64\pi^6a^4}=-{5.056\times10^{-5}\over a^4},
\end{equation}
which needs only to be multiplied by 5 for graviton fluctuations.
The general case of scalars on ${\cal S}=S^N$, $N$ odd, was considered
by Candelas and Weinberg \cite{candandwein}, who found that the Casimir energy
was positive for $3\le N\le 19$, with a maximum at $N=13$ of
$u_C=1.374\times 10^{-3}/a^4$. 
 The even dimensional case was much
more subtle, because it was divergent.  Kantowski and Milton \cite{kantowski}
showed that the coefficient of the logarithmic divergence was unique,
and adopting the Planck length as the natural cutoff, found
\begin{equation}
S^N, \,\,N \mbox{ even}: \quad u^N_C={\alpha_N\over a^4}\ln{a\over L_{\rm Pl}},
\label{evenccas}
\end{equation} 
but $\alpha_N$ was always negative for scalars.
In a second paper \cite{kantowski2} we extended the analysis to vectors, tensors,
fermions, and to massive particles, among which cases positive values of the
(divergent) Casimir energy could be found.   In an unsuccessful 
 attempt to find stable configurations, the analysis was extended
to cases where the internal space was the product of spheres \cite{birkanmil}.

It is important to recognize that these Casimir energies correspond to a
cosmological constant in our $3+1$ dimensional world, not in the extra
compactified dimensions or ``bulk.''  They constitute an effective source
term in the 4-dimensional Einstein equations.  
Note that because the scale $a$ makes no reference to four-dimensional
space, the total free energy of the universe (of volume $V$)
arising from this source
is $F=V u_c$, so as required for dark energy or a cosmological constant,
\begin{equation}
p=-\frac{\partial}{\partial V}F=-u_c\,\quad T^{\mu\nu}=-u_cg^{\mu\nu},
\quad \mbox{i.e.,}\quad w=-1.
\end{equation}

The goal, of course, in all these investigations was to include graviton
fluctuations.  However, it immediately became apparent that
the results
were gauge- and reparameterization-depen\-dent unless the DeWitt-Vilkovisky
formalism was adopted \cite{vilkovisky,vilk,barvinsky,dewitt2}.
 This was an extraordinarily difficult
task. Among the earlier papers in which the unique effective action
is given in simple cases we cite Ref.~\cite{Odintsov:1991yx}; see references
therein.
 Only in 2000 did the general analysis for gravity 
appear, with results
for a few special geometries \cite{cho}.
Cho and Kantowski obtain the unique
divergent part of the effective action for ${\cal S}=S^2$, $S^4$, and $S^6$,
as polynomials in $\Lambda a^2$. (Unfortunately, once again, 
they are unable to find any stable configurations.)

 The results for the coefficient $\alpha_N$ in (\ref{evenccas}) are
$\alpha_2=1.70\times10^{-2}$, $\alpha_4=-0.489$, and $\alpha_6=5.10$,
for $\Lambda a^2\sim G/a^2\ll1$.
Graviton fluctuations dominate matter fluctuations,
except in the case of a large number of matter fields in a small number
of dimensions.
 Of course, it would be very interesting to know the
graviton fluctuation results for odd-dimensional
spaces, but that seems to be a more difficult calculation; it is far easier to
compute the divergent part, which appears as a heat kernel coefficient,
 than the finite part,  which is all there is in odd-dimensional spaces.

These generic results may be applied to recent popular scenarios.  For example,
in the ADD scheme  only gravity propagates in the bulk, while the
RS approach  has other bulk fields in a single extra dimension.

Let us now perform some simple
estimates of the cosmological constant in these models.  The data require a
positive cosmological constant, so we can exclude those cases where the
Casimir energy is negative. If we use the divergent results for even
dimensions, merely requiring that this be less than the critical density 
$\rho_c$ implies the inequality ($\alpha>0$)
\begin{equation}
a>[\alpha\ln(a/L_{\rm Pl})]^{1/4}80\,\mu\mbox{m},
\label{alphalim}
\end{equation}
where we can approximate $(\ln a/L_{\rm Pl})^{1/4}\approx2.9$.
The absence of deviations from Newton's law above 100 $\mu$m 
rules out all but one of the gravity 
cases ($S^2$) given by Cho and Kantowski \cite{cho}. For matter
fluctuations only, excluded are $N>14$ for a single
vector field and $N>6$ for a single tensor field.  (Fermions always have
a negative Casimir energy in even dimensions.)
Of course, it is possible to achieve cancellations by including
various matter fields and gravity.
In general the Casimir energy is obtained
by summing over the species of field which propagate in the extra dimensions,
 \begin{equation}
 u_{\rm tot}={1\over a^4}
\sum_i\left[\alpha_i
\ln(a/L_{\rm Pl})+\beta_i\right]\approx{\beta_{\rm eff}\over a^4},
\end{equation}
which leads to a lower limit analogous to (\ref{alphalim}).
Presumably, if  exact supersymmetry 
held in the extra dimensions (including
supersymmetric boundary conditions), the Casimir energy would vanish, but
this  would seem to be difficult to achieve with {\em large\/} extra
dimensions  (1 mm corresponds to $2\times 10^{-4}$ eV).  (See, for example,
Ref.~\cite{dolgov}.)

That there is a correlation 
between the currently favored value of the cosmological constant and
submillimeter-sized extra dimensions has been noted qualitatively 
before \cite{Kaplan:2000hh,Banks:1988je,Beane:1995sk,Beane:1997it}.
A first attempt to calculate the cosmological constant in terms of Casimir
energies in the context of deconstructed extra dimensions is given
by Bauer, Lindner, and Seidl \cite{Bauer:2003mh}.

In summary,
we have proposed the following scenario to explain the predominance
of dark energy in the universe.
\begin{enumerate}
\item Quantum fluctuations of gravity/matter fields in extra dimensions
give rise to a dark energy, or cosmological constant, $\propto1/{a^4}$
where $a$ is the size of the extra dimensions.
\item The dark energy will be too large unless $a>10-300$ $\mu$m.
\item Laboratory (Cavendish) tests of Newton's law require $a<100$ $\mu$m.
\item Thus, extra dimensions may be on the verge of discovery.  If
serious limits on the validity of Newtonian gravity can be extended down to
10 $\mu$m, then we would have to conclude that
\item Extra dimensions probably do not exist, and
dark energy has another origin, for example quintessence
\cite{Steinhardt:2003st}.  However, the fact that the rapidly improving
data favor the cosmological constant interpretation of dark energy
\cite{Wang:2004gq,Wang:2004py}, makes alternatives disfavored, since they
would generally exhibit time-dependence.
 \end{enumerate}
 
\section{Future Prospects and Perspectives}
In this review we have attempted to present a personal perspective on
the progress in understanding quantum vacuum energy and its physical
implication in the past four years. The primary stimulus for the development
of the subject has been the tremendous progress on the experimental front.
This has brought to the fore issues that were regarded as arcane, such as
the temperature dependence of the Casimir forces between metal plates,
the meaning of infinities encountered in calculations of quantum vacuum
energy, and the source of the cosmological dark energy, which it is hard
to believe does not have something to do with quantum fluctuations, yet
is remarkably small.  At this point, no definitive resolution of any of
these issues has been given; yet, progress is rapid, and I hope that this
status report may help sharpen issues and contribute in some small way to
the solution of outstanding problems.

The reader will have noted that this document is far from even-handed.
I have continued to focus on the use of Green's function techniques as
expounded in my earlier monograph \cite{miltonbook}.  I do not mean to
disparage in any way the valuable progress made using other techniques,
including use of zeta-function methods, Jost functions, worldline approaches,
and scattering phase-shift formalisms; although I do continue to
believe that the Green's function approach is the most physical.
 I also have focused on topics
that are of personal interest, so if I have slighted important subjects
and researchers, I beg forbearance.  

I want to close this review by briefly mentioning a few topics that do
not seem to fit in the sections above.  For example, there has been
important work in the subject of the Casimir effect in critical
systems by Krech and collaborators \cite{krechanddont,krech},
in which they consider massless excitations caused by critical fluctuations
of the order parameter of a condensed-matter system about the critical
temperature $1/\beta_c$.  For $d$ transverse dimensions, that force is
\be
\beta_c\mathcal{F}=(d-1)\frac\Delta{a},
\ee
where $\Delta$ is universal.  For an application to thinning of superfluid
helium films, see Ref.~\cite{zandi}.  Williams \cite{williams} shows
that vortex excitations are the source of the critical fluctuations that
give rise to the critical Casimir force in this situation.

An acoustic analog of the Casimir effect has been discussed 
\cite{larraza,larraza2,barcenas}.

There have been many extremely interesting contributions to cosmological
and brane-world models.  For example, Dowker \cite{dowker04} considers
the Casimir effect in nontrivial cosmological topologies.  A condensate
of the metric tensor may stabilize Euclidean Einstein gravity in a manner
not unrelated to the Casimir effect \cite{polonyi}.  And Brevik has
questioned the meaning of the Cardy-Verlinde formula expressing a 
bound on the entropy \cite{brevikcv}.  (See also Ref.~\cite{entropy}.)
Mazur and Mottola \cite{mottolaqfext,mottola01} have suggested that
dark energy is quantum vacuum energy due to a causal boundary effect at the
cosmological horizon -- namely, that instead of a black hole, there are
three regions due to a quantum phase transition (perhaps due to the trace
anomaly) in spacetime itself:
exterior (Schwarzchild) where $\rho=p=0$; interior (de Sitter) where $\rho=-p$;
and a thin boundary shell where $\rho=p$.  Details of this proposal are
still vague; without a detailed calculation one cannot tell whether even the
vacuum energy will emerge correctly.

There have been many contributions on Casimir effects in brane-world
scenarios, for example, Refs.~\cite{nojiri,Ichinose:2004mt}.

Finally, we note that everything we have considered in this review
has been at the one-loop level. Radiative corrections to the Casimir
effect have, in fact, been considered by several authors.  Most of the
calculations have been in situations in which there is only one significant
direction: For QED, see Ref.~\cite{bordag,robaschik}, and for $\lambda\phi^4$
theory, see Ref.~\cite{Barone:2003nk,Barone:2003rn} and references therein.
There is an impressive calculation of the radiative correction to the Casimir
effect with a spherical shell boundary, 
perfectly conducting as far as electromagnetism
is concerned, but transparent to electrons by Bordag and Lindig \cite{radbord}.

So we leave the subject of Casimir phenomena as a work in progress.
It is clear that quantum fluctuation forces are vitally important both
in the very large and the very small domains, and that they will play
increasingly central roles in engineering applications.  Thus, the
subject is an exciting interdisciplinary topic, with both fundamental
and technological spinoffs.  Thus the uncertainty principle is not
just about atomic and subatomic physics, but it may control our future,
in many senses.

\ack
I am grateful to the US Department of Energy for partial funding of the
research reported here.  I thank many colleagues, particularly Carl Bender,
Michael Bordag, Iver Brevik, Ines Cavero-Pelaez,
Steve Fulling, Johan H\o ye, Galina Klimchitskaya,
 Vladimir Mostepanenko, Jack Ng, and Yun Wang
 for collaborations and many helpful conversations
over the years.

\section*{References}

\begin{thebibliography}{100}

\bibitem{dirac34}
P.~A.~M. Dirac.
\newblock {\em Proc. Camb. Phil. Soc.}, 30:150, 1934.

\bibitem{diracbook}
P.~A.~M. Dirac.
\newblock {\em The Principles of Quantum Mechanics}.
\newblock Oxford University Press, fourth edition, 1958.

\bibitem{straumann02}
N.~Straumann.
\newblock Invited talk at the XVIIIth IAP Colloquium: Observational and
  theoretical results on the accelerating universe, July 1-5 2002, Paris.
\newblock {[arXiv:gr-qc/0208027]}.

\bibitem{straumann02a}
N.~Straumann.
\newblock Invited lecture at the first {\it S\'eminaire Poincar\'e}, Paris,
  March 2002.
\newblock {[arXiv:astro-ph/0203330]}.

\bibitem{weincc}
S.~Weinberg.
\newblock {\em Rev. Mod. Phys.}, 61:1, 1989.

\bibitem{weinccc}
S.~Weinberg.
\newblock In {\em {Dark Matter 2000, Marina del Rey, CA}}.
\newblock {[arXiv:astro-ph/0005265]}.

\bibitem{Schwinger:1948iu}
J.~Schwinger.
\newblock {\em Phys. Rev.}, 73:416, 1948.

\bibitem{Feynman:1948fi}
R.~P. Feynman.
\newblock {\em Phys. Rev.}, 74:1430, 1948.

\bibitem{casimir}
H.~B.~G. Casimir.
\newblock {\em Proc. Kon. Ned. Akad. Wetensch.}, 51:793, 1948.

\bibitem{newreview}
M.~Bordag, U.~Mohideen, and V.~M. Mostepanenko.
\newblock {\em Phys. Rept.}, 353:1, 2001.
\newblock [arXiv:quant-ph/0106045].

\bibitem{lifshitz}
E.~M. Lifshitz.
\newblock {\em Zh. Eksp. Teor. Fiz.}, 29:94, 1956.
\newblock [English transl.: {\it Soviet Phys. JETP\/} 2:73, 1956].

\bibitem{dzyaloshinskii0}
I.~D. Dzyaloshinskii, E.~M. Lifshitz, and L.~P. Pitaevskii.
\newblock {\em Zh. Eksp. Teor. Fiz.}, 37:229, 1959.
\newblock [English transl.: {\it Soviet Phys. JETP\/} 10:161, 1960].

\bibitem{dzyaloshinskii}
I.~D. Dzyaloshinskii, E.~M. Lifshitz, and L.~P. Pitaevskii.
\newblock {\em Usp.~Fiz.~Nauk}, 73:381, 1961.
\newblock [English transl.: {\it Soviet Phys. Usp.\/} 4:153, 1961].

\bibitem{landauandlifshitz}
L.~D. Landau and E.~M. Lifshitz.
\newblock {\em Electrodynamics of Continuous Media}.
\newblock Pergamon, Oxford, 1960.

\bibitem{sparnaayrev}
M.~J. Sparnaay.
\newblock In A.~Sarlemijn and M.~J. Sparnaay, editors, {\em {Physics in the
  Making: Essays on Developments in 20th Century Physics in Honour of H.B.G.
  Casimir on the Occasion of his 80th Birthday}}, page 235, Amsterdam, 1989.
  {North-Holland}.

\bibitem{casimirandpolder}
H.~B.~G. Casimir and D.~Polder.
\newblock {\em Phys. Rev.}, 73:360, 1948.

\bibitem{casimir50}
H.~B.~G. Casimir.
\newblock In M.~Bordag, editor, {\em {The Casimir Effect 50 Years Later: The
  Proceedings of the Fourth Workshop on Quantum Field Theory Under the
  Influence of External Conditions, Leipzig, 1998}}, page~3, Singapore, 1999.
  {World Scientific}.

\bibitem{sabisky}
E.~S. Sabisky and C.~H. Anderson.
\newblock {\em Phys. Rev. A}, 7:790, 1973.

\bibitem{boyersphere}
T.~H. Boyer.
\newblock {\em Phys. Rev.}, 174:1764, 1968.

\bibitem{casimir2}
H.~B.~G. Casimir.
\newblock {\em Physica}, 19:846, 1956.

\bibitem{davis}
B.~Davies.
\newblock {\em J. Math. Phys.}, 13:1324, 1972.

\bibitem{balian}
R.~Balian and B.~Duplantier.
\newblock {\em Ann. Phys. (N.Y.)}, 112:165, 1978.

\bibitem{Jaffe:2003mb}
R.~L. Jaffe and A.~Scardicchio.
\newblock {\em Phys. Rev. Lett.}, 92:070402, 2004.
\newblock [arXiv:quant-ph/0310194].

\bibitem{mildersch}
K.~A. Milton, {L. L. DeRaad, Jr.}, and J.~Schwinger.
\newblock {\em Ann. Phys. (N.Y.)}, 115:388, 1978.

\bibitem{schwinger}
J.~Schwinger.
\newblock {\em Lett. Math. Phys.}, 1:43, 1975.

\bibitem{johnsondpf}
K.~Johnson.
\newblock In B.~Margolis and D.~G. Stairs, editors, {\em Particles and Fields
  1979}, page 353, New York, 1980. AIP.

\bibitem{miltonbag}
K.~A. Milton.
\newblock {\em Phys. Rev. D}, 22:1441, 1980.

\bibitem{miltoncond}
K.~A. Milton.
\newblock {\em Phys. Lett. B}, 104:49, 1981.

\bibitem{miltonbook}
K.~A. Milton.
\newblock {\em The Casimir Effect: Physical Manifestations of Zero-Point
  Energy}.
\newblock World Scientific, Singapore, 2001.

\bibitem{Graham:2003ib}
N.~Graham, R.L. Jaffe, V.~Khemani, M.~Quandt, O.~Schroeder, and H.~Weigel.
\newblock {\em Nucl. Phys. B}, 677:379, 2004.
\newblock [arXiv:hep-th/0309130].

\bibitem{sauer}
F.~Sauer.
\newblock PhD thesis, {G\"ottingen}, 1962.

\bibitem{mehra}
J.~Mehra.
\newblock {\em Physica}, 37:145, 1967.

\bibitem{Svet}
{V. B. Svetovoy and M. V. Lokhanin}.
\newblock {\em Mod. Phys. Lett. A}, 15:1437, 2000.
\newblock [arXiv:quant-ph/0008074].

\bibitem{Svet2}
{V. B. Svetovoy and M. V. Lokhanin}.
\newblock {\em Phys. Lett. A}, 280:177, 2001.
\newblock [arXiv:quant-ph/0101124].

\bibitem{bostrom}
{M. Bostr\"om and Bo E. Sernelius}.
\newblock {\em Phys. Rev. Lett.}, 84:4757, 2000.

\bibitem{bostrom2}
{M. Bostr\"om and Bo E. Sernelius}.
\newblock {\em Phys. Rev. A}, 61:052703, 2000.

\bibitem{sernelius01}
B.~E. Sernelius.
\newblock {\em Phys. Rev. Lett.}, 87:139102, 2001.

\bibitem{sernelius01a}
{B. E. Sernelius and M. Bostr\"om}.
\newblock {\em Phys. Rev. Lett.}, 87:259101, 2001.

\bibitem{klim2}
{M. Bordag, B. Geyer, G. L. Klimchitskaya, and V. M. Mostepanenko}.
\newblock {\em Phys. Rev. Lett.}, 85:503, 2000.
\newblock [arXiv:quant-ph/0003021].

\bibitem{Bordag:2001as}
M.~Bordag, B.~Geyer, G.~L. Klimchitskaya, and V.~M. Mostepanenko.
\newblock {\em Phys. Rev. Lett.}, 87:259102, 2001.

\bibitem{genet}
C.~Genet, A.~Lambrecht, and S.~Reynaud.
\newblock {\em Phys. Lett. A}, 62:012110, 2000.
\newblock [arXiv:quant-ph/0002061].

\bibitem{lamoreaux01}
S.~Lamoreaux.
\newblock {\em Phys. Rev. Lett.}, 87:139101, 2001.

\bibitem{klim01}
G.~L. Klimchitskaya and V.~M. Mostepanenko.
\newblock {\em Phys. Rev. A}, 63:062108, 2001.
\newblock [arXiv:quant-ph/0101128].

\bibitem{Brevik:2002bi}
I.~Brevik, J.~B. Aarseth, and J.~S. {H\o ye}.
\newblock {\em Phys. Rev. E}, 66:026119, 2002.
\newblock [arXiv:quant-ph/0201137].

\bibitem{bezerra66}
V.~B. Bezerra, G.~L. Klimchitskaya, and V.~M. Mostepanenko.
\newblock {\em Phys. Rev. A}, 66:062112, 2002.
\newblock [arXiv:quant-ph/0210209].

\bibitem{genetijmpa}
C.~Genet, A.~Lambrecht, and S.~Reynaud.
\newblock {\em Int. J. Mod. Phys. A}, 17:761, 2002.
\newblock [arXiv:quant-ph/0111162].

\bibitem{reynaudqfext}
S.~Reynaud, A.~Lambrecht, and C.~Genet.
\newblock In K.~A. Milton, editor, {\em Proceedings of the 6th Workshop on
  Quantum Field Theory Under the Influence of External Conditions}, Paramus,
  NJ, 2004. Rinton Press.
\newblock [arXiv:quant-ph/0312224].

\bibitem{itamp}
{ITAMP Workshop 2002: Casimir Forces: Recent Developments in Experiment and
  Theory}.
\newblock http://itamp.harvard.edu/casimir.html.

\bibitem{QFEXT03}
K.~A. Milton, editor.
\newblock {\em Proceedings of the 6th Workshop on Quantum Field Theory Under
  the Influence of External Conditions}, Paramus, NJ, 2004. Rinton Press.

\bibitem{mostbook}
V.~M. Mostepanenko and N.~N. Trunov.
\newblock {\em The Casimir Effect and its Applications}.
\newblock Oxford Science Publications, Oxford, 1997.

\bibitem{krech}
M.~Krech.
\newblock {\em Casimir Effect in Critical Systems}.
\newblock World Scientific, Singapore, 1994.

\bibitem{milonni}
P.~Milonni.
\newblock {\em The Quantum Vacuum: An Introduction to Quantum Electrodynamics}.
\newblock Academic Press, Boston, 1994.

\bibitem{Od}
{E. Elizalde, S. D. Odintsov, A. Romeo, A. A. Bytsenko, and S. Zerbini}.
\newblock {\em Zeta Regularization Techniques with Applications}.
\newblock World Scientific, {Singapore}, 1994.

\bibitem{El}
{E. Elizalde}.
\newblock {\em Ten Physical Applications of Spectral Zeta Functions}.
\newblock Springer, {Berlin}, 1995.

\bibitem{kirstenbook}
K.~Kirsten.
\newblock {\em Spectral Functions in Mathematics and Physics}.
\newblock Chapman and Hall/CRC, Boca Raton, 2002.

\bibitem{Vassilevich:2003xt}
D.~V. Vassilevich.
\newblock {\em Phys. Rept.}, 388:279, 2003.
\newblock [arXiv:hep-th/0306138].

\bibitem{Milton:2002vm}
K.~A. Milton.
\newblock {\em Phys. Rev. D}, 68:065020, 2003.
\newblock [arXiv:hep-th/0210081].

\bibitem{hennig}
M.~Bordag, D.~Hennig, and D.~Robaschik.
\newblock {\em J. Phys. A}, 25:4483, 1992.

\bibitem{bkv}
{M. Bordag, K. Kirsten, and D. Vassilevich}.
\newblock {\em Phys. Rev. D}, 59:085011, 1999.
\newblock [arXiv:hep-th/9811015].

\bibitem{graham2}
N.~Graham, R.L. Jaffe, V.~Khemani, M.~Quandt, M.~Scandurra, and H.~Weigel.
\newblock {\em Nucl. Phys. B}, 645:49, 2002.
\newblock [arXiv:hep-th/0207120].

\bibitem{Graham:2002fw}
N.~Graham, R.L. Jaffe, V.~Khemani, M.~Quandt, M.~Scandurra, and H.~Weigel.
\newblock {\em Phys. Lett.}, B572:196, 2003.
\newblock [arXiv:hep-th/0207205].

\bibitem{Milton:2004vy}
K.~A. Milton.
\newblock {\em J. Phys. A}, 2004.
\newblock {in press. [arXiv:hep-th/0401090]}.

\bibitem{kantowski}
R.~Kantowski and K.~A. Milton.
\newblock {\em Phys. Rev. D}, 35:549, 1987.

\bibitem{Brevik:2000hk}
I.~Brevik, B.~Jensen, and K.~A. Milton.
\newblock {\em Phys. Rev. D}, 64:088701, 2001.
\newblock [arXiv:hep-th/0004041].

\bibitem{luscher}
{M. L\"uscher, K. Symanzik, and P. Weisz}.
\newblock {\em Nucl. Phys. B}, 173:365, 1980.

\bibitem{luscher2}
{M. L\"uscher}.
\newblock {\em Nucl. Phys. B}, 180:317, 1981.

\bibitem{sundberg}
P.~Sundberg and R.~L. Jaffe.
\newblock {\em Ann. Phys. (N.Y.)}, 309:442, 2004.
\newblock [arXiv:hep-th/0308010].

\bibitem{johnson}
K.~Johnson.
\newblock {\em Acta Phys. Pol.}, B6:865, 1975.

\bibitem{Jaffe:2003ji}
R.~L. Jaffe.
\newblock {\em AIP Conf. Proc.}, 687:3, 2003.
\newblock [arXiv:hep-th/0307014].

\bibitem{Fulling:2003zx}
S.~A. Fulling.
\newblock {\em J. Phys. A}, 36:6529, 2003.
\newblock [arXiv:quant-ph/0302117].

\bibitem{Weigel:2003tp}
H.~Weigel.
\newblock In K.~A. Milton, editor, {\em Proceedings of the 6th Workshop on
  Quantum Field Theory Under the Influence of External Conditions}, {Princeton,
  N.J.}, 2004. Rinton Press.
\newblock [arXiv:hep-th/0310301].

\bibitem{Graham:2002yr}
N.~Graham and K.~D. Olum.
\newblock {\em Phys. Rev. D}, 67:085014, 2003.
\newblock [arXiv:hep-th/0211244].

\bibitem{Olum:2002ra}
K.~D. Olum and N.~Graham.
\newblock {\em Phys. Lett. B}, 554:175, 2003.
\newblock [arXiv:gr-qc/0205134].

\bibitem{ccj}
C.~G. {Callan, Jr., S. Coleman, and R. Jackiw}.
\newblock {\em Ann. Phys. (N.Y.)}, 59:42, 1970.

\bibitem{ce}
{J. Schwinger, L. L. DeRaad, Jr., K. A. Milton, and W.-y. Tsai}.
\newblock {\em Classical Electrodynamics}.
\newblock Perseus Books, {Reading, Massachusetts}, 1998.

\bibitem{henkel}
C.~Henkel, K.~Joulain, J.-Ph. Mulet, and J.-J. Greffet.
\newblock {\em Phys. Rev. A}, 69:023808, 2004.
\newblock [arXiv:physics/0308095].

\bibitem{lambrecht}
{A. Lambrecht and S. Reynaud}.
\newblock {\em Eur. Phys. J. D}, 8:309, 2000.
\newblock [arXiv:quant-ph/9907105].

\bibitem{genetx}
C.~Genet, F.~Intravaia, A.~Lambrecht, and S.~Reynaud.
\newblock {\em Ann. Fond. L. de Broglie}, 29:311, 2004.
\newblock [arXiv:quant-ph/0302072].

\bibitem{schdermil}
J.~Schwinger, {L. L. DeRaad, Jr.}, and K.~A. Milton.
\newblock {\em Ann. Phys. (N.Y.)}, 115:1, 1978.

\bibitem{sonokm2}
K.~A. Milton and Y.~J. Ng.
\newblock {\em Phys. Rev. E}, 57:5504, 1998.
\newblock [arXiv:hep-th/9707122].

\bibitem{london}
F.~London.
\newblock {\em Z. Physik}, 63:245, 1930.

\bibitem{buhmann}
S.~Y. Buhmann, H.~T. Dung, L.~{Kn\"oll}, and D.~G. Walsh.
\newblock [arXiv:quant-ph/0403128].

\bibitem{sukenik}
C.~I. Sukenik, M.~G. Boshier, D.~Cho, V.~Sundoghar, and G.~A. Hinds.
\newblock {\em Phys. Rev. Lett.}, 70:560, 1993.

\bibitem{bartonatomplates}
{G. Barton}.
\newblock {\em Proc. Roy. Soc. London}, 410:175, 1987.

\bibitem{Bordag:2004dn}
M.~Bordag.
\newblock 2004.
\newblock [arXiv:hep-th/0403222].

\bibitem{hu04}
B.~L. Hu, A.~Roura, and S.~Shresta.
\newblock [arXiv:quant-ph/0401188].

\bibitem{babb04}
J.~F. Babb, G.~L. Klimchitskaya, and V.~M. Mostepanenko.
\newblock [arXiv:quant-ph/0405163].

\bibitem{noguez04}
C.~Noguez and C.~E. {Rom\'an-Vel\'azquez}.
\newblock [arXiv:quant-ph/0312009].

\bibitem{kampen}
{N. G. van Kampen, B. R. A. Nijboer, and K. Schram}.
\newblock {\em Phys. Lett. A}, 26:307, 1968.

\bibitem{gerlach}
E.~Gerlach.
\newblock {\em Phys. Rev. B}, 4:393, 1971.

\bibitem{noguezqfext}
C.~Noguez and C.~E. {Rom\'an-Vel\'azquez}.
\newblock In K.~A. Milton, editor, {\em Proceedings of the 6th Workshop on
  Quantum Field Theory Under the Influence of External Conditions}, Paramus,
  NJ, 2004. Rinton Press.
\newblock [arXiv:quant-ph/0312090].

\bibitem{noguez68}
C.~Noguez, C.~E. {Rom\'an-Vel\'azquez}, R.~Esquivel-Sirvent, and C.~Villareal.
\newblock [arXiv:quant-ph/0310068].

\bibitem{fordqfext}
L.~H. Ford and V.~Sopova.
\newblock In K.~A. Milton, editor, {\em Proceedings of the 6th Workshop on
  Quantum Field Theory Under the Influence of External Conditions}, Paramus,
  NJ, 2004. Rinton Press.
\newblock [arXiv:quant-ph/0204125].

\bibitem{fordpra}
L.~Ford.
\newblock {\em Phys. Rev. A}, 58:4279, 1998.
\newblock [arXiv:quant-ph/9804055].

\bibitem{chen04}
F.~Chen, G.~L. Klimchitskaya, U.~Mohideen, and V.~M Mostepanenko.
\newblock {\em Phys. Rev. A}, 69:022117, 2004.
\newblock [arXiv:quant-ph/0401153].

\bibitem{reynaudepl}
C.~Genet, A.~Lambrecht, P.~M. Neto, and S.~Reynaud.
\newblock {\em Europhys. Lett.}, 62:484, 2003.
\newblock [arXiv:quant-ph/0302071].

\bibitem{mohideen}
U.~Mohideen and A.~Roy.
\newblock {\em Phys. Rev. Lett.}, 81:4549, 1998.
\newblock [arXiv:physics/9805038].

\bibitem{mohideen2}
{A. Roy, C.-Y. Lin, and U. Mohideen}.
\newblock {\em Phys. Rev. D}, 60:R111101, 1999.
\newblock [arXiv:quant-ph/9906062].

\bibitem{mohideen2a}
{B. W. Harris, F. Chen, and U. Mohideen}.
\newblock {\em Phys. Rev. A}, 62:052109, 2000.
\newblock [arXiv:quant-ph/0005088].

\bibitem{decca03}
{R. S. Decca, D. L\'opez, E. Fischbach, and D. E. Krause}.
\newblock {\em Phys. Rev. Lett.}, 91:050402, 2003.
\newblock [arXiv:quant-ph/0306136].

\bibitem{hargreaves}
C.~M. Hargreaves.
\newblock {\em Proc. Kon. Ned. Akad. Wetensch. B}, 68:231, 1965.

\bibitem{mostturn}
V.~M. Mostepanenko and N.~N. Trunov.
\newblock {\em Sov. J. Nucl. Phys.}, 42:812, 1985.

\bibitem{bezerra00}
V.~B. Bezerra, G.~L. Klimchitskaya, and V.~M. Mostepanenko.
\newblock {\em Phys. Rev. A}, 62:014102, 2000.
\newblock [arXiv:quant-ph/9912090].

\bibitem{lambrecht2}
{A. Lambrecht, M.-T. Jaekel, and S. Reynaud}.
\newblock {\em Phys. Lett. A}, 225:188, 1997.
\newblock [arXiv:quant-ph/9801055].

\bibitem{Kenneth:2002ij}
O.~Kenneth, I.~Klich, A.~Mann, and M.~Revzen.
\newblock {\em Phys. Rev. Lett.}, 89:033001, 2002.
\newblock [arXiv:quant-ph/0202114].

\bibitem{boyerunusual}
T.~H. Boyer.
\newblock {\em Phys. Rev. A}, 9:2078, 1974.

\bibitem{commentonrepulsion}
D.~Iannuzzi and F.~Capasso.
\newblock {\em Phys. Rev. Lett.}, 91:029101, 2003.
\newblock [arXiv:quant-ph/0305065].

\bibitem{klichqfext}
I.~Klich.
\newblock In K.~A. Milton, editor, {\em Proceedings of the 6th Workshop on
  Quantum Field Theory Under the Influence of External Conditions}, Paramus,
  NJ, 2004. Rinton Press.

\bibitem{Hoye:2002at}
J.~S. {H\o ye}, I.~Brevik, J.~B. Aarseth, and K.~A. Milton.
\newblock {\em Phys. Rev. E}, 67:056116, 2003.
\newblock [arXiv:quant-ph/0212125].

\bibitem{Brevik:2003rg}
I.~Brevik, J.~B. Aarseth, J.~S. {H\o ye}, and K.~A. Milton.
\newblock In K.~A. Milton, editor, {\em Proceedings of the 6th Workshop on
  Quantum Field Theory Under the Influence of External Conditions}, Paramus,
  NJ, 2004. Rinton Press.
\newblock [arXiv:quant-ph/0311094].

\bibitem{ms}
P.~C. Martin and J.~Schwinger.
\newblock {\em Phys. Rev.}, 115:1342, 1959.

\bibitem{brevik02}
I.~Brevik, J.~B. Aarseth, and J.~S. {H\o ye}.
\newblock {\em Int. J. Mod. Phys. A}, 17:776, 2002.
\newblock [arXiv:quant-ph/0111037].

\bibitem{klim02}
G.~L. Klimchitskaya.
\newblock {\em Int. J. Mod. Phys. A}, 17:751, 2002.
\newblock [arXiv:quant-ph/0111023].

\bibitem{lamoreaux}
S.~K. Lamoreaux.
\newblock {\em Phys. Rev. Lett.}, 78:5, 1997.

\bibitem{lamoreauxqfext}
S.~Lamoreaux.
\newblock In K.~A. Milton, editor, {\em Proceedings of the 6th Workshop on
  Quantum Field Theory Under the Influence of External Conditions}, Paramus,
  NJ, 2004. Rinton Press.

\bibitem{genet03}
C.~Genet, A.~Lambrecht, and S.~Reynaud.
\newblock {\em Phys. Rev. A}, 67:043811, 2003.
\newblock [arXiv:quant-ph/0210174].

\bibitem{aarseth01}
J.~S. {H\o ye}, I.~Brevik, and J.~B. Aarseth.
\newblock {\em Phys. Rev. E}, 63:051101, 2001.

\bibitem{palik}
E.~D. Palik, editor.
\newblock {\em Handbook of Optical Constants of Solids}.
\newblock Academic Press, New York, 1998.

\bibitem{lambrecht3}
{A. Lambrecht and S. Reynaud}.
\newblock {\em Phys. Rev. Lett.}, 84:5672, 2000.
\newblock [arXiv:quant-ph/9912085].

\bibitem{handbook67}
E.~U. Condon and H.~Odishaw, editors.
\newblock {\em Handbook of Physics}.
\newblock McGraw-Hill, New York, 1967.
\newblock {Eq.~(6.12)}.

\bibitem{resist}
{M. Khoshenevisan, {W. P. Pratt, Jr.}, P. A. Schroeder, and S. D. Steenwyk}.
\newblock {\em Phys. Rev. B}, 19:3873, 1979.

\bibitem{sernelius03}
Bo~E. Sernelius and M.~{Bostr\"om}.
\newblock In K.~A. Milton, editor, {\em Proceedings of the 6th Workshop on
  Quantum Field Theory Under the Influence of External Conditions}, Paramus,
  NJ, 2004. Rinton Press.

\bibitem{bezerra65}
V.~B. Bezerra, G.~L. Klimchitskaya, and V.~M. Mostepanenko.
\newblock {\em Phys. Rev. A}, 65:052133, 2002.
\newblock [arXiv:quant-ph/0202018].

\bibitem{geyer03}
B.~Geyer, G.~L. Klimchitskaya, and V.~M. Mostepanenko.
\newblock {\em Phys. Rev. A}, 67:062102, 2003.
\newblock [arXiv:quant-ph/0306038].

\bibitem{mostqfext}
V.~M. Mostepanenko.
\newblock In K.~A. Milton, editor, {\em Proceedings of the 6th Workshop on
  Quantum Field Theory Under the Influence of External Conditions}, Paramus,
  NJ, 2004. Rinton Press.

\bibitem{klimqfext}
G.~L. Klimchitskaya.
\newblock In K.~A. Milton, editor, {\em Proceedings of the 6th Workshop on
  Quantum Field Theory Under the Influence of External Conditions}, Paramus,
  NJ, 2004. Rinton Press.

\bibitem{bezerra04}
V.~B. Bezerra, G.~L. Klimchitskaya, V.~M. Mostepanenko, and C.~Romero.
\newblock {\em Phys. Rev. A}, 69:022119, 2004.
\newblock [arXiv:quant-ph/0401138].

\bibitem{bezerra03}
V.~B. Bezerra, G.~L. Klimchitskaya, and V.~M. Mostepanenko.
\newblock 2003.
\newblock [arXiv:quant-ph/0306050].

\bibitem{bezerra02}
V.~B. Bezerra, G.~L. Klimchitskaya, and C.~Romero.
\newblock {\em Phys. Rev. A}, 65:012111, 2002.
\newblock [arXi:quant-ph/0110128].

\bibitem{svetovoy03}
V.~B. Svetovoy and M.~B. Lokhanin.
\newblock {\em Phys. Rev. A}, 67:022113, 2003.
\newblock [arXiv:quant-ph/0301035].

\bibitem{esquivel02}
{W. L. Moch\' an, C. Villareal, and R. Esquivel-Sirvent}.
\newblock {\em Rev. Mex. Fis.}, 48:339, 2002.
\newblock [arXiv:quant-ph/0206119].

\bibitem{svet174}
V.~B. Svetovoy.
\newblock [arXiv:quant-ph/0306174].

\bibitem{torgerson1}
J.~R. Torgerson and S.~K. Lamoreaux.
\newblock [arXiv:quant-ph/0309153].

\bibitem{torgerson}
J.~R. Torgerson and S.~K. Lamoreaux.
\newblock [arXiv:quant-ph/0208042].

\bibitem{esqsvet}
R.~Esquivel and V.~B. Svetovoy.
\newblock [arXiv:quant-ph/0404073].

\bibitem{lifandpit}
E.~M. Lifshitz and L.~P. Pitaevskii.
\newblock {\em Physical Kinetics}.
\newblock Pergamon Press, Oxford, 1981.

\bibitem{abrikosov}
A.~A. Abrikosov.
\newblock {\em Fundamentals of the Theory of Metals}.
\newblock North Holland, Amsterdam, 1988.

\bibitem{esquivel}
{R. Esquivel, C. Villareal, and M. L. Moch\'an}.
\newblock {\em Phys. Rev. A}, 68:052103, 2003.
\newblock [arXiv:quant-ph/0306139].

\bibitem{esquivelqfext}
R.~Esquivel-Sirvent and M.~L. {Moch\'an}.
\newblock In K.~A. Milton, editor, {\em Proceedings of the 6th Workshop on
  Quantum Field Theory Under the Influence of External Conditions}, Paramus,
  NJ, 2004. Rinton Press.

\bibitem{blocki}
{J. Blocki, J. Randrup, W. J. \'Swi\c{a}tecki, and C. F. Tsang}.
\newblock {\em Ann. Phys. (N.Y.)}, 105:427, 1977.

\bibitem{abrikosova}
{I. I. Abrikosova and B. V. Deriagin (Derjaguin)}.
\newblock {\em Dokl. Akad. Nauk SSSR}, 90:1055, 1953.

\bibitem{deriagin}
{B. V. Deriagin (Derjaguin) and I. I. Abrikosova}.
\newblock {\em Zh. Eksp. Teor. Fiz.}, 30:993, 1956.
\newblock [English transl.: {\it Soviet Phys. JETP\/} 3:819, 1957].

\bibitem{derjaguin}
B.~V. Derjaguin, I.I. Abrikosova, and E.~M. Lifshitz.
\newblock {\em Quart. Rev.}, 10:295, 1956.

\bibitem{derpt}
{B. V. Deryagin (Derjaguin)}.
\newblock {\em Kolloid Z.}, 69:155, 1934.

\bibitem{derpt2}
{B. V. Deryagin (Derjaguin) \etal}.
\newblock {\em J. Colloid. Interface Sci.}, 53:314, 1975.

\bibitem{lamoreaux2}
S.~K. Lamoreaux.
\newblock {\em Phys. Rev. Lett.}, 81:5475(E), 1998.

\bibitem{Emig:2002xz}
T.~Emig.
\newblock {\em Europhys. Lett.}, 62:466--472, 2003.
\newblock [arXiv:cond-mat/0206585].

\bibitem{emigqfext}
T.~Emig.
\newblock In K.~A. Milton, editor, {\em Quantum Field Theory Under the
  Influence of External Conditions}, Paramus, NJ, 2004. Rinton Press.
\newblock [arXiv:cond-mat/0311465].

\bibitem{Buscher:2004tb}
R.~{B\"uscher} and T.~Emig.
\newblock 2004.
\newblock [arXiv:cond-mat/0401451].

\bibitem{mohideen3}
A.~Roy and U.~Mohideen.
\newblock {\em Phys. Rev. Lett.}, 82:4380, 1999.

\bibitem{mohideenlat}
F.~Chen, U.~Mohideen, G.~L. Klimchitskaya, and V.~M. Mostepanenko.
\newblock {\em Phys. Rev. Lett.}, 88:101801, 2002.
\newblock [arXiv:quant-ph/0201087].

\bibitem{mohideenlat2}
F.Chen, U.~Mohideen, G.~L. Klimchitskaya, and V.~M. Mostepanenko.
\newblock {\em Phys. Rev. A}, 66:032113, 2002.
\newblock [arXiv:quant-ph/0209167].

\bibitem{schaden}
{M. Schaden, L. Spruch, and F. Zhou}.
\newblock {\em Phys. Rev. A}, 57:1108, 1998.

\bibitem{schaden1}
{M. Schaden and L. Spruch}.
\newblock {\em Phys. Rev. A}, 58:935, 1998.

\bibitem{schaden2}
{M. Schaden and L. Spruch}.
\newblock {\em Phys. Rev. Lett.}, 84:459, 2000.

\bibitem{gutzweiler}
M.~C. Gutzweiler.
\newblock {\em J. Math. Phys.}, 12:343, 1971.

\bibitem{gutzbook}
M.~C. Gutzweiler.
\newblock {\em Chaos in Classical and Quantum Mechanics}.
\newblock Springer, Berlin, 1990.

\bibitem{balianandbloch3}
R.~Balian and C.~Bloch.
\newblock {\em Ann. Phys. (N.Y.)}, 60:401, 1970.

\bibitem{balianandbloch4}
R.~Balian and C.~Bloch.
\newblock {\em Ann. Phys. (N.Y.)}, 63:592, 1971.

\bibitem{balianandbloch5}
R.~Balian and C.~Bloch.
\newblock {\em Ann. Phys. (N.Y.)}, 69:76, 1972.

\bibitem{fullingpo}
S.~A. Fulling.
\newblock {\em J. Phys. A}, 35:4049, 2002.
\newblock [arXiv:quant-ph/0012070].

\bibitem{fullingqfext}
S.~A. Fulling.
\newblock In K.~A. Milton, editor, {\em Proceedings of the 6th Workshop on
  Quantum Field Theory Under the Influence of External Conditions}, Paramus,
  NJ, 2004. Rinton Press.

\bibitem{gies}
H.~Gies, K.~Langfeld, and L.~Moyaerts.
\newblock {\em JHEP}, 0306:018, 2003.
\newblock [arXiv:hep-th/0303264].

\bibitem{moyaerts}
L.~Moyaerts, K.~Langeld, and H.~Gies.
\newblock In K.~A. Milton, editor, {\em Proceedings of the 6th Workshop on
  Quantum Field Theory Under the Influence of External Conditions}, {Paramus,
  NJ}, 2004. Rinton Press.
\newblock [arXiv:hep-th/0311168].

\bibitem{graham}
N.~Graham, R.~L. Jaffe, and H.~Weigel.
\newblock {\em Int. J. Mod. Phys.}, A17:846, 2002.
\newblock [arXiv:hep-th/0201148].

\bibitem{mazzitelli}
F.~D. Mazzitelli, M.~J. {S\'anchez}, N.~Scoccola, and J.~Von Stecher.
\newblock {\em Phys. Rev. A}, 67:013807, 2003.
\newblock [arXiv:quant-ph/0209097].

\bibitem{mazzitelliqfext}
F.~D. Mazzitelli.
\newblock In K.~A. Milton, editor, {\em Proceedings of the 6th Workshop on
  Quantum Field Theory Under the Influence of External Conditions}, Paramus,
  NJ, 2004. Rinton Press.
  
\bibitem{kitchener}
A.~Kitchener and A.~P. Prosser.
\newblock {\em Proc. Roy. Soc. (London) A}, 242:403, 1957.

\bibitem{sparnaay}
M.~Y. Sparnaay.
\newblock {\em Physica}, 24:751, 1958.

\bibitem{black}
W.~Black, J.~G.~V. de~Jongh, J.~Th.~G. Overbeck, and M.~J. Sparnaay.
\newblock {\em Trans. Faraday Soc.}, 56:1597, 1960.

\bibitem{silfhout}
A.~van Silfhout.
\newblock {\em Proc. Kon. Ned. Akad. Wetensch. B}, 69:501, 1966.

\bibitem{tabor0}
D.~Tabor and R.~H.~S. Winterton.
\newblock {\em Nature}, 219:1120, 1968.

\bibitem{tabor}
D.~Tabor and R.~H.~S. Winterton.
\newblock {\em Proc. Roy. Soc. (London) A}, 312:435, 1969.

\bibitem{winterton}
R.~H.~S. Winterton.
\newblock {\em Contemp. Phys.}, 11:559, 1970.

\bibitem{israelachivili}
J.~N. Israelachivili and D.~Tabor.
\newblock {\em Proc. Roy. Soc. (London) A}, 331:19, 1972.

\bibitem{israelrev}
J.~N. Israelachivili.
\newblock {\em Intermolecular and Surface Forces}.
\newblock Academic, London, 1991.

\bibitem{mohcom}
{U. Mohideen and A. Roy}.
\newblock {\em Phys. Rev. Lett.}, 83:3341, 1999.

\bibitem{lamoreaux3}
S.~K. Lamoreaux.
\newblock {\em Phys. Rev. A}, 59:R3149, 1999.

\bibitem{lamreply}
{S. K. Lamoreaux}.
\newblock {\em Phys. Rev. Lett.}, 84:5673, 2000.

\bibitem{ederth}
T.~Ederth.
\newblock {\em Phys. Rev. A}, 62:062104, 2000.
\newblock [arXiv:quant-ph/0008009].

\bibitem{klim4}
{M. Bordag, B. Geyer, G. L. Klimchitskaya, and V. M. Mostepanenko}.
\newblock {\em Phys. Rev. D}, 58:075003, 1998.
\newblock [arXiv:hep-ph/9804223].

\bibitem{klim}
{G. L. Klimchitskaya, A. Roy, U. Mohideen, and V. M. Mostepanenko}.
\newblock {\em Phys. Rev. A}, 60:3487, 1999.
\newblock [arXiv:quant-ph/9906033].

\bibitem{belllabs}
H.~B. Chan, V.~A. Aksyuk, R.~N. Kleiman, D.~J. Bishop, and F.~Capasso.
\newblock {\em Science}, 291:1941, 2001.
\newblock (10.1126/science.1057984).

\bibitem{belllabs2}
{H. B. Chan, V. A. Aksyuk, R. N. Kleiman, D. J. Bishop, and F. Capasso}.
\newblock {\em Phys. Rev. Lett.}, 87:211801, 2001.
\newblock [arXiv:quant-ph/0109046].

\bibitem{klim5}
{G. L. Klimchitskaya, U. Mohideen, and V. M. Mostepanenko}.
\newblock {\em Phys. Rev. A}, 61:062107, 2000.
\newblock [arXiv:quant-ph/0003093].

\bibitem{maradudin}
A.~A. Maradudin and P.~Mazur.
\newblock {\em Phys. Rev. B}, 22:1677, 1980.

\bibitem{bezerra}
{V. B. Bezerra, G. L. Klimchitskaya, and C. Romero}.
\newblock {\em Mod. Phys. Lett. A}, 12:2613, 1997.

\bibitem{bressi}
{G. Bressi, G. Carugno, R. Onofrio, and G. Ruoso}.
\newblock {\em Phys. Rev. Lett.}, 88:041804, 2002.
\newblock [arXiv:quant-ph/0203002].

\bibitem{onofrioqfext}
R.~Onofrio.
\newblock In K.~A. Milton, editor, {\em Proceedings of the 6th Workshop on
  Quantum Field Theory Under the Influence of External Conditions}, Paramus,
  NJ, 2004. Rinton Press.

\bibitem{decca03a}
{R. S. Decca, E. Fischbach, G. L. Klimchitskaya, D. E. Krause, D. L\'opez, and
  V. M. Mostepanenko}.
\newblock {\em Phys. Rev. D}, 68:116003, 2003.
\newblock [arXiv:hep-ph/0310157].

\bibitem{iannuzzi}
D.~Iannuzzi.
\newblock In K.~A. Milton, editor, {\em Quantum Field Theory Under the
  Influence of External Conditions}, New Jersey. Rinton Press.
\newblock [arXiv:quant-ph/0312043].

\bibitem{harvard}
D.~Iannuzzi, M.~Lisanti, and F.~Capasso.
\newblock {\em Proc. Nat. Acad. Sci. USA}, 101:4019, 2004.
\newblock [arXiv:quant-ph/0403142].

\bibitem{bostrom99}
M.~{Bostr\"om} and Bo~E. Sernelius.
\newblock {\em Phys. Rev. B}, 61:2204, 1999.

\bibitem{sernelius00}
{M. Bostr\"om and Bo E. Sernelius}.
\newblock {\em Phys. Rev. A}, 61:046101, 2000.

\bibitem{svetovoyqfext}
V.~B. Svetovoy.
\newblock In K.~A. Milton, editor, {\em Quantum Field Theory Under the
  Influence of External Conditions}, {Paramus, New Jersey}, 2004. Rinton Press.
\newblock [arXiv:cond-mat/0401562].

\bibitem{chenqfext}
F.~Chen, U.~Mohideen, and P.~W. Milonni.
\newblock In K.~A. Milton, editor, {\em Proceedings of the 6th Workshop on
  Quantum Field Theory Under the Influence of External Conditions}, Paramus,
  NJ, 2004. Rinton Press.

\bibitem{chen03}
F.~Chen, G.~L. Klimchitskaya, U.~Mohideen, and V.~M. Mostepanenko.
\newblock {\em Phys. Rev. Lett.}, 90:160404, 2003.
\newblock [arXiv:quant-ph/0302149].

\bibitem{krauseqfext}
D.~E.~Krause \etal.
\newblock In K.~A. Milton, editor, {\em Proceedings of the 6th Workshop on
  Quantum Field Theory Under the Influence of External Conditions}, Paramus,
  NJ, 2004. Rinton Press.

\bibitem{deccaqfext}
R.~S. Decca.
\newblock In K.~A. Milton, editor, {\em Proceedings of the 6th Workshop on
  Quantum Field Theory Under the Influence of External Conditions}, Paramus,
  NJ, 2004. Rinton Press.

\bibitem{baessler}
S.~Baessler.
\newblock {Invited talk given at DPF 2004, May 1, 2004, abstract B9.004}.

\bibitem{nesvizhevsky}
V.~V.~Nesvizhevsky \etal.
\newblock {\em Phys. Rev. D}, 67:102002, 2003.
\newblock [arXiv:hep-ph/0306198].

\bibitem{Nesvizhevsky:2004qb}
V.~V. Nesvizhevsky and K.~V. Protasov.
\newblock 2004.
\newblock [arXiv:hep-ph/0401179].

\bibitem{chumak}
A.~A. Chumak, P.~W. Milonni, and G.~P. Berman.
\newblock [arXiv:cond-mat/0310081].

\bibitem{stipe}
B.~L. Stipe, H.~J. Mamin, T.~D. Stowe, T.~W. Kenny, and D.~Rugar.
\newblock {\em Phys. Rev. Lett.}, 87:096801, 2001.

\bibitem{lin}
{Y.-j.} Lin, I.~Teper, C.~Chin, and V.~{Vuleti\'c}.
\newblock {\em Phys. Rev. Lett.}, 92:050404, 2004.
\newblock [arXiv:cond-mat/0308457].

\bibitem{Romeo:2000wt}
A.~Romeo and A.~A. Saharian.
\newblock {\em J. Phys. A}, 35:1297, 2002.
\newblock [arXiv:hep-th/0007242].

\bibitem{saharian2}
A.~A. Saharian.
\newblock {\em Phys. Rev. D}, 63:125007, 2001.
\newblock [arXiv:hep-th/0012185].

\bibitem{saharian3}
A.~Romeo and A.~A. Saharian.
\newblock {\em Phys. Rev. D}, 63:105019, 2001.
\newblock [arXiv:hep-th/0101155].

\bibitem{brown}
L.~S. Brown and G.~J. Maclay.
\newblock {\em Phys. Rev.}, 184:1272, 1969.

\bibitem{Actor:1996zj}
A.~A. Actor and I.~Bender.
\newblock {\em Fortsch. Phys.}, 44:281, 1996.

\bibitem{dowkerandkennedy}
J.~S. Dowker and G.~Kennedy.
\newblock {\em J. Phys. A}, 11:895, 1978.

\bibitem{deutsch}
D.~Deutsch and P.~Candelas.
\newblock {\em Phys. Rev. D}, 20:3063, 1979.

\bibitem{brevly}
I.~Brevik and M.~Lygren.
\newblock {\em {Ann. Phys. (N.Y.)}}, 251:157, 1996.

\bibitem{sopovaqfext}
V.~Sopova and L.~H. Ford.
\newblock In K.~A. Milton, editor, {\em Proceedings of the 6th Workshop on
  Quantum Field Theory Under the Influence of External Conditions}, Paramus,
  NJ, 2004. Rinton Press.

\bibitem{grahamqfext}
N.~Graham.
\newblock In K.~A. Milton, editor, {\em Proceedings of the 6th Workshop on
  Quantum Field Theory Under the Influence of External Conditions}, Paramus,
  NJ, 2004. Rinton Press.

\bibitem{miltonballs}
K.~A. Milton.
\newblock {\em Ann. Phys. (N.Y.)}, 127:49, 1980.

\bibitem{candelas}
P.~Candelas.
\newblock {\em Ann. Phys. (N.Y.)}, 143:241, 1982.

\bibitem{candelas2}
P.~Candelas.
\newblock {\em Ann. Phys. (N.Y.)}, 167:257, 1986.

\bibitem{Bernasconi}
F.~Bernasconi, G.M. Graf, and D.~Hasler.
\newblock {\em Ann. Henri {Poincar\'e}}, 4:1001, 2003.
\newblock [arXiv:math-ph/0302035].

\bibitem{sen}
S.~Sen.
\newblock {\em Phys. Rev. D}, 24:869, 1981.

\bibitem{sen2}
S.~Sen.
\newblock {\em J. Math. Phys.}, 22:2968, 1981.

\bibitem{barton03}
G~Barton.
\newblock {\em J. Phys. A}, 37:1011, 2004.

\bibitem{Scandurra:1998xa}
M.~Scandurra.
\newblock {\em J. Phys. A}, 32:5679, 1999.
\newblock [arXiv:hep-th/9811164].

\bibitem{benmil}
C.~M. Bender and K.~A. Milton.
\newblock {\em Phys. Rev. D}, 50:6547, 1994.
\newblock [arXiv:hep-th/9406048].

\bibitem{lesed1}
S.~Leseduarte and A.~Romeo.
\newblock {\em Europhys. Lett.}, 34:79, 1996.

\bibitem{lesed2}
S.~Leseduarte and A.~Romeo.
\newblock {\em Ann. Phys. (N.Y.)}, 250:448, 1996.
\newblock [arXiv:hep-th/9605022].

\bibitem{klich}
I.~Klich.
\newblock {\em Phys. Rev. D}, 61:025004, 2000.
\newblock [arXiv:hep-th/9908101].

\bibitem{Bordag:2004rx}
M.~Bordag and D.~V. Vassilevich.
\newblock 2004.
\newblock [arXiv:hep-th/0404069].

\bibitem{miltonfermion}
K.~A. Milton.
\newblock {\em Phys. Rev. D}, 22:1444, 1980.

\bibitem{mildim}
K.~A. Milton.
\newblock {\em Phys. Rev. D}, 55:4940, 1997.
\newblock [arXiv:hep-th/9611078].

\bibitem{lesedflux}
S.~Leseduarte and A.~Romeo.
\newblock {\em Commun. Math. Phys.}, 193:317, 1998.
\newblock [arXiv:hep-th/9612116].

\bibitem{sonokm}
K.~A. Milton and Y.~J. Ng.
\newblock {\em Phys. Rev. E}, 55:4207, 1997.
\newblock [arXiv:hep-th/9607186].

\bibitem{bmm}
{I. Brevik, V. N. Marachevsky, and K. A. Milton}.
\newblock {\em Phys. Rev. Lett.}, 82:3948, 1999.
\newblock [arXiv:hep-th/9810062].

\bibitem{barton}
G.~Barton.
\newblock {\em J. Phys. A}, 32:525, 1999.

\bibitem{hb}
{J. S. H\o ye and I. Brevik}.
\newblock {\em J. Stat. Phys.}, 100:223, 2000.
\newblock [arXiv:quant-ph/9903086].

\bibitem{sonorev02}
M.~P. Brenner, S.~Hilgenfeldt, and D.~Lohse.
\newblock {\em Rev. Mod. Phys.}, 74:425, 2002.

\bibitem{scar}
{V. V. Nesterenko and G. Lambiase and and G. Scarpetta}.
\newblock {\em Phys. Rev.D}, 64:025013, 2001.
\newblock [arXiv:hep-th/0006121].

\bibitem{scar2}
V.~V. Nesterenko, G.~Lambiase, and G.~Scarpetta.
\newblock {\em Int. J. Mod. Phys. A}, 17:790, 2002.
\newblock [arXiv:hep-th/0111242].

\bibitem{vvn}
V.~V. Nesterenko.
\newblock In {\em {Proceedings of the International Conference `I. Ya.
  Pomeranchuk and Physics at the Turn of Centuries,' Moscow, January 24--28,
  2003}}, Singapore. World Scientific.
\newblock [arXiv:hep-th/0310041].

\bibitem{lukosz}
W.~Lukosz.
\newblock {\em Physica}, 56:109, 1971.

\bibitem{lukosz1}
W.~Lukosz.
\newblock {\em Z. Phys.}, 258:99, 1973.

\bibitem{lukosz2}
W.~Lukosz.
\newblock {\em Z. Phys.}, 262:327, 1973.

\bibitem{ruggiero}
J.~R. Ruggiero, A.~H. Zimerman, and A.~Villani.
\newblock {\em Rev. Bras. Fis.}, 7:663, 1977.

\bibitem{ruggiero2}
J.~R. Ruggiero, A.~H. Zimerman, and A.~Villani.
\newblock {\em J. Phys. A}, 13:761, 1980.

\bibitem{ambjorn}
J.~Ambj{\o}rn and S.~Wolfram.
\newblock {\em Ann. Phys. (N.Y.)}, 147:1, 1983.

\bibitem{caruso2}
{F. Caruso and N. P. Neto and B. F. Svaiter and N. F. Svaiter}.
\newblock {\em Phys. Rev. D}, 43:1300, 1991.

\bibitem{caruso}
{F. Caruso, R. De Paola, N. F. Svaiter}.
\newblock {\em Int. J. Mod. Phys. A}, 14:2077, 1999.
\newblock [arXiv:hep-th/9807043].

\bibitem{actor3}
A.~A. Actor.
\newblock {\em Ann. Phys. (N.Y.)}, 230:303, 1994.

\bibitem{actor}
A.~A. Actor and I.~Bender.
\newblock {\em Phys. Rev. D}, 52:3581, 1995.

\bibitem{li}
X.~Li, H.~Cheng, and X.~Zhai.
\newblock {\em Phys. Rev. D}, 56:2155, 1997.

\bibitem{queiroz}
H.~Queiroz, J.~C. da~Silva, F.C. Khanna, M.~Revzen, and A.~E. Santana.
\newblock [arXiv:hep-th/0311246].

\bibitem{stratton}
J.~A. Stratton.
\newblock {\em Electromagnetic Theory}.
\newblock McGraw-Hill, New York, 1941.

\bibitem{deraadcyl}
{L. L. DeRaad, Jr. and K. A. Milton}.
\newblock {\em Ann. Phys. (N.Y.)}, 136:229, 1981.

\bibitem{nestcyl}
V.~V. Nesterenko and I.~G. Pirozhenko.
\newblock {\em J. Math. Phys.}, 41:4521, 2000.
\newblock [arXiv:hep-th/9910097].

\bibitem{scancyl}
M.~Scandurra.
\newblock {\em J. Phys. A}, 33:5707, 2000.
\newblock [arXiv:hep-th/0004051].

\bibitem{borpiro}
M.~Bordag and I.~G. Pirozhenko.
\newblock {\em Phys. Rev. D}, 64:025019, 2001.
\newblock [arXiv:hep-th/0102193].

\bibitem{dicyl}
{K. A. Milton, A. V. Nesterenko, and V. V. Nesterenko}.
\newblock {\em Phys. Rev. D}, 59:105009, 1999.
\newblock [arXiv:hep-th/9711168, v3].

\bibitem{jskmer}
J.~Schwinger and K.~A. Milton.
\newblock {\em Electromagnetic Radiation}.
\newblock Springer-Verlag, Berlin, 2004.
\newblock in preparation.

\bibitem{brevny}
I.~Brevik and G.~H. Nyland.
\newblock {\em {Ann. Phys. (N.Y.)}}, 230:321, 1994.

\bibitem{gos}
P.~Gosdzinsky and A.~Romeo.
\newblock {\em Phys. Lett. B}, 441:265, 1998.
\newblock [arXiv:hep-th/9809199].

\bibitem{klichromeo}
I.~Klich and A.~Romeo.
\newblock {\em Phys. Lett. B}, 476:369, 2000.
\newblock [arXiv:hep-th/9912223].

\bibitem{romeocomm}
A.~Romeo.
\newblock {\em private communication}, 1998.

\bibitem{chodos}
{A. Chodos, R. L. Jaffe, K. Johnson, C. B. Thorn, and V. Weisskopf}.
\newblock {\em Phys. Rev. D}, 9:3471, 1974.

\bibitem{chodos2}
A.~Chodos, R.~L. Jaffe, K.~Johnson, and C.~B. Thorn.
\newblock {\em Phys. Rev. D}, 10:2599, 1974.

\bibitem{chodosthorn}
A.~Chodos and C.~B. Thorn.
\newblock {\em Phys. Rev. D}, 12:2733, 1975.

\bibitem{degrand}
T.~DeGrand, R.~L. Jaffe, K.~Johnson, and J.~Kiskis.
\newblock {\em Phys. Rev. D}, 12:2060, 1975.

\bibitem{donoghue2}
J.~F. Donoghue, E.~Golowich, and B.~R. Holstein.
\newblock {\em Phys. Rev. D}, 12:2875, 1975.

\bibitem{schrock}
R.~E. Schrock and S.~B. Treiman.
\newblock {\em Phys. Rev. D}, 19:2148, 1979.

\bibitem{Adler:1978xi}
S.~L. Adler.
\newblock {\em Phys. Rev. D}, 17:3212, 1978.

\bibitem{Adler:1979we}
S.~L. Adler.
\newblock {\em Phys. Lett.}, B86:203, 1979.

\bibitem{Adler:1980ex}
S.~L. Adler.
\newblock {\em Phys. Rev.}, D21:550, 1980.

\bibitem{Adler:1981as}
S.~L. Adler.
\newblock {\em Phys. Rev.}, D23:2905, 1981.

\bibitem{Adler:1982pj}
S.~L. Adler and T.~Piran.
\newblock {\em Phys. Lett.}, B117:91, 1982.

\bibitem{Adler:1982rk}
S.~L. Adler and T.~Piran.
\newblock {\em Phys. Lett.}, B113:405, 1982.

\bibitem{savvidy}
{G. K. Savvidy}.
\newblock {\em Phys. Lett. B}, 71:113, 1977.

\bibitem{miltonfinite}
K.~A. Milton.
\newblock {\em Phys. Rev. D}, 27:439, 1983.

\bibitem{Rebhan:2004vz}
A.~Rebhan, P.~van Nieuwenhuizen, and R.~Wimmer.
\newblock In K.~A. Milton, editor, {\em Proceedings of the 6th Workshop on
  Quantum Field Theory Under the Influence of External Conditions}, Paramus,
  NJ, 2004. Rinton Press.
\newblock [arXiv:hep-th/0401127].

\bibitem{Goldhaber:2004kn}
A.~S. Goldhaber, A.~Rebhan, P.~van Nieuwenhuizen, and R.~Wimmer.
\newblock 2004.
\newblock [arXiv:hep-th/0401152].

\bibitem{rebhan04}
R.~Wimmer A.~Rebhan, P. van~Nieuwenhuizen.
\newblock [arXiv:hep-th/0404223].

\bibitem{graham0}
N.~Graham, R.~L. Jaffe, M.~Quandt, and H.~Weigel.
\newblock {\em Phys. Rev. Lett.}, 87:131601, 2001.
\newblock [arXiv:hep-th/0103010].

\bibitem{quandtqfext}
M.~Quandt.
\newblock In K.~A. Milton, editor, {\em Proceedings of the 6th Workshop on
  Quantum Field Theory Under the Influence of External Conditions}, Paramus,
  NJ, 2004. Rinton Press.
\newblock [arXiv:hep-th/0311094].

\bibitem{fahri2}
{E. Fahri, N. Graham, R. L. Jaffe, V. Khemani, and H. Weigel}.
\newblock {\em Nucl. Phys. B}, 665:623, 2003.
\newblock [arXiv:hep-th/0303159].

\bibitem{khemaniqfext}
V.~Khemani.
\newblock In K.~A. Milton, editor, {\em Proceedings of the 6th Workshop on
  Quantum Field Theory Under the Influence of External Conditions}, Paramus,
  NJ, 2004. Rinton Press.

\bibitem{bordagqfext}
M.~Bordag.
\newblock In K.~A. Milton, editor, {\em Proceedings of the 6th Workshop on
  Quantum Field Theory Under the Influence of External Conditions}, Paramus,
  NJ, 2004. Rinton Press.
\newblock [arXiv:hep-th/0310249].

\bibitem{nielsenolesen}
H.~B. Nielsen and P.~Olesen.
\newblock {\em Nucl. Phys. B}, 61:45, 1973.

\bibitem{juge}
K.~J. Juge, J.~Kuti, and C.~Morningstar.
\newblock In {\em International Conference on Color Confinement and Hadrons in
  Quantum Chromodynamics (Confinement 2003)}, Tokyo, 2003.
\newblock [arXiv:hep-lat/0401032].

\bibitem{juge2}
K.~J. Juge, J.~Kuti, and C.~Morningstar.
\newblock In {\em International Conference on Color Confinement and Hadrons in
  Quantum Chromodynamics (Confinement 2003)}, Tokyo, 2003.
\newblock [arXiv:hep-lat/0312019].

\bibitem{Luscher:2002qv}
M.~Luscher and P.~Weisz.
\newblock {\em JHEP}, 07:049, 2002.
\newblock [arXiv:hep-lat/0207003].

\bibitem{js}
J.~Schwinger.
\newblock {\em Proc. Natl. Acad. Sci. USA}, 90:958, 2105, 4505, 7285, 1993.

\bibitem{js2}
J.~Schwinger.
\newblock {\em Proc. Natl. Acad. Sci. USA}, 91:6473, 1994.

\bibitem{eberlein}
C.~Eberlein.
\newblock {\em Phys. Rev. A}, 53:2772, 1996.
\newblock [arXiv:quant-ph/9506024].

\bibitem{eberlein2}
C.~Eberlein.
\newblock {\em Phys. Rev. Lett.}, 76:3842, 1996.
\newblock [arXiv:quant-ph/9506023].

\bibitem{chodossono}
A.~Chodos.
\newblock In B.~Kursonolglu, S.~Mintz, and A.~Perlmutter, editors, {\em {Orbis
  Scientiae 1996, Miami Beach}}, New York, 1996. Plenum.
\newblock [arXiv:hep-ph/9604368].

\bibitem{chodossono2}
A.~Chodos and S.~Groff.
\newblock {\em Phys. Rev. E}, 59:3001, 1999.
\newblock [arXiv:hep-ph/9807512].

\bibitem{carlson}
{C. E. Carlson, C. Molina-Par\'\i s, J. P\'erez-Mercader, and M. Visser}.
\newblock {\em Phys. Lett. B}, 395:76, 1997.
\newblock [arXiv:hep-th/9609195].

\bibitem{carlson2}
{C. E. Carlson, C. Molina-Par\'\i s, J. P\'erez-Mercader, and M. Visser}.
\newblock {\em Phys. Rev. D}, 56:1262, 1997.
\newblock [arXiv:hep-th/9702007].

\bibitem{cmp}
{C. Molina-Par\'\i s and M. Visser}.
\newblock {\em Phys. Rev. D}, 56:6629, 1997.
\newblock [arXiv:hep-th/9707073].

\bibitem{liberati}
{ M. Visser, S. Liberati, F. Belgiorno, and D. W. Sciama}.
\newblock {\em Phys. Rev. Lett.}, 83:678, 1999.
\newblock [arXiv:quant-ph/9805023].

\bibitem{liberati2}
{S. Liberati, M. Visser, F. Belgiorno, and D. W. Sciama}.
\newblock {\em Phys. Rev. D}, 61:085023, 2000.
\newblock [arXiv:quant-ph/9904013].

\bibitem{liberati3}
{S. Liberati, M. Visser, F. Belgiorno, and D. W. Sciama}.
\newblock {\em Phys. Rev. D}, 61:085024, 2000.
\newblock [arXiv:quant-ph/9905034].

\bibitem{liberati4}
{S. Liberati, F. Belgiorno, M. Visser, and D. W. Sciama}.
\newblock {\em J. Phys. A}, 33:2251, 2000.
\newblock [arXiv:quant-ph/9805031].

\bibitem{sonorev}
{B. P. Barber, R. A. Hiller, R. L\"ofstedt, S. J. Putterman, and K. Weniger}.
\newblock {\em Phys. Rep.}, 281:65, 1997.

\bibitem{moore}
{G. T. Moore}.
\newblock {\em J. Math. Phys.}, 11:2679, 1970.

\bibitem{fulldavies}
{S. A. Fulling and P. C. W. Davies}.
\newblock {\em Proc. R. Soc. London, Ser. A}, 348:393, 1976.

\bibitem{daviesfull}
P.~C.~W. Davies and S.~A. Fulling.
\newblock {\em Proc. R. Soc. London, Ser. A}, 356:237, 1977.

\bibitem{unruh}
{W. G. Unruh}.
\newblock {\em Phys. Rev. D}, 14:870, 1976.

\bibitem{birrell}
N.~D. Birrell and P.~C.~W. Davies.
\newblock {\em Quantum Fields in Curved Space}.
\newblock Cambridge University Press, Cambridge, 1982.

\bibitem{hawking}
S.~W. Hawking.
\newblock {\em Nature}, 248:30, 1974.

\bibitem{scully03}
M.~O. Scully, V.~A. Kocharovsky, A.~Belyanin, E.~Fry, and F.~Capasso.
\newblock [arXiv:quant-ph/0305178].

\bibitem{hu04a}
B.~L. Hu and A.~Roura.
\newblock [arXiv:quant-ph/0402088].

\bibitem{huqfext}
C.~R. Galley, B.~L. Hu, and P.~R. Johnson.
\newblock In K.~A. Milton, editor, {\em Proceedings of the 6th Workshop on
  Quantum Field Theory Under the Influence of External Conditions}, Paramus,
  NJ, 2004. Rinton Press.
\newblock [arXiv:quant-ph/0402002].

\bibitem{crocce}
M.~Crocce, D.~A.~R. Dalvit, F.~C. Lombardo, and F.~D. Mazzitelli.
\newblock 2004.
\newblock [arXiv:quant-ph/0404135].

\bibitem{wegrzyn}
P.~{W\c egrzyn}.
\newblock 2004.
\newblock [arXiv:quant-ph/0312219].

\bibitem{wegrzyn1}
P.~{W\c egrzyn}.
\newblock {\em Mod. Phys. Lett. A}, 19:769, 2004.
\newblock [arXiv:quant-ph/0312220].

\bibitem{dodonov}
V.~V. Dodonov and A.~B. Klimov.
\newblock {\em Phys. Lett. A}, 167:309, 1992.

\bibitem{dodonov1}
V.~V. Dodonov, A.~B. Klimov, and D.~E. Kikonov.
\newblock {\em J. Math. Phys.}, 34:2742, 1993.

\bibitem{Antunes}
N.~D. Antunes.
\newblock [arXiv:hep-ph/0310131].

\bibitem{uhlmann}
M.~Uhlmann, G.~Plunien, R.~{Sch\"utzhold}, and G.~Soff.
\newblock 2004.
\newblock [arXiv:quant-ph/0404157].

\bibitem{weincccc}
S.~Weinberg.
\newblock {\em Phys. Rev. D}, 61:103505, 2000.
\newblock [arXiv: astro-ph/0002387].

\bibitem{zeldovich}
Ya.~B. Zeldovich.
\newblock {\em Uspekhi Fiz. Nauk}, 95:209, 1968.

\bibitem{riess}
{A. G. Riess \etal}.
\newblock {\em Astron. J.}, 116:1009, 1998.
\newblock [arXiv:astro-ph/9805201].

\bibitem{perlmutter}
{S. Perlmutter \etal}.
\newblock {\em Astrophys. J.}, 517:565, 1999.
\newblock [arXiv:astro-ph/9812133].

\bibitem{sn}
{R. A. Knop \etal}.
\newblock {\em Astrophys. J.}, 598:102, 2003.
\newblock {[arXiv:astro-ph/0309368]}.

\bibitem{Tonry:2003zg}
{J. L. Tonry \etal}.
\newblock {\em Astrophys. J.}, 594:1, 2003.
\newblock [arXiv:astro-ph/0305008].

\bibitem{Riess:2004nr}
{A. G. Riess \etal}.
\newblock 2004.
\newblock [arXiv:astro-ph/0402512].

\bibitem{wmap}
{D.N. Spergel \etal (WMAP)}.
\newblock {\em Astrophys. J. Suppl.}, 148:175, 2003.
\newblock [arXiv:astro-ph/0302209].

\bibitem{Tegmark:2003ud}
{M. Tegmark \etal}.
\newblock {\em Phys. Rev. D}, 69:103501, 2003.
\newblock [arXiv:astro-ph/0310723].

\bibitem{Boughn:2004zm}
S.~P. Boughn and R.~G. Crittenden.
\newblock {\em Nature}, 427:45, 2004.
\newblock {[arXiv:astro-ph/0404470] and references therein}.

\bibitem{Wang:2004gq}
Yun Wang.
\newblock 2004.
\newblock [arXiv:astro-ph/0404484].

\bibitem{Wang:2004py}
Yun Wang and M.~Tegmark.
\newblock 2004.
\newblock [arXiv:astro-ph/0403292].

\bibitem{pad}
T.~Padmanabhan.
\newblock {\em Phys. Rep.}, 380:235, 2003.
\newblock [arXiv:gr-qc/0311036].

\bibitem{peebles}
P.~J.~E. Peebles and B.~Ratra.
\newblock {\em Rev. Mod. Phys.}, 75:559, 2003.
\newblock [arXiv:astro-ph/0207347].

\bibitem{dolgov}
A.~D. Dolgov.
\newblock [arXiv:hep-ph/0405089].

\bibitem{Milton:2001np}
K.~A. Milton, R.~Kantowski, C.~Kao, and Y.~Wang.
\newblock {\em Mod. Phys. Lett.}, A16:2281, 2001.
\newblock [arXiv:hep-ph/0105250].

\bibitem{Milton:2002hx}
K.~A. Milton.
\newblock {\em Grav. Cosmol.}, 9:66, 2003.
\newblock [arXiv:hep-ph/0210170].

\bibitem{kaluza}
T.~Kaluza.
\newblock {\em Sitz. Preuss. Akad. Wiss. Phys. Math.}, K1:966, 1921.

\bibitem{klein}
O.~Klein.
\newblock {\em Nature}, 118:516, 1926.

\bibitem{klein2}
O.~Klein.
\newblock {\em Z. Phys.}, 37:895, 1926.

\bibitem{add}
N.~Arkani-Hamed, S.~Dimopoulos, and G.~Dvali.
\newblock {\em Phys. Lett. B}, 429:263, 1998.
\newblock [arXiv:hep-ph/9803315].

\bibitem{add2}
N.~Arkani-Hamed, S.~Dimopoulos, G.~Dvali, and N.~Kaloper.
\newblock {\em Phys. Rev. Lett.}, 84:586, 2000.
\newblock [arXiv:hep-th/9907209].

\bibitem{rsundrum}
L.~Randall and R.~Sundrum.
\newblock {\em Phys. Rev. Lett.}, 83:4690, 1999.
\newblock [arXiv:hep-th/9906064].

\bibitem{long}
{J. C. Long, H. W. Chan, and J. C. Price}.
\newblock {\em Nucl. Phys. B}, 539:23, 1999.
\newblock [arXiv:hep-ph/9805217].

\bibitem{long2}
{J. C. Long, A. B. Churnside, and J. C. Price}.
\newblock In {\em {Proceedings of 9th Marcel Grossmann Meeting on Recent
  Developments in Theoretical and Experimental General Relativity, Gravitation
  and Relativistic Field Theories (MG 9), Rome, Italy, 2--9 July, 2000}}.
\newblock [arXiv:hep-ph/0009062].

\bibitem{hoyle}
{C. D. Hoyle, U. Schmidt, B. R. Heckel, E. G. Adelberger, J. H. Gundlach, D. J.
  Kapner, and H. E. Swanson}.
\newblock {\em Phys. Rev. Lett.}, 86:1418, 2001.
\newblock [arXiv:hep-ph/0011014].

\bibitem{Adelberger:2003zx}
E.~G. Adelberger, B.~R. Heckel, and A.~E. Nelson.
\newblock {\em Ann. Rev. Nucl. Part. Sci.}, 53:77, 2003.
\newblock [arXiv:hep-ph/0307284].

\bibitem{Long:2003ta}
J.~C. Long and J.~C. Price.
\newblock {\em Comptes Rendus Physique}, 4:337, 2003.
\newblock [arXiv:hep-ph/0303057].

\bibitem{boulder}
J.~C. Long, H.~W. Chan, A.~B. Churnside, E.~A. Gulbis, M.~C.~M. Varney, and
  J.~C. Price.
\newblock {\em Nature}, 421:922, 2003.
\newblock [arXiv:hep-ph/0210004].

\bibitem{appel}
T.~Appelquist and A.~Chodos.
\newblock {\em Phys. Rev. Lett.}, 50:141, 1983.

\bibitem{appel2}
T.~Appelquist and A.~Chodos.
\newblock {\em Phys. Rev. D}, 28:772, 1983.

\bibitem{candandwein}
P.~Candelas and S.~Weinberg.
\newblock {\em Nucl. Phys. B}, 237:397, 1984.

\bibitem{kantowski2}
R.~Kantowski and K.~A. Milton.
\newblock {\em Phys. Rev. D}, 36:3712, 1987.

\bibitem{birkanmil}
D.~Birmingham, R.~Kantowski, and K.~A. Milton.
\newblock {\em Phys. Rev. D}, 38:1809, 1988.

\bibitem{vilkovisky}
G.~A. Vilkovisky.
\newblock {\em Nucl. Phys. B}, 234:125, 1984.

\bibitem{vilk}
G.~A. Vilkovisky.
\newblock In S.~C. Christensen, editor, {\em Quantum Theory of Gravity},
  Bristol, England, 1984. Hilger.

\bibitem{barvinsky}
A.~O. Barvinsky and G.~A. Vilkovisky.
\newblock {\em Phys. Rep.}, 119:1, 1985.

\bibitem{dewitt2}
B.~S. DeWitt.
\newblock In I.~A. Batalin, C.~J. Isham, and G.~A. Vilkovisky, editors, {\em
  Quantum Field Theory and Quantum Statistics}, Bristol, England, 1987. Hilger.

\bibitem{Odintsov:1991yx}
S.~D. Odintsov.
\newblock {\em Phys. Lett.}, B262:394--397, 1991.

\bibitem{cho}
H.~T. Cho and R.~Kantowski.
\newblock {\em Phys. Rev. D}, 62:124003, 2000.
\newblock [hep-th/0004082].

\bibitem{Kaplan:2000hh}
D.~B. Kaplan and M.~B. Wise.
\newblock {\em JHEP}, 08:037, 2000.
\newblock [arXiv:hep-ph/0008116].

\bibitem{Banks:1988je}
T.~Banks.
\newblock {\em Nucl. Phys.}, B309:493, 1988.

\bibitem{Beane:1995sk}
S.~R. Beane.
\newblock {\em Phys. Lett.}, B358:203, 1995.
\newblock [arXiv:hep-ph/9502226].

\bibitem{Beane:1997it}
S.~R. Beane.
\newblock {\em Gen. Rel. Grav.}, 29:945--951, 1997.
\newblock [arXiv:hep-ph/9702419].

\bibitem{Bauer:2003mh}
F.~Bauer, M.~Lindner, and G.~Seidl.
\newblock 2003.
\newblock [arXiv:hep-th/0309200].

\bibitem{Steinhardt:2003st}
P.~J. Steinhardt.
\newblock {\em Phil. Trans. Roy. Soc. Lond.}, A361:2497, 2003.

\bibitem{krechanddont}
D.~Dantchev and M.~Krech.
\newblock [arXiv:cond-mat/0402238].

\bibitem{zandi}
R.~Zandi, J.~Rudnick, and M.~Kardar.
\newblock [arXiv:cond-mat/0404309].

\bibitem{williams}
G.~A. Williams.
\newblock {\em Phys. Rev. Lett.}, 92:197003, 2004.
\newblock [arXiv:cond-mat/0307125].

\bibitem{larraza}
A.~Larraza and B.~Denardo.
\newblock {\em Phys. Lett. A}, 248:151, 1998.

\bibitem{larraza2}
A.~Larraza, C.~D. Holmes, R.~T. Susbilla, and B.~Denardo.
\newblock {\em J. Acoust. Soc. Am.}, 103:276, 1998.

\bibitem{barcenas}
J.~{B\'arcenas}, L.~Reyes, and R.~Esquivel-Sirvent.
\newblock [arXiv:quant-ph/0405106].

\bibitem{dowker04}
J.~S. Dowker.
\newblock [arXiv:hep-th/0404093].

\bibitem{polonyi}
J.~Polonyi and E.~{Reg\"os}.
\newblock [arXiv:hep-th/0404185].

\bibitem{brevikcv}
I.~Brevik.
\newblock [arXiv:gr-qc/0404095].

\bibitem{entropy}
I.~Brevik, K.~A. Milton, and S.~D. Odintsov.
\newblock {\em Ann. Phys. (N.Y.)}, 302:120, 2002.
\newblock [arXiv:hep-th/0202048].

\bibitem{mottolaqfext}
P.~O. Mazur and E.~Mottola.
\newblock In K.~A. Milton, editor, {\em Proceedings of the 6th Workshop on
  Quantum Field Theory Under the Influence of External Conditions}, Paramus,
  NJ, 2004. Rinton Press.
\newblock [arXiv:gr-qc/0405111].

\bibitem{mottola01}
P.~O. Mazur and E.~Mottola.
\newblock [arXiv:gr-qc/0109035].

\bibitem{nojiri}
I.~Brevik, K.~A. Milton, S.~Nojiri, and S.~D. Odintsov.
\newblock {\em Nucl. Phys. B}, 599:305, 2001.
\newblock [arXiv:hep-th/0010205].

\bibitem{Ichinose:2004mt}
S.~Ichinose and A.~Murayama.
\newblock 2004.
\newblock [arXiv:hep-th/0401015].

\bibitem{bordag}
M.~Bordag, D.~Robaschik, and E.~Wieczorek.
\newblock {\em Ann. Phys. (N.Y.)}, 165:192, 1985.

\bibitem{robaschik}
{D. Robaschik, K Scharnhorst, and E. Wieczorek}.
\newblock {\em Ann. Phys. (N.Y.)}, 174:401, 1987.

\bibitem{Barone:2003nk}
F.~A. Barone, R.~M. Cavalcanti, and C.~Farina.
\newblock 2003.
\newblock [arXiv:hep-th/0312169].

\bibitem{Barone:2003rn}
F.~A. Barone, R.~M. Cavalcanti, and C.~Farina.
\newblock {\em Nucl. Phys. Proc. Suppl.}, 127:118, 2004.
\newblock [arXiv:hep-th/0306011].

\bibitem{radbord}
M.~Bordag and J.~Lindig.
\newblock {\em Phys. Rev. D}, 58:045003, 1998.
\newblock [arXiv:hep-th/9801129].

\end{thebibliography}

\end{document}